
\input harvmac
\input amssym.def
\input amssym
\baselineskip 14pt
\magnification\magstep1
\font\bigcmsy=cmsy10 scaled 1500
\parskip 6pt
\newdimen\itemindent \itemindent=32pt
\def\textindent#1{\parindent=\itemindent\let\par=\resetpar%
\indent\llap{#1\enspace}\ignorespaces}

\let\oldpar=\par
\def\resetpar{\oldpar\parindent=20pt\let\par=\oldpar}

\font\ninerm=cmr9 \font\ninesy=cmsy9
\font\eightrm=cmr8 \font\sixrm=cmr6
\font\eighti=cmmi8 \font\sixi=cmmi6
\font\eightsy=cmsy8 \font\sixsy=cmsy6
\font\eightbf=cmbx8 \font\sixbf=cmbx6
\font\eightit=cmti8
\def\eightpoint{\def\rm{\fam0\eightrm}
  \textfont0=\eightrm \scriptfont0=\sixrm \scriptscriptfont0=\fiverm
  \textfont1=\eighti  \scriptfont1=\sixi  \scriptscriptfont1=\fivei
  \textfont2=\eightsy \scriptfont2=\sixsy \scriptscriptfont2=\fivesy
  \textfont3=\tenex   \scriptfont3=\tenex \scriptscriptfont3=\tenex
  \textfont\itfam=\eightit  \def\it{\fam\itfam\eightit}%
  \textfont\bffam=\eightbf  \scriptfont\bffam=\sixbf
  \scriptscriptfont\bffam=\fivebf  \def\bf{\fam\bffam\eightbf}%
  \normalbaselineskip=9pt
  \setbox\strutbox=\hbox{\vrule height7pt depth2pt width0pt}%
  \let\big=\eightbig  \normalbaselines\rm}
\catcode`@=11 %
\def\eightbig#1{{\hbox{$\textfont0=\ninerm\textfont2=\ninesy
  \left#1\vbox to6.5pt{}\right.\n@@space$}}}
\def\vfootnote#1{\insert\footins\bgroup\eightpoint
  \interlinepenalty=\interfootnotelinepenalty
  \splittopskip=\ht\strutbox %
  \splitmaxdepth=\dp\strutbox %
  \leftskip=0pt \rightskip=0pt \spaceskip=0pt \xspaceskip=0pt
  \textindent{#1}\footstrut\futurelet\next\fo@t}
\catcode`@=12 %
\def \de{\delta}
\def \De{\Delta}
\def \si{\sigma}

\def \ga{\gamma}

\def \pr{\partial}
\def \tr{{\rm tr }}
\def \ta{{\tilde a}}
\def \tf{{\tilde f}}

\def \hht{{\hat t}}

\def \J{{\rm J}}

\def \X{{\rm X}}

\def \bta{{\bar \eta}}

\def \l{\big \langle}
\def \r{\big \rangle}
\def \ep{\epsilon}
\def \vep{\varepsilon}
\def \half{{\textstyle {1 \over 2}}}

\def \quar{{\textstyle {1 \over 4}}}
\def \ts{\textstyle}

\def \b{{\rm b}}
\def \d{{\rm d}}

\def \x{{\rm x}}
\def \y{{\rm y}}
\def \ta{{\tilde {\rm a}}}
\def \tb{{\tilde {\rm b}}}
\def \tv{{\tilde {\rm v}}}
\def \tx{{\tilde {\rm x}}}

\def \tT{{\tilde T}}
\def \A{{\cal A}}
\def \B{{\cal B}}
\def \C{{\cal C}}
\def \D{{\cal D}}
\def \E{{\cal E}}
\def \F{{\cal F}}
\def \G{{\cal G}}

\def \I{{\cal I}}
\def \J{{\cal J}}

\def \N{{\cal N}}
\def \O{{\cal O}}

\def \tb{{\tilde {\rm b}}}
\def \tx{{\tilde {\rm x}}}

\def \tsi{{\tilde \sigma}}

\def \b{{\rm b}}

\def \x{{\rm x}}
\def \y{{\rm y}}

\def \1{{\bar 1}}
\def \2{{\bar 2}}
\def \3{{\bar 3}}
\def \bth{{\bar\theta}}
\def \vphi{{\varphi}}
\def \Bsw{\!\mathrel{\hbox{\bigcmsy\char'056}}\!}
\def \Bse{\!\mathrel{\hbox{\bigcmsy\char'046}}\!}
\def \bchi{{\overline \chi}}
\def \bpsi{{\overline \psi}}
\def \bPsi{{\overline \Psi}}
\def \bPhi{{\overline \Phi}}
\def \brho{{\overline \rho}}
\def \btau{{\overline \tau}}
\def \bvphi{{\overline \vphi}}
\def \bep{{\bar \epsilon}}
\def \hep{{\hat {\epsilon}}}
\def \hla{{\hat {\lambda}}}
\def \hbep{{\hat {\bep}}}
\def \blam{{\overline \lambda}}
\def \bet{{\overline \eta}}
\def \of{{\overline f}}
\def \dal{{\dot \alpha}}
\def \dbe{{\dot \beta}}
\def \dga{{\dot \gamma}}
\def \dde{{\dot \delta}}
\def \dep{{\dot \epsilon}}
\def \deta{{\dot \eta}}
\def \bga{{\bar \gamma}}
\def \tx{{\tilde {\rm x}}}
\def \hom{{\hat{\omega}}}
\def \bom{{\bar{\omega}}}
\def \hbom{{\hat{\bom}}}
\def \bsi{\bar \sigma}
\def \hN{{\hat N}}
\def \oD{{\overline D}}
\hyphenation{non-renorm-alization}
\font \bigbf=cmbx10 scaled \magstep1

\lref\Olive{C. Montonen and D. Olive, {\it Magnetic Monopoles as Gauge 
Particles?}, Phys. Lett. B72 (1977) 117.}
\lref\hughtwo{J. Erdmenger and H. Osborn, {\it Conserved Currents and the 
Energy Momentum Tensor in Conformally Invariant Theories for General 
Dimensions}, Nucl. Phys. B483 (1997) 431, hep-th/9605009.}
\lref\hughone{H. Osborn and A. Petkou, {\it Implications of Conformal 
Invariance for Quantum Field Theories in $d>2$}
Ann. Phys. (N.Y.) {231} (1994) 311, hep-th/9307010.}
\lref\WCon{B.P. Conlong and P.C. West, {\it Anomalous dimensions of 
fields in a supersymmetric quantum field theory at a renormalization 
group fixed point}, J. Phys. A 26 (1993) 3325.}
\lref\HO{H. Osborn, {\it $\N=1$ Superconformal Symmetry in Four-Dimensional
Quantum Field Theory}, Ann. Phys. (N.Y.) 272 (1999) 243, hep-th/9808041.}
\lref\Sei{S. Lee, S. Minwalla, M. Rangamani and N. Seiberg, {\it Three-Point
Functions of Chiral Operators in $D=4$, $\N=4$ SYM at Large $N$},
Adv. Theor. Math. Phys.  2 (1998) 697, hep-th/9806074.}
\lref\LW{K. Lang and W. R\"uhl, Nucl. Phys. {B402} (1993) 573.}
\lref\Pet{A.C. Petkou, Ann. Phys. (N.Y.) 249 (1996) 180, hep-th/9410093.}
\lref\Fone{S. Ferrara, A.F. Grillo, R. Gatto and G. Parisi, Nucl. Phys. 
B49 (1972) 77\semi
S. Ferrara, A.F. Grillo, R. Gatto and G. Parisi, Nuovo Cimento 19A
(1974) 667.}
\lref\Ftwo{S. Ferrara, A.F. Grillo and R. Gatto, Ann. Phys. 76 (1973) 161.}
\lref\Dob{V.K. Dobrev, V.B. Petkova, S.G. Petrova and I.T. Todorov,
Phys. Rev. D13 (1976) 887.}
\lref\West{E. D'Hoker, D.Z. Freedman, S.D. Mathur, A. Matusis,
L. Rastelli, {\it in} The Many Faces of the Superworld, ed. M.A. Shifman,
hep-th/9908160.\semi
B. Eden, P.S. Howe, C. Schubert, E. Sokatchev and P.C. West,
Phys. Lett. B472 (2000) 323, hep-th/9910150\semi
B. Eden, P.S. Howe, E. Sokatchev and P.C. West, hep-th/0004102\semi
M. Bianchi and S. Kovacs, Phys. Lett. B468 (1999) 102, hep-th/9910016\semi
J. Erdmenger and M. P\'erez-Victoria, Phys. Rev. D62 (2000) 045008,
Nucl. Phys. B589 (2000) 3, hep-th/9912250\semi
B. Eden, C. Schubert and E. Sokatchev, Phys. Lett. B482 (2000) 309, 
hep-th/0003096\semi
E. D'Hoker, J. Erdmenger, D.Z. Freedman and M. P\'erez-Victoria, 
hep-th/0003218.}
\lref\Free{D.Z. Freedman, S.D. Mathur, A. Matsusis and L. Rastelli, 
Nucl. Phys. B546 (1999) 96, hep-th/9804058\semi
D.Z. Freedman, S.D. Mathur, A. Matsusis and L. Rastelli, 
Phys. Lett. B452 (1999) 61, hep-th/9808006\semi
E. D'Hoker and D.Z. Freedman, Nucl. Phys. B550 (1999) 261, hep-th/9811257\semi
E. D'Hoker and D.Z. Freedman, Nucl. Phys. B544 (1999) 612, hep-th/9809179.}
\lref\FreeI{E. D'Hoker, D.Z. Freedman and L. Rastelli, 
Nucl. Phys. B562 (1999) 395, hep-th/9905049.}
\lref\FreeD{E. D'Hoker, D.Z. Freedman, S.D. Mathur, A. Matsusis and 
L. Rastelli, Nucl. Phys. B562 (1999) 353, hep-th/9903196.}
\lref\AF{L. Andrianopoli and S. Ferrara, {\it On short and long $SU(2,2/4)$
multiplets in the AdS/CFT correspondence}, Lett. Math. Phys. 48 (1999) 145, 
hep-th/9812067.}
\lref\Short{S. Ferrara and A. Zaffaroni, {\it Superconformal Field Theories,
Multiplet Shortening, and the AdS${}_5$/SCFT${}_4$ Correspondence},
Proceedings of the Conf\'erence Mosh\'e Flato 1999, vol. 1, ed. G. Dito and 
D. Sternheimer, Kluwer Academic Publishers (2000), hep-th/9908163.}
\lref\Gonz{F. Gonzalez-Rey, I. Park and K. Schalm, {\it A note on four-point
functions of conformal operators in $N=4$ Super-Yang Mills},
Phys. Lett. B448 (1999) 37, hep-th/9811155.}
\lref\Eloop{B. Eden, P.S. Howe, C. Schubert, E. Sokatchev and P.C. West,
{\it Four-point functions in $N=4$ supersymmetric Yang-Mills theory at
two loops}, Nucl. Phys. B557 (1999) 355, hep-th/9811172; {\it Simplifications
of four-point functions in $N=4$ supersymmetric Yang-Mills theory at
two loops}, Phys. Lett. B466 (1999) 20, hep-th/9906051.}
\lref\Bia{M. Bianchi, S. Kovacs, G. Rossi and Y.S. Stanev, {\it On the
logarithmic behaviour in $\N=4$ SYM theory}, JHEP 9908 (1999) 020, 
hep-th/9906188.} 
\lref\Arut{G. Arutyunov and S. Frolov, {\it Four-point Functions of Lowest
Weight CPOs in $\N=4$ SYM${}_4$ in Supergravity Approximation},
Phys. Rev. D62 (2000) 064016, hep-th/0002170.}
\lref\Three{G. Arutyunov and S. Frolov, {\it Three-point function of the 
stress-tensor in the AdS/CFT correspondence}, Phys. Rev. D60 (1999) 026004,
hep-th/9901121.}
\lref\OPEN{G. Arutyunov, S. Frolov and A.C. Petkou, {\it Operator Product
Expansion of the Lowest Weight CPOs in $\N=4$ SYM${}_4$ at Strong Coupling},
Nucl. Phys. B586 (2000) 547, hep-th/0005182; (E) Nucl. Phys. B609 (2001) 539.}
\lref\OPEW{G. Arutyunov, S. Frolov and A.C. Petkou, {\it Perturbative and
instanton corrections to the OPE of CPOs in $\N=4$ SYM${}_4$},
Nucl. Phys. B602 (2001) 238, hep-th/0010137; (E) Nucl. Phys. B609 (2001) 540.}
\lref\one{F.A. Dolan and H. Osborn, {\it Implications of $\N=1$ Superconformal
Symmetry for Chiral Fields}, Nucl. Phys. B593 (2001) 599, hep-th/0006098.}
\lref\Dos{F.A. Dolan and H. Osborn, {\it Conformal four point functions
and the operator product expansion}, Nucl. Phys. B599 (2001) 459, 
hep-th/0011040.}
\lref\Hokone{E. D'Hoker, S.D. Mathur, A. Matsusis and L. Rastelli, {\it The 
Operator Product Expansion of $N=4$ SYM and the 4-point Functions of 
Supergravity}, Nucl. Phys. B589 (2000) 38, hep-th/9911222.}
\lref\Hoff{L.C. Hoffmann, A.C. Petkou and W. R\"uhl, 
Phys. Lett. B478 (2000) 320, hep-th/0002025\semi
L.C. Hoffmann, A.C. Petkou and W. R\"uhl, hep-th/0002154.}
\lref\Witt{E. Witten, Adv. Theor. Math. Phys. 2 (1998) 253, hep-th/9802150.}
\lref\HoffR{L. Hoffmann, L. Mesref and W. R\"uhl, Nucl. Phys. B589 (2000) 337.}
\lref\Howe{P.S. Howe, E. Sokatchev and P.C. West, {\it 3-Point Functions in 
$N=4$ Yang-Mills}, Phys. Lett. B444 (1998) 341, hep-th/9808162.}
\lref\Eden{B. Eden, P.S. Howe, A. Pickering, E. Sokatchev and P.C. West,
{\it Four-point functions in $N=2$ superconformal field theories},
Nucl. Phys. B581 (2000) 523, hep-th/0001138.}
\lref\Edent{B. Eden, A.C. Petkou, C. Schubert and E. Sokatchev, {\it Partial
non-renormalisation of the stress-tensor four-point function in $N=4$
SYM and AdS/CFT}, Nucl. Phys. B607 (2001) 191, hep-th/0009106.}
\lref\Ans{D. Anselmi, {\it The $N=4$ Quantum Conformal Algebra},
Nucl. Phys. B541 (1999) 369,  hep-th/9809192.}
\lref\HMR{L. Hoffmann, L. Mesref and W. R\"uhl, {\it Conformal partial wave
analysis of AdS amplitudes for dilaton-axion four-point functions},
Nucl. Phys. B608 (2001) 177,
hep-th/0012153.}
\lref\Class{M. G\"unaydin and N. Marcus, {\it The spectrum of the $S^5$
compactification of the chiral $N=2$, $D=10$ supergravity and the unitary
supermultiplets of $U(2,2/4)$}, Class. and Quantum Gravity, 2 (1985) L11.}
\lref\Intr{K. Intriligator, {\it Bonus Symmetries of ${\cal N} =4$ 
Super-Yang-Mills Correlation Functions via AdS Duality}, 
Nucl. Phys. B551 (1999) 575, hep-th/9811047.}
\lref\Bon{K. Intriligator and W. Skiba, {\it Bonus Symmetry and the Operator
Product Expansion of ${\cal N} =4$ Super-Yang-Mills},
Nucl. Phys. B559 (1999) 165, hep-th/9905020.}
\lref\Pick{A. Pickering and P. West, {\it Chiral Green's Functions in 
Superconformal Field Theory}, Nucl. Phys. B569 (2000) 303,
hep-th/9904076.}
\lref\Non{G. Arutyunov, B. Eden and E. Sokatchev, {\it On Non-renormalization
and OPE in Superconformal Field Theories}, Nucl. Phys. B619 (2001) 359, 
hep-th/0105254.}
\lref\bpsN{B. Eden and E. Sokatchev, {\it On the OPE of 1/2 BPS Short
Operators in $N=4$ SCFT${}_4$}, Nucl. Phys. B618 (2001) 259, hep-th/0106249.}
\lref\Hes{P.J. Heslop and P.S. Howe, {\it OPEs and 3-point correlators of
protected operators in $N=4$ SYM}, Nucl. Phys. B626 (2002) 265, hep-th/0107212.}
\lref\Except{G. Arutyunov, B. Eden, A.C. Petkou and E. Sokatchev, 
{\it Exceptional non-renorm-alization properties and OPE analysis of
chiral four-point functions in $\N=4$ SYM${}_4$}, Nucl. Phys. B620 (2002) 380, 
hep-th/0103230.}
\lref\BKRS{M. Bianchi, S. Kovacs, G. Rossi and Y.S. Stanev, {\it Properties
of the Konishi multiplet in  $\N=4$ SYM theory}, JHEP 0105 (2001) 042,
hep-th/0104016.}
\lref\Pen{S. Penati and A. Santambrogio, {\it Superspace approach to anomalous 
dimensions in ${\N}=4$ SYM}, Nucl. Phys. B614 (2001) 367, hep-th/0107071.}
\lref\Hok{A.V. Ryzhov, {\it Quarter BPS Operators in $\N=4$ SYM}, 
JHEP 0111 (2001) 046, hep-th/0109064\semi
E. D'Hoker and A.V. Ryzhov, {\it Three Point Functions of Quarter BPS 
Operators in $\N=4$ SYM}, JHEP 0202 (2002) 047, hep-th/0109065.}
\lref\HH{P.J. Heslop and P.S. Howe, {\it A note on composite operators in
$N=4$ SYM}, Phys. Lett. 516B (2001) 367, hep-th/0106238.}
\lref\Soh{M.F. Sohnius, {\it The multiplet of currents for $N=2$ extended
supersymmetry}, Phys. Lett. 81B (1979) 8.}

{\nopagenumbers
\rightline{DAMTP/01-82}
\rightline{hep-th/0112251}
\vskip 1.5truecm
\centerline {\bigbf Superconformal Symmetry, Correlation Functions}
\vskip  6pt
\centerline {\bigbf and the Operator Product Expansion}
\vskip 1.5 true cm
\centerline {F.A. Dolan and H. Osborn${}^\dagger$}

\vskip 12pt
\centerline {\ Department of Applied Mathematics and Theoretical Physics,}
\centerline {Silver Street, Cambridge, CB3 9EW, England}
\vskip 1.5 true cm

{\eightpoint
\parindent 1.5cm{

{\narrower\smallskip\parindent 0pt

Superconformal transformations are derived for the $\N=2,4$ 
supermultiplets corresponding to the simplest chiral primary operators.
These are applied to two, three and four point correlation functions.
When $\N=4$, results are obtained for the three point function of
various descendant operators, including the energy momentum tensor and
$SU(4)$ current. For both $\N=2$ or $4$ superconformal identities are 
derived for the functions of the two conformal invariants appearing in
the four point function for the chiral primary operator. These are solved 
in terms of a single arbitrary  function of the two conformal invariants 
and one or three single variable functions. The results are applied to the 
operator product expansion using the exact formula for the contribution 
of an operator in the operator product expansion in four dimensions
to a scalar four point function. Explicit expressions representing exactly
the contribution of both long and possible short supermultiplets to the 
chiral primary four point function are obtained. These are applied to
give the leading perturbative and large $N$ corrections to the scale 
dimensions of long supermultiplets.

PACS no: 11.25.Hf

Keywords: Superconformal symmetry, Chiral primary operators, 
Correlation functions, Operator product expansion.

\narrower}}

\vfill
\line{${}^\dagger$ 
address for correspondence: Trinity College, Cambridge, CB2 1TQ, England\hfill}
\line{\hskip0.2cm emails:
{{\tt fad20@damtp.cam.ac.uk} and \tt ho@damtp.cam.ac.uk}\hfill}
}

\eject}
\pageno=1

\newsec{Introduction}

Amongst four dimensional quantum field theories the $\N=4$ supersymmetric 
theories for any non abelian gauge group enjoy a very special status. As was 
surmised from their advent they are finite and have the maximal possible 
supersymmetry if gravity is excluded. For a gauge group $SU(n)$ the theory is 
parameterised by the coupling $g$ together with the  associated
vacuum angle $\theta$ and the moduli corresponding to non zero expectation
values of various scalar fields. In this case there are exact duality 
symmetries as was also conjectured long ago by Montonen and Olive \Olive. 
Conversely at the special point where all expectation values are zero and there
are no mass scales the theory has  $\N=4$ superconformal symmetry.

Despite its intrinsic interest a full analysis of the consequences of
$\N=4$ superconformal symmetry in four dimensions, which corresponds to the 
supergroup $PSU(2,2|4)$, has yet to be completed, primarily due to the
lack of a straightforward superfield formalism. Primarily for $\N=2$, 
superconformal identities for correlation functions of various operators
\refs{\Howe,\Eden,\Non,\bpsN,\Hes} have been obtained using harmonic 
superspace methods and such results have also been extended to the $\N=4$ case.
It is particularly desirable to explore fully the implications for four point
correlation functions since in this case the ordinary conformal group leaves
these undetermined up to one or more arbitrary functions of two 
conformal invariants $u,v$.

Previously \one, see also \Pick, we discussed the four point function of 
four $\N=1$ chiral superfields when eight linear differential equations
involving four functions of $u,v$ were derived. Taking into account
crossing symmetry constraints there was a unique solution up to an
overall constant. However this four point function was unphysical in that
for it to be non zero the scale dimensions of the fields had to be below
their unitarity bound. In general superconformal symmetry relates
correlation functions for the different fields in each supermultiplet which
are expressed in terms of  the correlation functions for the lowest weight 
(or dimension) operators in each supermultiplet.
For $\N=2,4$ superconformal symmetry there are potential
additional constraints on correlation functions of the lowest weight
operators themselves when some of these fields belong
to short supermultiplets of the superconformal group. Any
supermultiplet forming a representation of the superconformal group may be 
generated by the successive action of superconformal
transformations on a lowest weight operator with the scale dimension 
increasing  by $\half$ between each level. Here operators belonging to level $n$
are obtained by $n$ superconformal transformations acting on the lowest
weight operators. The operators present at each level are naturally classified  
in terms of representations of the $R$ symmetry and spin group. If $d$ is 
the number of lowest weight operators  with scale dimension $\Delta$,
corresponding to an irreducible representation of the $R$-symmetry group and 
spin group with dimension $d$, then 
in a generic long multiplet, without any restriction on $\Delta$, there are
$2^8d$ or $2^{16}d$, for $\N=2$ or $\N=4$, operators in the supermultiplet and 
$n_{\rm max} =8$ or $16$ respectively.  For the simplest BPS-like short 
supermultiplets one of the representations, which would be present in a general
multiplet, is missing at the first level, so that there are less than $8d$
or $16d$, for $\N=2$ or $\N=4$, operators with dimension $\Delta +\half$. 
Under the action of superconformal transformations there are consequential
restrictions on the representation content at higher levels leaving then a 
truncated multiplet with a reduced maximum level. 
Such short multiplets are present only if the lowest weight operator belongs 
to particular representations and has a corresponding value of $\Delta$ \Short.  

For $\N=4$, the AdS/CFT correspondence  may be applied most simply to obtain 
correlation functions for chiral primary operators which are scalar
fields transforming under the $SU(4)$  $R$-symmetry according to the
representations with Dynkin labels $[0,p,0]$, $p=2,3,\dots$, and having 
scale dimension $\Delta=p$. These are the lowest weight operators for short
supermultiplets whose total dimension is proportional to $2^8$ and 
$n_{\rm max} = 4$.  The existence of constraints at the first level 
under superconformal transformations leads to integrability conditions for four
point correlation functions of such chiral primary operators which become linear 
differential relations on the associated invariant functions of $u,v$.

In this paper we mostly consider $\N=4$ superconformal symmetry for
correlation functions involving the short supermultiplet for $p=2$, when
the representation $[0,2,0]$ has dimension 20, which is the simplest case 
of relevance. This example is of particular significance since this
supermultiplet contains the energy momentum tensor as well as the 
$R$-symmetry current. Even in this case the short multiplets are quite
large so in some cases we consider also examples with $\N=2$ and even
$\N=1$ superconformal symmetry. Our discussion is based on using directly
the superconformal transformation properties of the component fields in
each supermultiplet without any need for considering superspace. Although
in general superspace techniques provide considerable calculational
simplifications this is perhaps less evident for $\N=4$. The crucial
integrability conditions arise just by considering the action of
superconformal transformations on the lowest weight operators. However
for illustration we give the full superconformal transformations for
all fields in the $p=2$ supermultiplet since this allows us to obtain
some results for correlation functions involving the energy momentum
tensor itself.

For the four point function of four chiral primary operators with $p=2$
the integrability conditions necessary for superconformal symmetry 
give six linear differential relations on the initially six independent
functions of $u,v$ corresponding to the number of irreducible representations 
appearing in the tensor product $[0,2,0] \otimes [0,2,0]$. 
The corresponding result for $\N=2$ using harmonic
superspace was given in \Eden\ and this was extended to $\N=4$ for
the four point function of interest here in \Edent. Without imposing
any conditions following from crossing symmetry the linear partial
differential equations may be solved in terms of one arbitrary function $\G$
of $u,v$ as well as some single variable functions. Here we express these
as functions of $z,x$ which are related to $u,v$ by $u=zx, \, v=(1-z)(1-x)$.
The essential dynamics is then contained in $\G(u,v)$. 

It is our aim to relate the general analysis of superconformal invariance
for the four point function to the operator product expansion. Previous
discussions, both for weak coupling using ${\rm O}(g^2)$ perturbative
results \OPEW\ and also for strong coupling for large $N$ using results
from the AdS/CFT correspondence \OPEN, have considered the contributions in 
the operator product expansion to the four point function of
operators belonging to different $SU(4)$ representations independently. The 
various operators in the same supermultiplet must of course
have related scale dimensions and hence for a long supermultiplet
the same anomalous dimensions. However for operators of free dimension 4 
and higher in $\N=4$ supersymmetric gauge theories there are in general 
several with same  spin and belonging to the same $SU(4)$ representation. 
In such situations  there can then be a complicated mixing problem to solve 
before the independent anomalous dimensions can be disentangled. 
In this work we consider
the contribution of complete supermultiplets, in both long and short
representations, to the operator product expansion which thereby avoids
having to consider operator mixing unless there are degenerate
supermultiplets in the free theory. The contribution of a supermultiplet
in the operator product expansion to the four point correlation
function satisfies the necessary superconformal identities, although
not the crossing symmetry properties, of the full correlation function.
Using a basis for the six invariant functions of $u,v$, denoted by
$A_R(u,v)$, which correspond to contributions of operators in the operator
product expansion belonging to $SU(4)$ representations  labelled by their
dimensions $R=(1,15,20,84,105,175)$, then constraints arise since they each
may be expressed in terms of just the single function  $\G(u,v)$.

For application to long supermultiplets the essential observation is that, 
if the lowest weight operator is a $SU(4)$ singlet,
there is just one operator in the supermultiplet belonging to the
representation of dimension 105. It is convenient to therefore take
$A_{105}(u,v) = u^4 \G(u,v)$ which then gives the remaining $A_R(u,v)$ in
terms of $\G(u,v)$ multiplied by simple polynomials in $u,v$.
For a single long supermultiplet we take $A_{105}(u,v) = 
u^{{1\over 2}(\Delta+4-\ell)}G_{\Delta+4}^{(\ell)}(u,v)$, 
which represents the contribution of a single operator of dimension $\Delta+4$
and spin $\ell$ (so that the $SO(3,1)$ representation is $(j,j)$ where
$\ell=2j$), together with its derivatives, in the operator product expansion. 
Explicit simple formulae,
in four dimensions, for  $G_\Delta^{(\ell)}(u,v)$ were found by us \Dos\
earlier which are naturally expressed in terms of the variables $z,x$
mentioned above.  Using various recurrence relations, which may be derived
from this explicit result, the  other $A_R(u,v)$ may then be expressed as
a sum of contributions involving $G_{\Delta'}^{({\ell'})}(u,v)$,
for suitable $\Delta',\ell'$, corresponding to operators in the operator
product expansion with dimension $\Delta'$ and spin $\ell'$. Depending on
$R$, $\Delta'$ ranges from $\Delta$ to $\Delta+8$ and $\ell'$ from 
$\ell-4$ to $\ell+4$.  The associated operators
correspond exactly with those expected in a long supermultiplet where
the lowest weight operator is a singlet of dimension $\Delta$ and spin $\ell$.
A crucial consistency check is that the coefficients of all
$G_{\Delta'}^{({\ell'})}(u,v)$ are positive, as required by unitarity.

We also consider the role of the single variable functions of $z,x$ which
are present in the general solution of the superconformal identities as
well as $\G(u,v)$.
There are two such functions for the case of interest here and they are
related to the operator product expansion for various short multiplets which
have no anomalous dimensions.
This ties in nicely with the argument \Edent\ that these functions receive no
perturbative corrections beyond the free theory form. 

An alternative approach to discussing the compatibility of the operator product
expansion with superconformal symmetry, as expressed in the reduction of
the operator product contributions for a representation $R$ to the four 
point function $A_R$ in terms of a single unknown conformal invariant
$\G(u,v)$ were given in \Except. Our results are in accord with 
\Except\ but go further in
that they identify a special role for $A_{105}$ and make extensive use
of the explicit form for $G_\Delta^{(\ell)}(u,v)$ obtained in \Dos. In
our approach, besides the contributions of the well known \Short\ short
multiplets with lowest weight operators scalar fields belonging to the 
20, 105 and 84 dimensional representations, there are also in general
supermultiplets with lowest weight operators in the 20 and 15-representations
which have protected scale dimensions $\Delta=\ell+4$, for spin
$\ell=0,2, \dots$ and $\ell=1,3,\dots$ respectively. The existence of
a protected scalar operator with $\Delta=4$ contributing to the operator
product expansion for the four point function was shown in \refs{\OPEN,\OPEW}
and discussed further in \refs{\Except,\BKRS,\Pen}.

With the aid of this formalism it is easy to rederive and extend results
for both weak and strong coupling. Thus the Konishi supermultiplet and
its higher spin partners, which in free field theory have scale dimension
$\Delta = \ell+2$, $\ell=0,2\dots$, develop an anomalous dimension to 
first order in $g^2$,
\eqn\Konl{
\eta_{\ell} = 
{g^2N\over 2\pi^2} \sum_{j=1}^{\ell+2} {1\over j} \, ,
}
which extends the results of Anselmi \Ans\ for the first three cases. Our
results allow this to be extended to cases when the free field dimension
is $\Delta=\ell +2t, \, t=2,3, \dots$ when the anomalous dimensions are
${\rm O}(g^2N/N^2)$ for large $N$. Corresponding results are for large 
$g^2N$ and large $N$ using results \Arut\ from supergravity calculations
in the AdS/CFT correspondence for $\G(u,v)$. In this discussion we adopt
the simplest assumption of neglecting any possible mixing between
supermultiplets contributing to the operator product expansion having the 
same quantum numbers at zeroth order in $g^2$ or $1/N$. This requirement
is of course not valid for individual operators in 
$SU(N)$ $\N=4$ supersymmetric gauge theories but, without as yet a 
detailed analysis, does not seem to be invalid for complete supermultiplets.
If this does not hold, and further the assumption that the lowest
dimension operator in a long supermultiplet with $\Delta$ unrestricted is
a $SU(4)_R$ singlet needs modification, then analysis of the four point
function of chiral primary operators belonging to the $p=2$ supermultiplet
by itself is not sufficient to determine the full spectrum of operators
contributing to the operator product expansion. Our results are fully
consistent with $U(1)_Y$ bonus symmetry \refs{\Intr,\Bon}, where $U(1)_Y$
is an external automorphism of $PSU(2,2|4)$ which acts analogously to the
$U(1)_R$ $R$-symmetry group in $\N=1$ superconformal symmetry and also
to $U(1)\in U(2)_R$ for $\N=2$.

The arrangement of this paper is then as follows. In the next section,
following a simpler discussion for $\N=1,2$ superconformal symmetry
we list the fields belonging to the $p=2$ short supermultiplet containing
the energy momentum tensor and then describe their transformations under
$\N=4$ superconformal symmetry verifying the required closure of the
superconformal algebra. In section 3 we consider the two and three
point functions for this supermultiplet showing how these may be 
determined by using superconformal transformations in terms of the
two and three point functions for the lowest dimension chiral primary
operators. By successive superconformal transformations the three point
function of the energy momentum tensor is derived starting from the basic
three point function for three chiral scalars belonging to the 20 dimensional
representation. In section 4 the analysis is extended to the case when one of
the operators belongs to a long supermultiplet. We recover from
the superconformal transformation rule for a  lowest weight scalar field
belonging to a $SU(4)_R$ $[q,p,q]$ representation the usual conditions 
for multiplet shortening by determining the values of the dimension $\Delta$
when superconformal transformations on the lowest weight operator have
a non trivial cokernel at the first level.  Singlet operators
may contribute to the three point function for arbitrary scale dimension $\Delta$
but we show that for non singlet operators $\Delta$ is constrained.
In section 5 we obtain the necessary integrability
conditions on the four point function of chiral primary operators
which follow from $\N=2$ and $\N=4$ superconformal symmetry. The resulting
linear differential equations are solved in section 6 and used to
express the different $A_R$, which correspond to contributions from 
representation $R$ to the four point function, in terms of a single 
invariant function $\G(u,v)$. Sections 7 and 8 then describe the analysis
in terms of the operator product expansion making clear how the contributions
of long and various short supermultiplets may be separated. In accord
with section 4 we find that there are contributions associated with
operators belonging to the 20 and 15 dimensional representations where
$\Delta = 4 + \ell$, for $\ell$ even or odd respectively. These results
are applied in the $\N=4$ case for weak coupling and formulae for
anomalous dimensions are obtained including \Konl. Where appropriate
they are identical with previous results. 

Several technical details are contained in four appendices. 
Appendix A lists our conventions for
$SU(4)$ gamma matrices together with other relevant notation and also contains 
some of the algebraic details necessary for the derivation of the
integrability conditions for the $\N=4$ four point function in section 5. 
Appendix B describes the short multiplet based on a $[0,p,0]$ chiral primary
operator and also discusses the shortening conditions for multiplets
whose lowest weight operator belongs to the $(j,0)$ spin representation,
extending the results of section 4. 
Appendix C exhibits the result obtained in \Dos\ for the
contribution of an operator $\Delta$ and spin $\ell$ in the operator
product expansion for the four point function and then derives 
recurrence relations, using properties of standard hypergeometric functions,
for $\half(1\pm v)G_\Delta^{(\ell)}(u,v)$ in terms of 
$G_{\Delta'}^{({\ell'})}(u,v)$ for suitable $\Delta',\ell'$. These relations,
which are analogous in the ordinary partial wave expansion to results such
as $(2\ell+1)zP_\ell(z) = (\ell+1)P_{\ell+1}(z) - \ell P_{\ell-1}(z)$ for
Legendre polynomials, are required to obtain the operator 
product expansion results compatible with superconformal symmetry in 
sections 7 and 8. Finally in appendix D we use general relations for
AdS/CFT integrals to show how the supergravity results in \Arut\ may
be reduced to a form in agreement with superconformal symmetry and also
significantly simplified to a single $\oD$ function defined by integrals
on AdS${}_5$.

\newsec{Superconformal Transformations on Fields}

The action of an infinitesimal conformal transformation on fields with
a  finite number of components is straightforward. A quasi-primary field 
$O(x)$,  following \hughtwo, transforms as
\eqn\dO{
\de O = - v{\cdot \pr} O - \half \hom^{ab}s_{ab} O -
\Delta \hla \, O \, ,
}
where $\Delta$ is the scale dimension and $s_{ab}=-s_{ba}$ are the appropriate 
spin matrices, obeying the algebra of $Sl(2,\Bbb C)$ or $O(3,1)$, acting
on $O$. For a  general conformal transformation we have
\eqn\vcon{\eqalign{
v^a(x) = {}& a^a + \omega^a{}_{\! b} x^b + \lambda x^a + b^a x^2 - 
2 b{\cdot x}\, x^a \, , \cr
\hla(x) = {}& \lambda - 2b{\cdot x} \, , \qquad
\hom^{ab}(x) = \omega^{ab} + 4 b^{[a} x^{b]} \, , \cr}
}
where $v^a$ is a conformal Killing vector and $\hla,\hom^{ab}$ represent
associated local infinitesimal scale transformations, rotations.
In four dimensions for extension to supersymmetry it is more useful
to rewrite in terms of spinor notation\foot{Thus $4$-vectors
are identified with $2\times 2$ matrices using the hermitian $\si$-matrices
$\si_a, \, \tsi_a, \, \si_{(a} \tsi_{b)} = - \eta_{ab}1$,
$x^a \to \x_{\alpha\dal} =
x^a (\si_a)_{\alpha\dal}, \ \tx^{\dal\alpha} = x^a (\tsi_a)^{\dal\alpha}
= \ep^{\alpha\beta}\ep^{\smash {\dal \dbe}} \x_{\smash {\beta \dbe}}$,
with inverse $x^a = - {1\over 2}{\rm tr}(\si^a \tx)$. We have $x{\cdot y}
= x^a y_a = - {1\over 2}{\rm tr}(\tx \y)$.}
\eqn\vtwo{
\tv(x) = \ta + \bom \tx - \tx \omega + \lambda \, \tx + \tx \b \tx \, ,
}
where $\omega^a{}_{\! b} \to \omega_\alpha{}^\beta = -\quar \omega^{ab}
(\si_a \tsi_b)_\alpha{}^\beta, \ \bom{}^\dal{}_{\smash\dbe} = 
-\quar \omega^{ab} (\tsi_a \si_b)^\dal{}_{\smash\dbe}$. With the definition
in \vcon\ we then have 
\eqn\wwh{
\hom_\alpha{}^\beta(x) = \omega_\alpha{}^\beta + \half
(\x\, \tb - \b \, \tx )_\alpha{}^\beta \, , \qquad 
\hbom{}^\dal{}_{\smash\dbe}(x) = \bom{}^\dal{}_{\smash\dbe} 
+ \half ( \tx \,\b - \tb \, \x) {}^\dal{}_{\smash\dbe}\, .
}

For the superconformal group $SU(2,2|\N)$ in four dimensions the conformal 
group $SU(2,2)$ is extended by the usual Grassman supertranslations
$\ep_i{}^{\! \alpha}, \ \bep{}^{\, i \dal}$ and also their conformal
extensions $\bet{}_{i\dal}, \ \eta^i{}_{\! \alpha}$, $i=1,\dots \N$,
as well as the appropriate $R$-symmetry group. The critical cases are
of course $\N=2$ and $\N=4$, when the superconformal group reduces to
$PSU(2,2|4)$ by removal of an ideal, although for orientation we briefly
consider $\N=1$. The formulae for infinitesimal supertransformations
are naturally written in terms of Killing spinors,
\eqn\scep{
\hep_i{}^{\! \alpha}(x)= \ep_i{}^{\! \alpha} - i \, \bet{}_{i\dal} 
\tx^{\dal\alpha} \, , \qquad \hbep{}^{\, i \dal}(x) = \bep{}^{\, i \dal}
+ i \, \tx^{\dal\alpha} \eta^i{}_{\! \alpha} \, .
}

For $\N=1$ we consider just the complex component fields, 
$\vphi,\, \psi_\alpha,\, F$ belonging to a chiral scalar superfield 
for which the crucial transformation rules, dropping the index $i$ for
this case, are
\eqn\deone{\eqalign{
\de \vphi ={}& \hep^\alpha \psi_\alpha \, , \cr
\de \psi_\alpha ={}& 2i\, \pr_{\alpha\dal}\vphi \, \hbep{}^{\, \dal} + 4q \,
\vphi\, \eta_\alpha + \vep_{\alpha\beta} \hep{}^{\beta}\, F \, \cr
\de F ={}& 2i\, \vep_{\smash{\dal\dbe}} \pr^{\dbe\beta} \psi_\beta \,
\hbep{}^{\,\dal} - 4(q-1)\, \vep^{\alpha\beta} \psi_\beta \, \eta_\alpha\,,\cr}
}
The algebra is readily seen to close on an ordinary conformal transformation
together with a $U(1)_R$ transformation,
\eqn\clone{\eqalign{
[\de_2 , \de_1] \vphi ={}& - v{\cdot \pr} \vphi - q \, \si \vphi \, ,\cr
[\de_2 , \de_1] \psi_\alpha ={}& - v{\cdot \pr} \psi_\alpha +
\hom_\alpha{}^\beta \psi_\beta - \big ( (q-\half) \si + \bsi \big )
\psi_\alpha \, , \cr
[\de_2 , \de_1] F ={}&  - v{\cdot \pr} F - 
\big ( (q-1) \si + 2 \bsi \big ) F \, , \cr}
}
for
\eqn\vone{
\tv = 4i ( \hbep_1 \hep_2 - \hbep_2 \hep_1 ) \, ,
}
and
\eqn\sone{
\si = 4( \hep_2 \eta_1 - \hep_1 \eta_2 ) \, , \quad
\bsi = 4 ( \bet_1 \hbep_2 - \bet_2 \hbep_1 \big ) \, , \qquad
\hom_\alpha{}^\beta - \half \de_\alpha{}^{\beta} \si = 
4( \eta_{1\alpha} \hep_2{}^{\! \beta} - 1 \leftrightarrow 2 ) \, . 
}
It is easy to see that \vone\ can be expanded in the form \vtwo\ showing
that the right hand side of \clone\ is of the form of a conformal 
transformation as in \dO\ with $\hla = \half(\si + \bsi)$, and with the
coefficient of $\si-\bsi$ corresponding to the $U(1)_R$ charge.
In \deone\ $q$ is then the scale dimension of $\vphi$, for free fields $q=1$,
but as is well known \WCon, it also determines the $U(1)_R$ charge. 

For $\N=2$ we consider the simplest gauge invariant short multiplet
involving scalar fields $\vphi^{ij}=\vphi^{ji}, \, \rho, \, \btau$, 
spinors $\psi^i{}_{\!\alpha}, \, \bchi_{i\dal}$ and a conserved vector
$J_{\alpha\dal}, \, \pr^{\dal\alpha}J_{\alpha\dal}=0$, with $i,j=1,2$ 
here $SU(2)$ indices. The action of 
supersymmetry transformations generating the short
supermultiplet from $\vphi^{ij}$ is conveniently summarised diagrammatically,
with $\ep$ and $\bep$ corresponding to $\swarrow$ and $\searrow$
respectively, as follows
\eqn\Ntwo{\def\normalbaselines{\baselineskip16pt\lineskip3pt
 \lineskiplimit3pt}
\matrix{&&&&\vphi^{ij}&&&&\cr
&&&\Bsw&&\Bse&&&\cr
&&\hidewidth\psi^i{}_{\!\alpha}&&&&\bchi_{i\dal}\hidewidth&&\cr
&\Bsw~~&&\Bse&&\Bsw&&~~\Bse&\cr
\rho\hidewidth&&&&J_{\alpha\dal}&&&&\hidewidth\btau\cr}
}
In detail the transformation formulae are
\eqn\detwo{\eqalign{
\de \vphi^{ij} = {}& \hep{}^{\,(i}\, \psi^{j)} + \vep^{ik} \vep^{jl}\, 
\bchi_{(k} \, \hbep_{l)} \, , \cr
\de \psi^i_{\,\alpha} ={}& 2i \, \pr_{\alpha\dal} \vphi^{ij} \, 
\hbep{}_j^{\,\dal}
+ 8\, \vphi^{ij}\, \eta_{j\alpha} + \rho \,\vep_{\alpha\beta} \hep{}^{\,i\beta}
+ \vep^{ij} \, J_{\alpha\dal} \hbep{}_j^{\, \dal}  \, , \cr
\de \bchi_{\, i\dal} = {}& -2i \, \hep^{j\alpha} \pr_{\alpha\dal} \, \vep_{ik}
\vep_{jl} \vphi^{kl} + 8 \, \bet{}^j_{\, \dal} \, \vep_{ik}\vep_{jl} \vphi^{kl} 
+ \vep_{\smash{\dal\dbe}}\hbep{}_i^{\, \dbe} \, \btau
- \vep_{ij} \, \hep{}^{\,j\alpha} J_{\alpha\dal}\, , \cr
\de \rho ={}& 2i \, \vep_{\smash{\dal\dbe}} \pr^{\dbe \beta} \psi^j_{\,\beta}\, 
\hbep{}_j^{\, \dal}
- 4 \, \vep^{\alpha\beta}\psi^j_{\,\beta} \, \eta_{j\alpha} \, , \cr
\de \btau = {}& - 2i \,  \hep{}^{\, j \alpha} \, \vep_{\alpha\beta}\, 
\bchi_{\smash{j\dbe}}
{\overleftarrow \pr}{}^{\dbe \beta} - 4 \, \bet^j_{\,\dal} \vep^{\dal\dbe}\, 
\bchi_{\smash{j\dbe}} \, , \cr
\de J_{\alpha\dal} ={}& -i \, \hep^{\, i\beta}\, \pr_{\alpha\dal} \vep_{ij}
\psi^j_{\,\beta} + 2i \, \vep_{\alpha\beta}\vep_{\smash{\dal\dbe}}  
\hep^{\, i\beta}\, \pr^{\dbe\gamma} \vep_{ij} \psi^j_{\,\gamma} - 
6\, \vep_{ij} \psi^j_{\, \alpha} \, \bet^i_{\, \dal} \cr
{}& - i\, \vep^{ij} \, \bchi_{\smash{j\dbe}} \! 
{\overleftarrow \pr}_{\!\!\alpha\dal} \, \hbep{}_i^{\,\, \dbe} + 
2i\, \vep_{\alpha\beta}\vep_{\smash{\dal\dbe}}\vep^{ij} \, 
\bchi_{j\dga} \!{\overleftarrow \pr}{}^{\dga\beta} \, \hbep{}_i^{\,\, \dbe}
+ 6\, \eta_{i\alpha} \, \vep^{ij} \bchi_{\, j \dal} \, .  \cr}
}
The coefficients are determined by closure of the algebra to give
\eqn\cltwo{\eqalign{
[\de_2 , \de_1] \vphi^{ij} ={}& - v{\cdot \pr} \vphi^{ij} -
(\si + \bsi) \vphi^{ij} + \hht^i{}_{\! k} \vphi^{kj} +
\hht^i{}_{\! k} \vphi^{ik} \, , \cr 
[\de_2 , \de_1] \psi^i{}_{\! \alpha} = {}& - v{\cdot \pr} \psi^i{}_{\! \alpha}
+ \hom_\alpha{}^\beta \psi^i{}_{\! \beta} - \big 
(\si +{\ts{3\over 2}} \bsi \big ) \psi^i{}_{\! \alpha} 
+ \hht^i{}_{\! j} \psi^j{}_{\! \alpha} \, , \cr
[\de_2 , \de_1]\bchi_{i\dal} = {}& - v{\cdot \pr} \bchi_{i\dal} -
\bchi_{\smash{i\dbe}}\, \hbom{}^\dbe{}_{\dal} - \big ( {\ts{3\over 2}} \si +
\bsi \big ) \bchi_{i\dal} - \bchi_{j\dal}  \hht^j{}_{\! i}\, , \cr
[\de_2 , \de_1] \rho = {}& - v{\cdot \pr} \rho  - (\si + 2\bsi) \rho \, ,\cr
[\de_2 , \de_1] \btau = {}& - v{\cdot \pr} \btau - (2\si + \bsi ) \btau \,,\cr
[\de_2 , \de_1] J_{\alpha\dal} = {}& - v{\cdot \pr} J_{\alpha\dal}
+ \hom_\alpha{}^\beta J_{\beta\dal} - J_{\smash{\alpha\dbe}}\,
\hbom{}^\dbe{}_{\dal} - {\ts{3\over 2}}( \si + \bsi) J_{\alpha\dal} \, . \cr}
}
Here, in addition to \vone\ and \sone\ with implicit summation over
the $SU(2)$ indices, we have
\eqn\otwo{\eqalign{
\hbom{}^\dal{}_{\smash{\!\dbe}} + \half \de^\dal{}_{\smash{\!\dbe}} \bsi ={}&
4\big ( \hbep_1{}^{\! \dal}\bet{}_{\smash{ 2 \dbe}}  - 1 \leftrightarrow 2 
\big ) \, , \cr
\hht^i{}_{\! j} + \de^i{}_{\! j} ( \bsi - \si) ={}& 4 
\big ( \hep_1{}^{\!i\alpha} \eta_{2j\alpha} + \bet{}_1{}^{\!i}{}_{\!\smash \dal}
\hbep{}_2{}^{\!\dal}{}_{\smash \!\! j} - 1 \leftrightarrow 2 \big )\,. \cr}
}
$\hht^i{}_{\! j}$, $\hht^i{}_{\! i}=0$, represents an infinitesimal 
$SU(2)$ transformation of the fields which is part of the $U(2)$ $R$-symmetry
group in this case.
In \detwo\ the result for $\de J_{\alpha\dal}$ is also compatible
with the conservation equation $\pr^{\dal\alpha}J_{\alpha\dal}=0$, which
is necessary to obtain the result for $[\de_2 , \de_1] J_{\alpha\dal}$.
For each field \cltwo\ is of the required form
\eqn\Ctw{
[\de_2 , \de_1] O = - v{\cdot \pr} O - \Delta \half (\si + \bsi) O + r
\half (\si - \bsi) O - \half \hom^{ab} s_{ab} O + \hht_a R_a O \, ,
}
where $r$ is the $U(1)_R$ charge, $R_a$ are the appropriate generators of
$SU(2)_R$ with  $\hht_a = {\rm tr}(\hht \tau_a)$, for $\tau_a$ the usual
Pauli matrices.

For $\N=4$ the fields in the corresponding self-conjugate short supermultiplet
proliferate. For the appropriate $SU(4)$ $R$-symmetry  and spin representations 
they are listed, with the necessary constraints, in the table below. 

\hskip -1.2cm
\vbox{\offinterlineskip
\hrule
\halign{&\vrule#&\strut\ \ \hfil#\  \cr
height2pt&\omit&&\omit&&\omit&&\omit&&\omit&&\omit&\cr
&$SU(4)\,$rep \hfil && $SU(4)\,$dim && $(j_1,j_2)$ && field 
&& field constraints\qquad&&field dim&\cr 
height2pt&\omit&&\omit&&\omit&&\omit&&\omit&&\omit&&\omit&\cr
\noalign{\hrule}
height2pt&\omit&&\omit&&\omit&&\omit&&\omit&&\omit&&\omit&\cr
& [0,2,0]\quad && 20\qquad && $(0,0)\ $  && $\vphi_{rs}$ &
& $\vphi_{rs}=\vphi_{(rs)}, \ \vphi_{rr}=0$&& 20\quad &\cr
& [0,1,1]\quad && 20\qquad && $(\half,0)\ $ 
&&$\psi_{ir\alpha}$&& $\ga_r \psi_r =0$&&40\quad & \cr
& [1,1,0]\quad && 20\qquad && $(0,\half)\ $  
&&$\bpsi{}^i{}_{\!r\dal}$&&
$\bpsi_r\bga_r =0$&&40\quad & \cr
& [0,1,0]\quad && 6\qquad && $(1,0)\ $ &&$f_{r\alpha\beta}$&& $f_{r\alpha\beta}
=f_{r(\alpha\beta)}$&&18\quad & \cr
& [0,1,0]\quad && 6\qquad && $(0,1)\ $ &&$\of{}_{\smash{r\dal\dbe}}$&
& $\of{}_{\smash{r\dal\dbe}} =\of{}_{\smash{r(\dal\dbe)}}$&&18\quad & \cr
& [0,0,2]\quad && 10\qquad && $(0,0)\ $ &&$\rho_{ij}$&&$\rho_{ij}=\rho_{(ij)}$&
&10\quad & \cr
& [2,0,0]\quad && 10\qquad && $(0,0)\ $ &&$\brho{}^{\, ij}$&
&$\brho{}^{\,ij}=\brho{}^{(\,ij)}$&&10\quad & \cr
& [1,0,1]\quad && 15\qquad && $(\half,\half)\ $ &&$J_{rs\alpha\dal}$&
&$J_{rs\alpha\dal} = J_{[rs]\alpha\dal},\, \pr^{\dal\alpha}J_{rs\alpha\dal}=0$&
&45\quad & \cr
& [0,0,1]\quad && 4\qquad && $(\half,0)\ $ 
&&$\lambda_{i\alpha}$&&  &&8\quad & \cr
& [1,0,0]\quad && 4\qquad && $(0,\half)\ $ 
&&$\blam{}^i_{\,\dal}$&&  &&8\quad & \cr
& [1,0,0]\quad && 4\qquad && $(1,\half)\ $ 
&&$\Psi^i_{\alpha\beta\dal}$&
&$\Psi_{\alpha\beta\dal}=\Psi_{(\alpha\beta)\dal},\,
\pr^{\dal\alpha}\Psi_{\alpha\beta\dal}=0$&&16\quad & \cr
& [0,0,1]\quad && 4\qquad && $(\half,1)\ $ 
&&$\bPsi{}_{\smash{i\alpha\dal\dbe}}$&&$\bPsi{}_{\smash{\alpha\dal\dbe}}
=\bPsi{}_{\smash{\alpha(\dal\dbe)}},\,
\pr^{\dal\alpha}\bPsi{}_{\smash{\alpha\dal\dbe}}=0$&&16\quad & \cr
& [0,0,0]\quad && 1\qquad && $(0,0)\ $ && $\Phi$ && && 1\quad &\cr
& [0,0,0]\quad && 1\qquad && $(0,0)\ $ && $\bPhi$ && && 1\quad &\cr
& [0,0,0]\quad && 1\qquad && $(1,1)\ $ && $T_{\smash{\alpha\beta\dal\dbe}}$&
&$T_{\smash{\alpha\beta\dal\dbe}}=T_{\smash{(\alpha\beta)(\dal\dbe)}},\, 
\pr^{\dal\alpha}T_{\smash{\alpha\beta\dal\dbe}}=0$&&5\quad & \cr
height2pt&\omit&&\omit&&\omit&&\omit&&\omit&&\omit&\cr}
\hrule}

\noindent
Here $i,j=1,\dots 4$ are $SU(4)$ indices and $r,s=1,\dots 6$ correspond
to $SO(6)$, other details of the notation are given in appendix A. 
The matrices $\ga_r^{ij}= - \ga_r^{ji}$ and $\bga_{rij}=-\bga_{rji}$
give the explicit identification of the 6-dimensional representation
of $SO(6)$ with the antisymmetric tensor products $4\times 4$ and 
${\bar 4} \times {\bar 4}$ for $SU(4)$. From the last column it is easy to     
see that there are 128 bosonic and also 128 fermionic degrees of freedom.
This is the simplest gauge invariant supermultiplet arising in $\N=4$
supersymmetric gauge theories, besides the $SU(4)$ current it of course
contains the energy momentum tensor. For $\N=4$ gauge theories
$\Phi \propto F^2 + iF{\tilde F}$, where $F$ is the field strength and
$\tilde F$ is its dual.

In a similar fashion to \Ntwo\ the supersymmetry transformations on this 
multiplet may be represented diagrammatically by
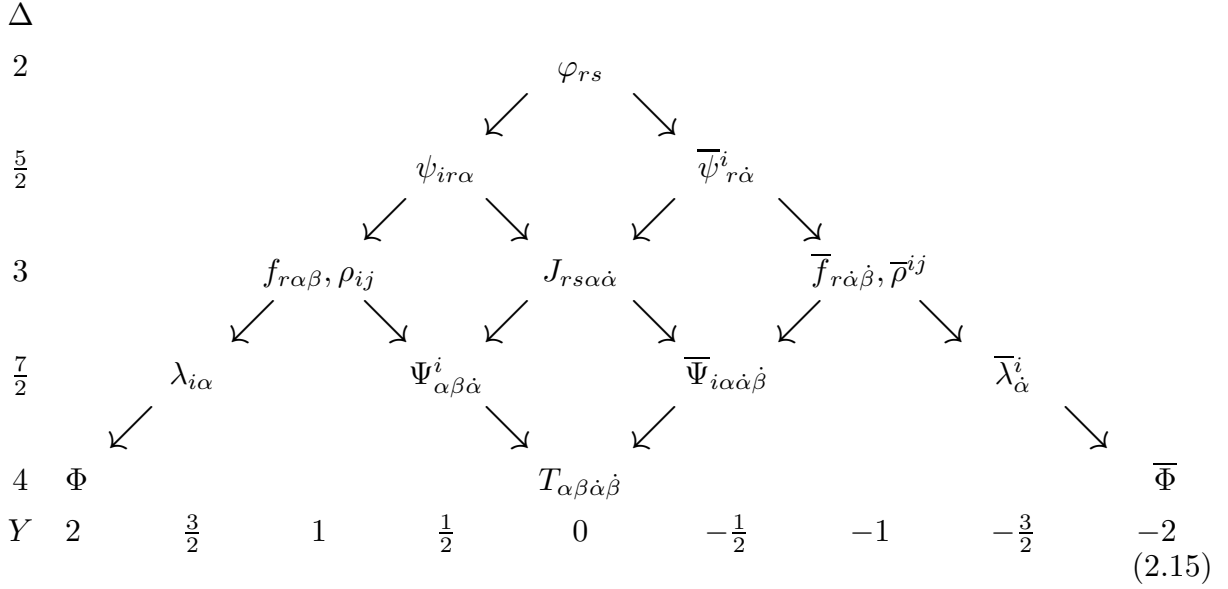
\eqn\Nfour{\def\normalbaselines{\baselineskip16pt\lineskip3pt
 \lineskiplimit3pt}
\matrix{\Delta&&&{~~}&&{~~~}&&{~~}&&{~~~}&&{~~}&&{~~~}&&{~~}&&\cr
2&&&{~~}&&{~~~}&&{~~}&&\vphi_{rs}&&{~~}&&{~~~}&&{~~}&&\cr
&&&&&&&&\Bsw&&\Bse&&&&&&&\cr
{\ts{5\over 2}}&&&&&&&\hidewidth\psi_{ir\alpha}\hidewidth&&&&
\hidewidth\bpsi{}^i{}_{\!r\dal}\hidewidth&&&&&&\cr
&&&&&&\Bsw&&\Bse&&\Bsw&&\Bse&&&&&\cr
3&&&&&\hidewidth f_{r\alpha\beta},\rho_{ij}\hidewidth&&&&
\hidewidth J_{rs\alpha\dal}\hidewidth&&&&
\hidewidth\of{}_{\smash{r\dal\dbe}},\brho{}^{ij}\hidewidth&&&&\cr
&&&&\Bsw&&\Bse&&\Bsw&&\Bse&&\Bsw&&\Bse&&&\cr
{\ts {7\over 2}}&&&\hidewidth\lambda_{i\alpha}\hidewidth&&&
&\hidewidth\Psi^i_{\alpha\beta\dal}\hidewidth&&&
&\hidewidth\bPsi_{\smash{i\alpha\dal\dbe}}\hidewidth&&&
&\hidewidth\blam{}^i_\dal\hidewidth&&\cr
&&\Bsw&&&&&&\Bse&&\Bsw&&&&&&\Bse&\cr
4&\Phi\hidewidth&&&&&&&
&\hidewidth T_{\smash{\alpha\beta\dal\dbe}}\hidewidth&&&&&&&&\hidewidth\bPhi\cr
Y&2&&{\ts{3\over 2}}&&1&&{\ts{1\over 2}}&&0&&-{\ts{1\over 2}}&&-1&&
-{\ts{3\over 2}}&&-2\cr}
}
where we list the scale dimension $\Delta$ and also the $Y$-charge
corresponding to the $U(1)_Y$ bonus symmetry \Intr.
Disregarding those which may be obtained by conjugation the relevant
supersymmetry transformations are, suppressing explicit $SU(4)$ indices
when convenient by assuming for example $\psi_{r\alpha}$ is a column
vector with $\psi_{r\alpha}{}^{\!\!\! t}$ its transpose, then
\eqnn\defour
$$\eqalignno{
\de \vphi_{rs} = {}& - \hep \ga_{(r} \psi_{s)} + 
\bpsi{}_{(r} \bga_{s)} \hbep \, , \cr
\de \psi_{r\alpha} ={}& 
i \pr_{\alpha\dal} \vphi_{rs} \, \bga_s \hbep{}^{\, \dal}
+ 4 \, \vphi_{rs} \, \bga_s \eta_\alpha - f_{r\alpha\beta} \, \hep^{\beta \, t}
- {\ts{1\over 6}} \, f_{s\alpha\beta}\, \bga_r \ga_s \hep^{\beta \, t} \cr
& {}+ \rho \, \vep_{\alpha\beta}\hep^{\beta \, t} +
J_{\smash{rs\alpha\dal}} \, \bga_s \hbep{}^{\, \dal} + {\ts{1\over 6}}
J_{\smash{st\alpha\dal}} \, \bga_r \ga_s \bga_t \hbep{}^{\, \dal} \, , \cr
\de f_{r\alpha\beta} = {}& 2i \, \psi_{r(\alpha}^{\ t} \!
{\overleftarrow \pr}_{\!\! \beta)\dal}^{\vphantom t} \, \hbep{}^{\, \dal} + 
12 \, \psi_{r(\alpha}^{\ t} \, \eta_{\beta)}^{\vphantom t} - 
\vep_{(\alpha|\ga}\hep{}^{\, \ga\, t} \, \ga_r \, \lambda_{\beta)} + 
2 \, \Psi^{\, t}_{\alpha\beta\dal} \, \bga_r \, \hbep{}^{\, \dal} \, ,\cr
\de \rho_{ij} = {}& - i \, \vep_{\smash{\dal\dbe}} \hbep{}^{\, k \dbe}
\, \bga_{r k(i} \pr^{\dal\alpha} \psi_{rj)\alpha} + 2 \, \vep^{\alpha\beta}
\eta^k_{\,\beta}\, \bga_{r k(i} \, \psi_{rj)\alpha} + 
\hep_{(i} \, \lambda_{j)} \, , \cr
\de J_{\smash{rs\alpha\dal}} = {}& 2i\, \hep^\beta  \ga_{[r}  \psi_{s]\alpha}
\!{\overleftarrow \pr}_{\!\! \beta\dal} - i \, \hep^\beta \ga_{[r}  
\psi_{s]\beta}\! {\overleftarrow \pr}_{\!\! \alpha\dal} - 6 \,
\bet{}_{\dal} \, \ga_{[r}  \psi_{s]\alpha} \cr
{}& + 2i \, \pr_{\smash{\alpha\dbe}} \bpsi{}_{\smash{[r \dal}} \, 
\bga_{\smash{s]}} \, \hbep{}^{\, \dbe} - i \, 
\pr_{\smash{\alpha\dal}} \bpsi{}_{\smash{[r \dbe}} \, 
\bga_{s]} \, \hbep{}^{\, \dbe} + 
6 \, \bpsi{}_{\smash{[r \dal}} \, \bga_{\smash{s]}} \, \eta_\alpha \cr
{}& + \hep{}^\beta \ga_{[r} \bga_{s]} \, \Psi_{\smash{\beta\alpha\dal}} +
\bPsi_{\smash{\alpha\dal\dbe}}\, \ga_{[r} \bga_{s]} \, \hbep{}^{\, \dbe}\, ,\cr
\de \lambda_\alpha = {}& i \, \vep^{\beta\ga} f_{r\alpha\gamma} \!
{\overleftarrow \pr}_{\!\! \beta\dal} \, \bga_r \, \hbep{}^{\, \dal}
+ 2 \, \vep^{\beta\ga} f_{r\alpha\gamma} \bga_r \, \eta_\beta +
2i \, \pr_{\alpha\dal}\rho \, \hbep{}^{\, \dal} + 12 \, \rho \, \eta_\alpha
+ \Phi \, \vep_{\alpha\beta} \hep{}^{\beta \, t} \, , \cr
\de \Psi_{\alpha\beta\dal} = {}& -\half i \, f_{r\alpha\beta} \!
{\overleftarrow \pr}_{\!\! \gamma \dal} \, \ga_r \, \hep^{\ga \, t} +
{\ts {1\over 3}}i \, f_{r\gamma(\alpha} \!
{\overleftarrow \pr}_{\!\! \beta) \dal} \, \ga_r \, \hep^{\ga \, t} +
{\ts {4\over 3}} \, f_{r\alpha\beta} \, \ga_r \, \bet_\dal{}^{\! t} \cr
{}& - \quar i \, \pr_{\smash{(\alpha\dga}} \, J_{rs\beta)\dal} \,
\ga_r \bga_s \, \hbep{}^{\, \dga} + {\ts {1\over 12}}i \, 
\pr_{\smash{(\alpha\dal}} \, J_{rs\beta)\dga} \,\ga_r\bga_s \,\hbep{}^{\, \dga}
- {\ts {4\over 3}} \, J_{rs(\alpha\dal} \,\ga_r\bga_s \, \eta_{\beta)} \cr
{}& + T_{\smash{\alpha\beta\dal\dbe}} \, \hbep{}^{\, \dbe} \, , \cr
\de \Phi = {}& 2i \, \vep_{\smash{\dal\dbe}} \hbep{}^{\, \dbe t} 
\pr^{\dal\alpha} \lambda_\alpha - 8 \, \vep^{\alpha\beta} \eta_\beta{}^{\! t}
\lambda_\alpha \, , \cr
\de T_{\smash{\alpha\beta\dal\dbe}} = {}& 2i \, \hep^\gamma 
\pr_{\ga(\dal} \Psi_{\smash{\alpha\beta\dbe)}} - i \, \hep^\gamma 
\pr_{(\alpha(\dal} \Psi_{\smash{\ga|\beta)\dbe)}} -10 \, \bet_{(\dal} \,
\Psi_{\smash{\alpha\beta\dbe)}} \cr
{}& - 2i \, \bPsi_{\smash{(\alpha\dal\dbe}} \! {\overleftarrow \pr}_{\smash
{\!\!\beta)\dga}} \, \hbep{}^{\, \dga}+ i \, \bPsi_{\smash{(\alpha(\dal|\dga}} 
\! {\overleftarrow \pr}_{\smash {\!\!\beta)\dbe)}} \, \hbep{}^{\, \dga}
- 10 \, \bPsi_{\smash{(\alpha\dal\dbe}}\, \eta_{\beta)} \, .  & \defour \cr}
$$
These formulae are consistent with the conservation equations listed in
the table and lead to
\eqnn\clfour
$$\eqalignno{
[\de_2 , \de_1] \vphi_{rs} ={}& - v{\cdot \pr} \vphi_{rs} - 2
\hla\, \vphi_{rs} + \hht_{rt}\, \vphi_{ts} + \hht_{st}\, \vphi_{rt} \, , \cr 
[\de_2 , \de_1] \psi_{ir\alpha} = {}& - v{\cdot \pr} \psi_{ir\alpha}
+ \hom_\alpha{}^\beta \psi_{ir \beta} - {\ts {5\over 2}} \hla\, \psi_{ir \alpha} 
+ \hht_i{}^{j}\, \psi_{jr \alpha} + \hht_{rs} \, \psi_{is \alpha}\, , \cr
[\de_2 , \de_1]f_{r\alpha\beta} = {}& - v{\cdot \pr} f_{r\alpha\beta}
+ \hom_\alpha{}^\gamma f_{r\gamma\beta} +  \hom_\beta{}^\gamma f_{r\alpha\gamma}
- 3\hla\, f_{r\alpha\beta} + \hht_{rs} \, f_{s\alpha\beta} \, , \cr
[\de_2 , \de_1] \rho_{ij} = {}& - v{\cdot \pr} \rho_{ij}  - 3\hla\, \rho_{ij} 
+ \hht_i{}^k  \rho_{kj} + \hht_j{}^k \rho_{ik} \, ,\cr
[\de_2 , \de_1] J_{rs\alpha\dal} = {}& - v{\cdot \pr} J_{rs\alpha\dal}
+ \hom_\alpha{}^\beta J_{rs\beta\dal} - J_{\smash{rs\alpha\dbe}}\,
\hbom{}^\dbe{}_{\dal} - 3 \hla\, J_{\alpha\dal} 
+ \hht_{rt}\, J_{ts\alpha\dal} + \hht_{st}\, J_{rt\alpha\dal} \, ,\cr
[\de_2 , \de_1] \lambda_{i\alpha} = {}& - v{\cdot \pr} \lambda_{i\alpha}
+ \hom_\alpha{}^\beta \lambda_{i \beta} - {\ts {7\over 2}} \hla\, \lambda_{i\alpha} 
+ \hht_i{}^{j}\, \lambda_{j\alpha} \, , \cr
[\de_2 , \de_1] \Psi^i_{\alpha\beta\dal} = {}& - v{\cdot \pr} \Psi^i_{\alpha\beta\dal} 
+ \hom_\alpha{}^\gamma \Psi^i_{\gamma\beta\dal} + \hom_\beta{}^\gamma 
\Psi^i_{\alpha\gamma\dal} - \Psi^i_{\smash{\alpha\beta\dbe}}\,\hbom{}^\dbe{}_{\dal}
- {\ts {7\over 2}} \hla\, \Psi^i_{\smash{\alpha\beta\dal}}  
- \Psi^j_{\smash{\alpha\beta\dal}} \hht_j{}^i \, , \cr
[\de_2 , \de_1] T_{\smash{\alpha\beta\dal\dbe}} = {}& - v{\cdot \pr} 
T_{\smash{\alpha\beta\dal\dbe}} 
+ \hom_\alpha{}^\gamma T_{\smash{\gamma\beta\dal\dbe}} 
+ \hom_\beta{}^\gamma T_{\smash{\alpha\gamma\dal\dbe}} 
- T_{\smash{\alpha\beta\dga\dbe}}\,\hbom{}^\dga{}_{\dal}
- T_{\smash{\alpha\beta\dal\dga}}\,\hbom{}^\dga{}_{\smash\dbe} \cr
{}& - 4\hla\, T_{\smash{\alpha\beta\dal\dbe}}  \, , \cr
[\de_2 , \de_1] \Phi = {}& - v{\cdot \pr} \Phi - 4\hla\, \Phi \, . &\clfour
\cr}
$$
The definition of $v^a$ is as in \vone\ while
\eqn\lfour{\eqalign{
\hht_{rs} = {}& -2 \big ( \hep_1 \ga_{[r}\bga_{s]} \eta_2 - 
\bet_2 \ga_{[r}\bga_{s]} \hbep{}_1 - 1\leftrightarrow 2 \big ) \, , \cr
\hht_i{}^{j} = {}& 4\big ( \hep_{1i}\, \eta_2{}^{\! j} -
\bet{}_{2i}\, \hbep{}_1{}^{\! j}
- \hbox{trace}(ij) - 1\leftrightarrow 2 \big ) \, , \cr}
}
generate local $SU(4)$ transformations and
\eqn\defla{\eqalign{\!\!\!
\hom_\alpha{}^\beta  = {}& 4\big ( \eta_{1\alpha}{}^{\!\!\!\!i}\,\, 
\hep_{2i}{}^{\!\!\beta} -  \hbox{trace}(\alpha\beta)
- 1 \leftrightarrow 2 \big ) \, , \quad
\hbom{}^\dal{}_{\smash{\!\dbe}} =
4\big ( \hbep_1{}^{\!\! i \dal}\, \bet{}_{\smash{ 2i\dbe}}
-  \hbox{trace}(\dal\dbe)  - 1 \leftrightarrow 2 \big ) \, , \cr
&\qquad\qquad\qquad \hla=  2 \big( \hep_2 \eta_1 + \bet{}_1 \hbep{}_2 - 
1\leftrightarrow 2 \big ) \, , \cr}
}
represent infinitesimal local rotations and scale transformations.

\newsec{Two and Three Point Correlation Functions}

The superconformal transformations obtained in \deone, \detwo\ and 
\defour\ allow correlation functions of the fields in the relevant
supermultiplets to be related. This is shown most clearly in the 
case of two point functions which are a precursor to considering
higher point functions later.

For the simplest $\N=1$ case we start from the two point correlation
function for the scalar field $\vphi$ in the basic chiral supermultiplet
\eqn\onep{
\l \vphi(x_1)\, \bvphi(x_2) \r = {1\over r_{12}^{\,\, q}} \, ,
}
with a convenient choice of normalisation, and where
\eqn\defrij{
r_{ij} = (x_i-x_j)^2 \, .
}
The basic superconformal identity here requires
\eqn\superp{
\l \psi_\alpha(x_1) \, \de \bvphi(x_2) \r + \l \de \psi_\alpha(x_1) \, 
\bvphi(x_2) \r = 0 \, .
}
Using \deone\ in \superp\ with \onep\ then gives
\eqn\suprel{
\l \psi_\alpha(x_1) \, \bpsi{}_{\dal}(x_2) \r \, \hbep{}^{\, \dal}(x_2)
= - 2i \pr_{\alpha\dal} {1\over r_{12}^{\,\, q}} \, \hbep{}^{\, \dal}(x_1)
- 4q \, {1\over r_{12}^{\,\, q}} \, \eta_\alpha =
4iq \, {\x_{12\alpha\dal}\over r_{12}^{\,\, q+1}}\,\hbep{}^{\, \dal}(x_2)\, ,
}
where we use $\hbep{}^{\, \dal}(x_1) = 
\hbep{}^{\, \dal}(x_2) + i \tx_{12}^{\dal\alpha}\eta_\alpha$. Hence it is
clear that
\eqn\twopsi{
\l \psi_\alpha(x_1) \, \bpsi{}_{\dal}(x_2) \r = 
4iq \, {\x_{12\alpha\dal}\over r_{12}^{\,\, q+1}}\, , 
}
and in a similar fashion
\eqn\twoF{
\l F(x_1) \, {\bar F}(x_2)\r = 16q(q-1) \, {1\over r_{12}^{\,\, q+1}}\, .
}
Clearly positivity requires $q\ge 1$.

For $\N=2$ similar calculations give for the two point functions of
the fields in short multiplet shown in
\Ntwo\ together with their conjugates,
\eqn\twotwoP{\eqalign{
\l \vphi^{ij}(x_1)\, \bvphi{}_{kl}(x_2) \r = {}&
\de^{(i}{}_{\! k} \de^{j)}{}_{\! l} \, {1\over r_{12}^{\,\, 2}} \, , \cr
\l \psi^i{}_{\!\alpha}(x_1) \, \bpsi{}_{j \dal}(x_2) \r = 
\l \chi^i{}_{\!\alpha}(x_1) \, \bchi{}_{j \dal}(x_2) \r = {}&
8i \, \de^i{}_{\! j} \, {\x_{12\alpha\dal}\over r_{12}^{\,\, 3}}\, , \cr
\l \rho(x_1) \, {\overline \rho}(x_2) \r = \l \tau(x_1) \, \btau(x_2) \r =
{}& 32 \, {1\over r_{12}^{\,\, 3}}\, , \cr
\l J_{\alpha\dal}(x_1) \, {\bar J}{}_{\smash{\beta\dbe}} (x_2) \r ={}&
48 \, {\x_{\smash{12\alpha \dbe}}\, \x_{\smash{21 \beta\dal}} \over
r_{12}^{\,\, 4}} \, . \cr}
}

For $\N=4$ the two point functions are also determined for the short
multiplet given in the table in terms of that for $\vphi_{rs}$. Choosing
a normalisation coefficient $\hN$ we have
\eqn\twofourP{\eqalign{
\l \vphi_{rs}(x_1)\, \vphi{}_{uv}(x_2) \r = {}&
{1\over 2}\hN \Big ( \half\big ( \de_{ru}\de_{sv} + \de_{rv}\de_{su} \big ) -
{\ts {1\over 6}}\, \de_{rs}\de_{uv} \Big ) 
 {1\over r_{12}^{\,\, 2}} \, , \cr
\l \psi_{r \alpha}(x_1) \, \bpsi{}_{s \dal}(x_2) \r = {}&
2\hN i \,\big ( \de_{rs} 1 + {\ts{1\over 6}} \bga_r \ga_s
\big ) \, {\x_{12\alpha\dal}\over r_{12}^{\,\, 3}}\, , \cr
\l f_{r\alpha\beta}(x_1) \, \of{}_{\smash{s\dal\dbe}}(x_2) \r = {}&
- 24\hN  \, \de_{rs} \, {\x_{12(\alpha\dal}\, \x_{\smash{12\beta)\dbe}}
\over r_{12}^{\,\, 4}} \, , \cr
\l \rho_{ij} (x_1) \, \brho^{kl} (x_2) \r = {}& 4\hN \, \de_i{}^{\!(k}
\de_j{}^{\!l)} \, {1\over r_{12}^{\,\, 3}} \, , \cr
\l J_{rs \alpha\dal}(x_1) \, J_{\smash{uv \beta\dbe}} (x_2) \r ={}&
- 6\hN (\de_{ru}\de_{sv} - \de_{rv}\de_{su} ) \, 
{\x_{\smash{12\alpha\dbe}}\,\x_{\smash{21 \beta\dal}}
\over r_{12}^{\,\, 4}}\, , \cr
\l \lambda_\alpha (x_1) \, \blam_\dal(x_2) \r = {}& 48\hN i \, 1 \, 
{\x_{12\alpha\dal} \over r_{12}^{\,\, 4}}\, , \cr
\l \Psi_{\alpha\beta\dal} (x_1) \, \bPsi{}_{\smash{\gamma\dga{\dot \de}}}
(x_2) \r = {}& 16\hN i \, 1 \, {\x_{12(\alpha\dga}\, \x_{\smash{12\beta)\dde}}
\, \x_{21\gamma \dal} \over r_{12}^{\,\, 5}} \, , \cr
\l \Phi(x_1) \, \bPhi(x_2) \r = {}& 384\hN \,{1\over r_{12}^{\,\, 4}} \, , \cr
\l T_{\smash{\alpha\beta\dal\dbe}}(x_1)\, 
T_{\smash{\gamma\de\dga\dde}}(x_2) \r
={}& 160\hN \,{\x_{12(\alpha\dga}\,\x_{\smash{12\beta)\dde}}\,\x_{21(\gamma\dal}
\, \x_{\smash{21\de)\dbe}} \over r_{12}^{\,\, 6}} \, . \cr }
}

The associated three point functions contain the essential information 
necessary to obtain the operator product expansion. It is therefore
of interest to consider the three point functions for component fields
in the short multiplet listed in the table above. Using $\N=4$ superconformal
symmetry we show how various three point functions involving descendant fields
such as the $SU(4)_R$ symmetry current and the energy momentum tensor
may be uniquely obtained.  With notation
described in appendix A, so that $C^I{}_{\!\!\! rs}$ is a basis for symmetric
traceless tensors, and defining $\vphi^I\equiv C^I{}_{\!\!\! rs}\vphi_{rs}$
the starting point is
\eqn\threep{
\l \vphi^{I_1}(x_1) \, \vphi^{I_2}(x_2)\, \vphi^{I_3}(x_3) \r =
\hN \, { C^{I_1I_2I_3} \over r_{12}  \, r_{13} \, r_{23}} \, ,
}
where $C^{IJK}=\tr(C^IC^JC^K)$. In \threep\ we have taken into account
that the normalisation is fixed once that of the two point function is
given as a consequence of non-renormalisation theorems \Sei. For
a $SU(N)$ gauge theory with $\N=4$ superconformal symmetry then 
\twofourP\ and \threep\ are valid if $\hN=N^2-1$.

The superconformal transformations given
in \defour\ relate correlation functions of the component fields in 
which $\sum_i \Delta_i$ differ by one. At the first step we may
use the condition $\de \l \psi_{r\alpha}(x_1) \, \vphi^{I_2}(x_2)\,
\vphi^{I_3}(x_3) \r = 0$ and keep just the terms involving $\hbep$ from 
\defour\ which then gives
\eqnn\fith
$$\eqalignno{
& \l \psi_{r\alpha}(x_1) \, \bpsi{}_{s\dal}(x_2)\, \vphi^{I_3}(x_3) \r
C^{I_2}{}_{\!\!\! st} \bga_t \hbep{}^\dal(x_2) +
\l \psi_{r\alpha}(x_1) \, \vphi^{I_2}(x_2)\, \bpsi{}_{s\dal}(x_3) \r
C^{I_3}{}_{\!\!\! st} \bga_t \hbep{}^\dal(x_3) \cr
&{}+ \l J_{rs\alpha\dal}(x_1) \, \vphi^{I_2}(x_2)\, \vphi^{I_3}(x_3) \r
\bga_s \hbep{}^\dal(x_1)
+ {\ts{1\over 6}} \l J_{st\alpha\dal}(x_1) \, \vphi^{I_2}(x_2)\, 
\vphi^{I_3}(x_3) \r \, \bga_r \ga_s \bga_t \hbep{}^\dal(x_1) \cr
&\quad {}= 2\hN i \bigg ( {1\over r_{12}}\x_{12\alpha\dal} \, C^{I_2I_3}_{rs}
\bga_s \, \hbep{}^\dal(x_2) + {1\over r_{13}}\x_{13\alpha\dal} \, 
C^{I_2I_3}_{rs} \bga_s \, \hbep{}^\dal(x_3) \bigg ) \, , & \fith \cr}
$$
with $C^{IJ}_{rs}$ the symmetric traceless part of $(C^IC^J)_{rs}$. To 
solve this we introduce, for three points $x_i,x_j,x_k$, 
\eqn\defX{
\X_{i[jk]} = {\x_{ij}\tx_{jk}\x_{ki}\over r_{ij}\, r_{ik}} =
{1\over r_{ij}} \x_{ij} - {1\over r_{ik}} \x_{ik} \, ,
}
which transforms under conformal transformations as a vector at $x_i$ and
is antisymmetric in $jk$. It is important to note the relations
\eqn\Xep{\eqalign{
\X_{i[jk]\alpha\dal}\, \hbep{}^{\,\dal}(x_i) = {}& {1\over r_{ij}}
\x_{ij\alpha\dal} \hbep{}^{\,\dal}(x_j) - {1\over r_{ik}} 
\x_{ik\alpha\dal} \hbep{}^\dal(x_k) \, , \cr 
\hep{}^{\,\alpha}(x_i)\, \X_{i[jk]\alpha\dal} ={}& {1\over r_{ik}} 
\hep{}^{\,\alpha}(x_k)\, \x_{ki\alpha\dal} - 
{1\over r_{ij}}\hep{}^{\,\alpha}(x_j)\, \x_{ji\alpha\dal}\,.\cr}
}
With the aid of the definition \defX\ and \Xep\ we then find
\eqn\threeJp{\eqalign{
\l \psi_{r\alpha}(x_1) \, \bpsi{}_{s\dal}(x_2)\, \vphi^{I}(x_3) \r ={}&
2\hN i \, {\x_{12\alpha\dal} \over r_{12}^{\,\, 2}\, r_{13}^{\vphantom g}\, 
r_{23}^{\vphantom g}} \big (
C^I{}_{\!\!\! rs}1 +{\ts{1\over 6}} \bga_r \ga_t  C^I{}_{\!\!\! ts} +
{\ts{1\over 6}}  C^I{}_{\!\!\! rt} \bga_t \ga_s \big ) \, , \cr
\l J_{rs\alpha\dal}(x_1) \, \vphi^{I_2}(x_2)\, \vphi^{I_3}(x_3) \r ={}&
2\hN i \, {1 \over r_{12}  \, r_{13} \, r_{23}} \,
\X_{1[23] \alpha\dal}\, \big ( C^{I_2} C^{I_3} \big ){}_{[rs]} \, . \cr}
}
In a similar fashion other three point functions may be determined iteratively.
Thus from $\de \l J_{rs\alpha\dal}(x_1) \, \vphi^{I}(x_2) \, 
\bpsi{}_{\smash{v \dbe}} (x_3) \r = 0$ we find
\eqnn\threeJP
$$\eqalignno{
& \hep{}^{\,\beta}(x_1)\ga_{[r}\bga_{s]} \l \Psi_{\beta\alpha\dal}(x_1) \,
\vphi^I(x_2) \, \bpsi{}_{\smash{v\dbe}}(x_3) \r
- \hep{}^{\,\beta}(x_2)\ga_t C^I{}_{\!\!\! tu} \l J_{rs\alpha\dal}(x_1) \,
\psi_{u\beta}(x_2) \, \bpsi{}_{\smash{v\dbe}}(x_3) \r  \cr
&{}+ \hep{}^{\,\beta}(x_3) \ga_u \l J_{rs\alpha\dal}(x_1) \, \vphi^I(x_2) \, 
J_{\smash{uv \beta\dbe}} (x_3) \r + {\ts{1\over 6}}
\hep{}^{\,\beta}(x_3) \ga_t \bga_u \ga_v \l J_{rs\alpha\dal}(x_1) \,
\vphi^I(x_2) \, J_{\smash{tu \beta\dbe}} (x_3) \r \cr
& {}=  4\hN  
{1\over r_{12}^{\vphantom g}\, r_{13}^{\,\, 2}\, r_{23}^{\vphantom g}} 
\Big ( 6 \hep{}^{\,\beta}(x_3) \, \x_{31\beta\dal}
\x_{\smash{13\alpha\dbe}}{1\over r_{13}} - \hep{}^{\,\beta}(x_1) \big (
2 \X_{\smash{1[23] \alpha\dbe}} \x_{\smash{13\alpha\dbe}} -
\X_{\smash{1[23] \alpha\dal}} \x_{\smash{13\beta\dbe}} \big ) \Big ) \cr
& \qquad \qquad \qquad \qquad \qquad {} \times 
\ga_{[r} \big ( C^I{}_{\!\!\! s]v} + {\ts{1\over 6}}
\bga{}_{s]}\ga_t C^I{}_{\!\!\! tv} + {\ts{1\over 6}} C^I{}_{\!\!\! s]t}
\bga{}_t \ga_v \big ) \cr
&\quad {} + 2\hN  
{1\over r_{12}\, r_{13} \, r_{23}} 
\bigg ( \hep{}^{\,\beta}(x_3) \Big ( \x_{31 \beta\dal}
\x_{\smash{13\alpha\dbe}}{1\over 
r_{13}^{\,\, 2}} + \X_{\smash {3[12]\beta\dbe}} \X_{\smash{1[23] \alpha\dal}}
\Big )  + 2 \hep{}^{\,\beta}(x_1) \x_{\smash{13\beta\dbe}}
\X_{\smash{1[23] \alpha\dal}}{1\over r_{13}} \bigg ) \cr
& \qquad \qquad \qquad \qquad \qquad {} \times \ga_u \big (
C^I{}_{\!\!\! u[r} \de_{s]v} + C^I{}_{\!\!\! v[r} \de_{s]u}\big ) \, .  & 
\threeJP \cr}
$$
Using \Xep\ again this may be decomposed to give
\eqnn\JJP
$$\eqalignno{
\l J_{rs\alpha\dal}(x_1) \, \vphi^I(x_2) \, J_{\smash{uv \beta\dbe}} (x_3) \r
={}& 4\hN \, {1\over r_{12} \, r_{13} \, r_{23}} \Big ({1\over r_{13}^{\,\, 2}}
\x_{\smash{13\alpha\dbe}} \x_{31 \beta\dal} 
- \X_{\smash{1[23] \alpha\dal}} \X_{\smash {3[12]\beta\dbe}} \Big )
C^I{}_{\!\!\! [u[r} \de_{s]v]} \, , \cr
\l \Psi_{\alpha\beta\dal}(x_1) \,
\vphi^I(x_2) \, \bpsi{}_{\smash{v\dbe}}(x_3) \r = {}&
{4\over 3}\hN  \, 
{1\over r_{12}^{\vphantom g}\, r_{13}^{\,\, 2} \, r_{23}^{\vphantom g}}\,
\X_{\smash{1[23] (\alpha\dal}}\, \x_{\smash{13\beta)\dbe}} \,
\ga_u C^I{}_{\!\!\! uv} \, , & \JJP \cr}
$$
and also
\eqnn\JPP
$$\eqalignno{ \!\!\!\!
\l & J_{rs\alpha\dal}(x_1)\,\psi_{u\beta}(x_2)\,\bpsi{}_{\smash{v\dbe}}(x_3)\r 
\cr 
& {}
= - 2\hN \bigg \{2 \, {\X_{\smash{1[23]\alpha\dal}}\, \x_{\smash{23\beta\dbe}} 
\over r_{12}^{\vphantom g} \, r_{13}^{\vphantom g} \, r_{23}^{\,\, 2} } 
\big ( \de_{u[r}\de_{s]v}1 + {\ts{1\over 6}} \bga{}_u \ga_{[r} \de_{s]v} 
+ {\ts{1\over 6}} \de_{u[r} \bga{}_{s]} \ga_v + {\ts{1\over 36}} 
\bga{}_u\ga_{[r} \bga{}_{s]} \ga_v \big ) \cr
&\qquad\qquad {}+ {\x_{21\beta\dal}\,  \x_{\smash{13 \alpha\dbe}} \over
r_{12}^{\,\, 2} \, r_{13}^{\,\, 2} \, r_{23}^{\vphantom g} } \big ( \de_{uv} 
\bga_{[r} \ga{}_{s]} + {\ts{2\over 3}} \bga{}_u \ga_{[r} \de_{s]v} +
{\ts{2\over 3}} \de_{u[r} \bga{}_{s]} \ga_v + {\ts {5\over 18}}
\bga{}_u\ga_{[r} \bga{}_{s]} \ga_v \big )\bigg \}  \, .  & \JPP \cr}
$$
Similarly from $\de \l \Psi_{\alpha\beta\dal}(x_1) \, \vphi^{I_2} (x_2)\, 
\vphi^{I_3}(x_3) \r = 0$ we get
\eqn\threeppT{
\l T_{\smash{\alpha\beta\dal\dbe}}(x_1) \, \vphi^{I_2} (x_2)\, 
\vphi^{I_3}(x_3) \r = - {4\over 3}\hN \, 
{1\over r_{12} \, r_{13} \, r_{23}}\, 
\X_{\smash{1[23](\alpha\dal}} \X_{\smash{1[23]\beta)\dbe}}\, \de^{I_2 I_3} \, .
}

At the next stage, for $\sum \Delta_i =9$, 
we obtain from $\de \l J_{rs\alpha\dal}(x_1) \, 
J_{\smash{uv \beta\dbe}} (x_2) \, \bpsi{}_{w \dga} (x_3) \r =0$, after
some calculation, 
\eqna\threeJJJ
$$\eqalignno{ 
& \l J_{rs\alpha\dal}(x_1) \, \Psi_{\smash{\beta\gamma\dbe}} (x_2) \, 
\bpsi{}_{w \dga} (x_3) \r \cr
&{} = {8\over 3}\hN i \bigg \{
\Big ( {1\over r_{12}^{\,\, 2}}\x_{\smash{12\alpha\dbe}}\,
\x_{\smash{21(\beta\dal}} - \X_{\smash{1[23]\alpha\dal}}
\X_{\smash{2[31](\beta\dbe}}\Big ) \x_{\smash{23\gamma)\dga}} 
 + 2 {r_{23}\over r_{12}r_{13}}\, \x_{\smash{13\alpha\dga}} \,
\x_{\smash{21(\beta\dal}} \X_{\smash{2[31]\gamma)\dbe}}\bigg \} \cr
& \qquad \qquad\qquad\qquad \times
{1\over r_{12}^{\vphantom g}\, r_{13}^{\vphantom g} \, r_{23}^{\,\, 2} } 
\big ( \ga_{[r} \de_{s]w} + {\ts {1\over 6}} \ga_{[r} \bga_{s]} \ga_w \big)\, ,
& \threeJJJ{a} \cr
& \l J_{rs\alpha\dal}(x_1) \, J_{\smash{uv \beta\dbe}} (x_2) \,
 J_{\smash{tw \ga\dga}} (x_3) \r \cr
&\quad {}= 8\hN i \bigg \{ 5 \big ( 
\x_{\smash{12\alpha\dbe}}\, \x_{\smash{23\beta\dga}} \, \x_{\smash{31\ga\dal}} 
-\x_{\smash{13\alpha\dga}}\, \x_{\smash{32\ga\dbe}} \, \x_{\smash{21\beta\dal}} 
\big ) 
{1\over r_{12}\, r_{13} \, r_{23} }\cr
&\qquad\qquad\quad
{} - 2 \, \X_{\smash{1[23]\alpha\dal}} \, \X_{\smash{2[31]\beta\dbe}}
 \, \X_{\smash{3[12]\ga\dga}} \bigg \} {1\over r_{12} \, r_{13} \, r_{23}}\,
\de_{[t[r}\, \de_{s][u}\, \de_{v]w]} \cr 
&\qquad
{} + 4\hN \big ( 
x_{\smash{12\alpha\dbe}}\, \x_{\smash{23\beta\dga}} \, \x_{\smash{31\ga\dal}} 
+\x_{\smash{13\alpha\dga}}\, \x_{\smash{32\ga\dbe}} \, \x_{\smash{21\beta\dal}} 
\big ) 
{1\over r_{12}^{\,\, 2}\, r_{13}^{\,\, 2} \, r_{23}^{\,\, 2} }\, 
\vep_{rsuvtw}  \, .  & \threeJJJ{b} \cr}
$$
To achieve this form requires the use of the identity
\eqn\xxx{\eqalign{
\big (& \x_{\smash{12\alpha\dbe}}\, \x_{\smash{23\beta\dga}}
\, \x_{\smash{31\ga\dal}} - \x_{\smash{13\alpha\dga}}\, \x_{\smash{32\ga\dbe}} \,
\x_{\smash{21\beta\dal}} \big ) {1\over r_{12} \, r_{13} \, r_{23}} \cr
&{}= {1\over r_{12}^{\,\, 2}} \x_{\smash{12\alpha\dbe}}\, 
\x_{\smash{21\beta\dal}} \, \X_{\smash{3[12]\ga\dga}} + 
{1\over r_{13}^{\,\, 2}} \x_{\smash{13\alpha\dga}} \, \x_{\smash{31\ga\dal}}\,
\X_{\smash{2[31]\beta\dbe}} + {1\over r_{23}^{\,\, 2}}
\x_{\smash{23\beta\dga}}\, \x_{\smash{32\ga\dbe}}\,
\X_{\smash{1[23]\alpha\dal}} \cr
&\qquad {} + \X_{\smash{1[23]\alpha\dal}} \,
\X_{\smash{2[31]\beta\dbe}} \,\X_{\smash{3[12]\ga\dga}} \, . \cr}
}
Using also $\de \l T_{\smash{\alpha\beta\dal\dbe}}(x_1) \, \vphi_{rs} (x_2) \,
\bpsi{}_{\smash{u\dga}}(x_3) \r = 0$ we obtain
\eqn\Tpp{\eqalign{
\l & T_{\smash{\alpha\beta\dal\dbe}}(x_1) \, \psi_{r\gamma}(x_2) \,
\bpsi{}_{\smash{s\dga}}(x_3) \r \cr
&{} = - {8\over 3}\hN i \Big ( {1\over r_{23}}
\X_{\smash{1[23](\alpha\dal}} \X_{\smash{1[23]\beta)\dbe}} \, 
\x_{\smash{23\gamma \dga}}  - {3\over r_{12}\, r_{13}} 
\X_{\smash{1[23](\alpha(\dal}} \, \x_{\smash{13 \beta)\dga}}
\, \x_{\smash{21 \gamma \dbe)}} \Big ) \cr
& \qquad \qquad \quad {}\times {1\over r_{12} \, r_{13} \, r_{23}}\,  
\big ( \de_{rs} 1 + {\ts{1\over 6}} \bga_r \ga_s \big )\, . \cr}
}

For $\sum \Delta_i = 10$, 
using now $\de \langle T_{\smash{\alpha\beta\dal\dbe}}(x_1) \, 
J_{\smash{rs\gamma\dga}}(x_2) \,\bpsi{}_{\smash{v\dde}}(x_3) \rangle =0$, 
we obtain from \threeJJJ{a} and \Tpp
\eqn\TJJ{\eqalign{\!\!\!\!
\l & T_{\smash{\alpha\beta\dal\dbe}}(x_1) \, 
J_{\smash{rs\gamma\dga}}(x_2) \, J{}_{\smash{uv\de\dde}}(x_3) \r \cr
&{}= - {16\over 3}\hN \, {1\over  r_{12}\, r_{13} \, r_{23} }
\bigg \{ {2\over  r_{12}^{\,\, 2}\, r_{13}^{\,\, 2} }
\, \x_{\smash{12(\alpha\dga}}\, \x_{\smash{13\beta)\dde}}
\, \x_{\smash{21\ga(\dal}}\,  \x_{\smash{31\de\dbe)}} \cr
&\qquad\qquad {}+ \X_{\smash{1[23](\alpha\dal}}\, \X_{\smash{1[23]\beta)\dbe}} 
\Big( 3 \X_{\smash{2[31]\gamma\dga}} \,\X_{\smash{3[12]\de\dde}} +
{1\over r_{23}^{\,\, 2}} \x_{\smash{23\ga\dde}}\x_{\smash{32\de\dga}}\Big)\cr
& \qquad\quad 
{}- {8\over  r_{12}\, r_{13} \, r_{23} }\,\X_{\smash{1[23](\alpha(\dal}}
\big ( \x_{\smash{12\beta)\dga}}\, \x_{\smash{23\ga\dde}}
\, \x_{\smash{31\de\dbe)}} - \x_{\smash{13\beta)\dde}}\,\x_{\smash{32\de\dga}}\,
\x_{\smash{21\ga\dbe)}} \big ) \bigg \} 
\de_{u[r}\de_{s]v} \, . \cr}
}
Further from $\de \langle \Psi_{\smash{\alpha\delta\dal}} (x_1) \, 
J_{\smash{uv \beta \dbe}}(x_2) \, J_{\smash{tw \gamma \dga}}(x_3) \rangle =0$
we may determine
\eqnn\PJP
$$\eqalignno{
\l & \Psi_{\smash{\alpha\delta\dal}} (x_1) \, J_{\smash{uv \beta \dbe}}(x_2) \,
\bPsi_{\smash{\gamma \dga\dde}}(x_3) \r \cr
&{} = {8\over 9}\hN  \bigg \{
{1\over r_{12}^{\vphantom g}\, r_{13}^{\,\, 2} \, r_{23}^{\vphantom g}}
\bigg ( 4 \, \X_{\smash{1[23](\alpha\dal}}\,
\X_{\smash{2[31]\beta\dbe}} \, \X_{\smash{3[12]\gamma(\dga}} 
+ \X_{\smash{2[31]\beta\dbe}} \, \x_{\smash{13(\alpha (\dga}} \, 
\x_{\smash{31\ga\dal}} \, {1\over  r_{13}^{\,\, 2}} \cr
&\qquad \qquad{}+ \big ( 11\, \x_{\smash{13(\alpha(\dga}}\, 
\x_{\smash{32\ga\dbe}} \,
\x_{\smash{21\beta\dal}} - 4 \,  \x_{\smash{12(\alpha\dbe}}\,
\x_{\smash{23\beta(\dga}}  \, \x_{\smash{31\ga\dal}} \big )
{1 \over  r_{12}\, r_{13} \, r_{23}} \bigg ) \x_{\smash{ 13\de)\dde)}} \cr
&\qquad \qquad{}+  
{2 \over r_{12}^{\,\, 2}\, r_{13}^{\vphantom g} \, r_{23}^{\,\, 2}} \,  
\X_{\smash{1[23](\alpha\dal}}\, \X_{\smash{3[12]\gamma(\dga}}\,
\x_{\smash{12\de)\dbe}}\, \x_{\smash{23\beta\dde)}} \bigg \} \ga_{[u} \bga_{v]}\, .
& \PJP \cr}
$$

Considering now  $\sum \Delta_i = 11$ we analyse
$\de \langle T_{\smash{\alpha\beta\dal\dbe}}(x_1) \,
J_{\smash{rs\gamma\dga}}(x_2) \, 
\bPsi{}_{\smash{\ep\dep\deta}}(x_3) \rangle =0$ to obtain
\eqnn\TPP
$$\eqalignno{
& \l  T_{\smash{\alpha\beta\dal\dbe}}(x_1) \, 
\Psi_{\smash{\gamma\de\dga}}(x_2) \, \bPsi_{\smash{\ep\dep\deta}}(x_3) \r \cr
&{} = - {64\over 3}i \hN  \bigg \{
{1\over r_{12}^{\vphantom g}\, r_{13}^{\vphantom g} \, r_{23}^{\, \, 2}}\,
\X_{\smash{1[23](\alpha(\dal}}\, \bigg (\half\, 
\X_{\smash{1[23]\beta)\dbe)}} \, \x_{\smash{32\ep\dga}} \,
\x_{\smash{23(\gamma(\dep}} {1\over r_{23}^{\, \, 2}}  \cr
&\qquad \qquad \quad {}+  
\X_{\smash{2[31](\gamma\dga}} \, \x_{\smash{13\beta)(\dep}} \,
\x_{\smash{31\ep\dbe)}} {1\over r_{13}^{\, \, 2}} +
\X_{\smash{3[12]\ep(\dep}} \, \x_{\smash{12\beta)\dga}} \,
\x_{\smash{21(\ga\dbe)}} {1\over r_{12}^{\, \, 2}} \cr
&\qquad \qquad \quad {}+ \big ( {\ts{7\over 6}} \,
\x_{\smash{12\beta)\dga}}\,
\x_{\smash{23(\gamma(\dep}}  \, \x_{\smash{31\ep\dbe)}} 
- {\ts{5\over 3}} \, \x_{\smash{13\beta)(\dep}}\, \x_{\smash{32\ep\dga}} \,
\x_{\smash{21(\gamma\dbe)}} \big )
{1 \over  r_{12}\, r_{13} \, r_{23}} 
\bigg ) \x_{\smash{23 \de) \deta)}} \qquad\quad \cr 
&\quad {}+ 
{1\over r_{12}^{\, \, 2}\, r_{13}^{\, \, 2} \, r_{23}^{\vphantom g}}
\bigg ( {- {\ts{3\over 2}}}\, \X_{\smash{2[31](\gamma\dga}} \, 
\x_{\smash{13(\alpha(\dep}} \, \x_{\smash{31\ep(\dal}} {1\over r_{13}^{\, \, 2}}  
- {\ts{3\over 2}}\, \X_{\smash{3[12]\ep(\dep}} \, \x_{\smash{12(\alpha\dga}} \,
\x_{\smash{21(\ga(\dal}} {1\over r_{12}^{\, \, 2}} \cr
&{} \quad {}+ \big ( {\ts{1\over 3}}\, \x_{\smash{12(\alpha\dga}}\,
\x_{\smash{23(\gamma(\dep}}  \, \x_{\smash{31\ep(\dal}} 
+2  \, \x_{\smash{13(\alpha(\dep}}\, \x_{\smash{32\ep\dga}} \,
\x_{\smash{21(\gamma(\dal}} \big )
{1 \over  r_{12}\, r_{13} \, r_{23}} \bigg ) 
\x_{\smash{21\de)\dbe)}}\, \x_{\smash{13\beta)\deta)}}\! \bigg \}1\, ,
& \TPP \cr}
$$
where the symmetrisations act on pairs of indices at the same point.
To achieve the form \TPP\ we use the identity \xxx\ as well as
\eqn\xid{\eqalign{
\X_{\smash{1[23][\alpha\dal}} \, \X_{\smash{2[31]\gamma\dga}} \, 
\x_{\smash{13\beta]\dep}} {1\over r_{13}}
{}& - \X_{\smash{1[23][\alpha\dal}} \,  \x_{\smash{12\beta]\dga}} \,
\x_{\smash{23\gamma\dep}} \, {1\over r_{12}\, r_{23}} \cr
&{} +
\x_{\smash{12[\alpha\dga}} \, \x_{\smash{21\ga\dal}} \,
\x_{\smash{13\beta]\dep}}  \, 
{1 \over  r_{12}^{\, \, 2} \, r_{13}^{\vphantom g}} =0 \, , \cr} 
}
together with various permutations.

Finally from $\de \langle T_{\smash{\alpha\beta\dal\dbe}}(x_1) \,
T_{\smash{\gamma\de\dga\dde}}(x_2) \, 
\bPsi{}_{\smash{\ep\dep\deta}}(x_3) \rangle =0$ we determine the
energy momentum tensor three point function, with similar conventions on 
symmetrisations of indices,
\eqnn\TTT
$$\eqalignno{
& \l  T_{\smash{\alpha\beta\dal\dbe}}(x_1) \, 
T_{\smash{\gamma\de\dga\dde}}(x_2) \, T_{\smash{\ep\eta\dep\deta}}(x_3) \r \cr
&{} = {128\over 3} \hN  \bigg \{
{1\over r_{12}^{\vphantom g}\, r_{13}^{\vphantom g} \, r_{23}^{\vphantom g}}\,
\bigg ( {-3} \X_{\smash{1[23](\alpha\dal}}\,  \X_{\smash{1[23]\beta)\dbe}} \, 
\X_{\smash{2[31](\gamma\dga}} \, \X_{\smash{2[31]\de)\dde}} \,
\X_{\smash{3[12](\ep\dep}} \, \X_{\smash{3[12]\eta)\deta}} \cr
& \qquad\qquad\qquad\qquad\qquad{}+ \X_{\smash{1[23](\alpha\dal}}\,
\X_{\smash{1[23]\beta)\dbe}} \, \x_{\smash{23(\gamma\dep}}\, \x_{\smash{23\de)\deta}} 
\, \x_{\smash{32(\ep\dga}}\, \x_{\smash{32\eta)\dde}} \, {1\over r_{23}^{\, \, 4}}  \cr
& \qquad\qquad\qquad\qquad\qquad {}+ \X_{\smash{2[31](\gamma\dga}}\,
\X_{\smash{2[31]\de)\dde}} \, \x_{\smash{31(\ep\dal}}\, \x_{\smash{31\eta)\dbe}} \, 
\x_{\smash{13(\alpha\dep}}\, \x_{\smash{13\beta)\deta}}\, {1\over r_{13}^{\, \, 4}}  \cr
& \qquad\qquad\qquad\qquad\qquad {}+ \X_{\smash{3[12](\ep\dep}}\,
\X_{\smash{3[12]\eta)\deta}} \, \x_{\smash{12(\alpha\dga}}\, \x_{\smash{12\beta)\dde}} 
\, \x_{\smash{21(\gamma\dga}}\, \x_{\smash{21\de)\dbe}} \, {1\over r_{12}^{\, \, 4}} 
\, \bigg ) \cr
&\ \ {} + {13\over  r_{12}^{\, \, 2}\, r_{13}^{\, \, 2}\, r_{23}^{\, \, 2} }\,
\X_{\smash{1[23](\alpha(\dal}}\, \X_{\smash{2[31](\gamma(\dga}}\,
\X_{\smash{3[12](\ep(\dep}}\,
\big ( \x_{\smash{12\beta)\dde)}}\, \x_{\smash{23\de)\deta)}}
\, \x_{\smash{31\eta)\dbe)}} - \x_{\smash{13\beta)\deta)}}\,\x_{\smash{32\eta)\dde)}}\,
\x_{\smash{21\de)\dbe)}} \big ) \cr
&\ \ {} - {1\over  r_{12}^{\, \, 3}\, r_{13}^{\, \, 3}\, r_{23}^{\, \, 3} }\,
\Big ( {\ts{49\over 3}} \big ( \x_{\smash{12(\alpha\dga}}\, \x_{\smash{12\beta)\dde}}\, 
\x_{\smash{23(\gamma\dep}}\, \x_{\smash{23\de)\deta}}
\, \x_{\smash{31(\ep\dal}} \, \x_{\smash{31\eta)\dbe}} +
\x_{\smash{13(\alpha\dep}}\,\x_{\smash{13\beta)\deta}}\,\x_{\smash{32(\ep\dga}}\,
\x_{\smash{32\eta)\dde}}\, \x_{\smash{21(\gamma\dal}} \,
\x_{\smash{21\de)\dbe}} \big ) \cr
&\qquad \qquad \qquad\qquad\quad  {} -  {\ts{34\over 3}} \, 
\x_{\smash{12(\alpha(\dga}}\, \x_{\smash{21(\gamma(\dal}} \, \x_{\smash{23\de)(\dep}}
\, \x_{\smash{32(\eta\dde)}}\, \x_{\smash{31\eta)\dbe)}} \, \x_{\smash{13\beta)\deta)}}
\Big ) \bigg \}  .
& \TTT \cr}
$$
This satisfies the necessary symmetry requirements.
The same three point function has also been calculated directly using the AdS/CFT
correspondence in \Three.

As a check on the above results for two and three point functions
we have verified in each case the appropriate  Ward
identities which restrict the operator product expansions involving the $SU(4)$ current
and the energy momentum tensor. From \hughtwo\ we have
\eqn\OPEJ{
x^a J_{rs\, a}(x) \O(0) \sim i \, {1\over x^2}\, T_{rs} \O(0) + \dots \, ,\qquad
J_{rs\, a} = - \half (\tsi_a)^{\dal\alpha}J_{rs\, \alpha\dal} \, ,
}
where $T_{rs}=-T_{sr}$ are the generators of $SU(4)$ acting on $\O$ obeying
\eqn\algT{
[ T_{rs}, T_{uv}] = - \de_{ru} T_{sv} + \de_{rv} T_{su} + \de_{su} T_{rv}
- \de_{sv} T_{ru}  \, ,
}
(for a 6-vector $v_p$ $T_{rs}\to 2\de_{p[r}\de_{s]q}$ and for a spinor
$\psi_i$ $T_{rs}\to - \half (\bga_{[r}\ga_{s]})_i{}^{j}$) and also
\eqn\OPET{
x^a x^bT_{ab}(x) \O(0) \sim - {1\over x^2}\, \Delta \O(0) + \dots \, ,\qquad
T_{ab} = \quar (\tsi_a)^{\dal\alpha}(\tsi_b)^{\dbe\beta}
T_{\smash{\alpha\beta\dal\dbe}} \, .
}
In \OPEJ\ and \OPET\ $\dots $ denote terms which vanish on integration over
$\d \Omega_{\hat x}$. More generally for the operator product expansion of the
energy momentum tensor itself we have
\eqn\OPETT{\eqalign{
x^a x^bT_{ab}(x) T_{cd}(0) \sim - {1\over x^2} \Big (& A\, T_{cd}(0)
+ B \,{1\over x^2} x^e \big ( x_{(c} T_{d)e}(0) - \quar \eta_{cd} x^f
T_{ef}(0) \big ) \cr
&{} + C \, {1\over (x^2)^2} \big ( x_c x_d - \quar \eta_{cd} x^2
\big )  x^e x^f T_{ef}(0) \Big ) \, . \cr}
}
For compatibility with \OPET\ we must have
\eqn\sABC{
A + \quar B +{\ts {1\over 12}} C = 4 \, .
}
The coefficients $A, B , C$, along with the coefficient $C_T$ of the energy
momentum tensor two point function, determine fully the conformal three point
function of the energy momentum tensor\foot{In terms of the coefficients in
\hughtwo, with the normalisations here, $C_T A= {1\over 4}(\A-3\C)$,
$C_TB = 5\A + {1\over 2} \B - 6\C$, $C_TC = - {7\over 2}(\B-2\C)$. For free
scalars $A={7\over 9}, \, B = {88\over 9}, \, C= {28\over 9}$, for free
spin-${1\over 2}$ fields $A={3\over 2}, \, B=10, \, C=0$ and for free vectors
$A=4, \, B=C=0$.}. From \TTT\ we have, along with $C_T=40\hN$,
\eqn\ABCf{
A= {106\over 45} \, , \qquad B= {268\over 45} \, , \qquad C={28\over 15} \, .
}
These satisfy \sABC\ and also
\eqn\cABC{
A - {\ts {7\over 4}} B + {\ts {121\over 28}}\, C= 0 \, ,
}
which is a necessary condition for $\N=1$ superconformal symmetry \HO.
Furthermore in terms of the coefficients $a,c$ of the energy momentum
tensor trace on curved space we have, from the results in \hughtwo,
\eqn\ac{
{a\over c} = {1\over 288}(124 A - B + C) \, ,
}
and \ABCf\  gives $a=c$ as expected for $\N=4$ superconformal symmetry.

\newsec{Further Applications to Three Point Functions}

Superconformal symmetry leads to further constraints on the correlation
functions involving operators belonging to short multiplets. We here
discuss the conditions for a three point function for two chiral
primary operators $\vphi_{rs}$, whose superconformal transformation
properties were described in section two, and a third operator belonging 
to a long multiplet. For simplicity we consider first a self-conjugate
scalar operator $\Phi^I ={\bar \Phi}^I$ where $I$ is an index
for the $SU(4)$ representation space, the representation having
Dynkin labels $[q,p,q]$. Under a superconformal transformation we assume
\eqn\deP{
\de \Phi^I = \hep_i^\alpha \Psi_{\alpha}^{Ii} + {\bar \Psi}{}^I_{i\dal}
\hbep{}^{\, i\dal} \, ,
}
where in general $\Psi_{\alpha}^{Ii}, {\bar \Psi}{}^I_{i\dal}$ transform
under reducible $SU(4)$ representations. To ensure closure of the
algebra we take
\eqna\clP$$
\eqalignno{
\de_{\hbep} \Psi^I{}_{\!\! \alpha} = {} & i \pr_{\alpha\dal} \Phi^I \,
\hbep{}^{\,\dal} + 2\Delta\, \eta_\alpha \Phi^I - \ga_{[r}\bga_{s]} \eta_\alpha
\, (T_{rs})^I{}_{\! J} \Phi^J \cr
{}& - {1\over 2\Delta}\, (T_{rs})^I{}_{\! J}  \, i\pr_{\alpha\dal} \Phi^J
\, \ga_{[r}\bga_{s]} \hbep{}^{\,\dal} + \J^I{}_{\!\!\alpha\dal}\hbep{}^{\,\dal} 
\, , & \clP{a} \cr
\de_{\hep} {\bar \Psi}{}^I{}_{\!\! \dal}=  {}& - \hep^\alpha  i \pr_{\alpha\dal} 
\Phi^I  + 2\Delta \, \Phi^I \, \bet{}_\dal + (T_{rs})^I{}_{\! J} \Phi^J
\, \bet{}_\dal \ga_{[r}\bga_{s]} \cr
{}& - {1\over 2\Delta} \,   \hep^\alpha \ga_{[r}\bga_{s]} \, 
(T_{rs})^I{}_{\! J} \, i \pr_{\alpha\dal} \Phi^J + \hep^\alpha 
\J^I{}_{\!\!\alpha\dal} \, , & \clP{b} \cr }
$$
where $\J^{Ii}{}_{\!\!j\alpha\dal}$ is a new vector field. From \deP\ and
\clP{a,b} we obtain
\eqn\closP{
[\de_2, \de_1] \Phi^I = - v{\cdot \pr} \Phi^I - \Delta \, \hla \Phi^I 
+ \half \hht_{rs} (T_{rs})^I{}_{\! J} \Phi^J \, ,
}
where $\hla$ and $\hht_{rs}$ are given by \defla\ and \lfour\ and $v$
by \vone. The result \closP\ is just
as required by closure of the $\N=4$ superconformal algebra for 
$T_{rs}$ the appropriate generators of $SU(4)$ obeying \algT.
The terms in the second lines of \clP{a} and \clP{b} do not contribute
to the right hand side of \closP\ but are necessary for 
$\de_{\hbep} \Psi{}_{\alpha}{}^{\!\! I}$ and 
$\de_{\hep} {\bar \Psi}{}^I{}_{\!\! \dal}$ to transform covariantly. To
see this we consider the conformal transformations, following \dO,
\eqn\dPP{
\de_v \Phi^I =  - v{\cdot \pr} \Phi^I - \Delta \, \hla \Phi^I \, ,
\qquad \de_v  {\bar \Psi}{}^I{}_{\!\! \dal} = -
v{\cdot \pr} {\bar \Psi}{}^I{}_{\!\! \dal} - (\Delta +\half) \, \hla
{\bar \Psi}{}^I{}_{\!\! \dal} -  {\bar \Psi}{}^I{}_{\smash{\!\! \dbe}}\,
\hbom{}^\dbe{}_{\dal} \, ,
}
where $v, \, \hla$ and $\hbom{}^\dbe{}_{\dal}$ are as in \vcon\ and \wwh.
The additional terms involving $1/\Delta$ are necessary to achieve
\eqn\clve{
[\de_\hep, \de_v ] {\bar \Psi}{}^I{}_{\!\! \dal} = 
\de_{\hep'} {\bar \Psi}{}^I{}_{\!\! \dal} \, ,
}
where\foot{To verify \clve\ it is useful to note that $(\pr_{\alpha\dal}v)
{\cdot \pr} = \hbom{}^\dbe{}_{\dal} \pr_{\smash{\alpha\dbe}} -
\hom_\alpha{}^\beta \pr_{\beta\dal} + \hla \pr_{\alpha\dal}$.}
\eqn\epp{
\hep{}^{\, \prime\alpha} = - v{\cdot \pr} \hep{}^{\, \alpha} + \half
\hla \, \hep{}^{\, \alpha} - \hep{}^{\, \beta} \hom_\beta{}^\alpha \, ,
}
which may easily be decomposed as in \scep\ into $\eta'$ and
\eqn\betp{
\bet{}'{}_{\! \dal} = - \half \hla\, \bet{}_{\dal} - \bet{}_{\smash\dbe}\,
\hbom{}^\dbe{}_{\dal} + i \hep{}^{\, \alpha}b_{\alpha \dal} \, .
}

It is perhaps appropriate to remark that from \clP{a} it is straightforward
to recover the usual conditions for obtaining a short supermultiplet belonging
to the C series \Short\ with a scalar lowest dimension operator (the B
series is discussed in appendix B). The relevant requirement is that,
for a suitable $\Delta$, 
\eqn\sho{
2\Delta\, \eta\, \Phi^I - \ga_{[r}\bga_{s]} \eta\, 
(T_{rs})^I{}_{\! J} \Phi^J \, ,
}
should not, for arbitrary $\eta^i$, span the full representation space
corresponding to the direct product $[1,0,0]\otimes [q,p,q] = [q{+1},p,q]
\oplus [q{-1},p{+1},q]\oplus [q,p{-1},q{+1}] \oplus [q,p,q{-1}]$, so that
there is a non zero cokernel. To illustrate this we may consider the 
$[0,p,0]$ representation when we take 
\eqn\sPhi{
\Phi^I \to \vphi_{u_1  \dots  u_p}= \vphi_{(u_1  \dots  u_p)} \, , \quad
 \vphi_{u_1  \dots u_{p-2}uu}=0 \, .
}
Since $(T_{rs})^I{}_{\! J} \Phi^J \to
p(\de_{r(u_1} \vphi_{u_2\dots u_p)s} - \de_{s(u_1} \vphi_{r_2\dots r_p)r})$
we may obtain for this case
\eqn\shp{
\ga_{[r}\bga_{s]} \eta\, (T_{rs})^I{}_{\! J} \Phi^J \to
2p\, \ga_{(u_1} \vphi_{u_2 \dots u_p)s} \, \bga_s^{\vphantom g} \eta + 2p\,
\vphi_{u_1  \dots  u_p} \, \eta \, .
}
Choosing $\Delta = p$ then projects out in \sho\ the $[1,p,0]$
representation. In \clP{a} the terms involving $\hbep$ have an identical form
to \sho\ and therefore we may restrict, for $\Delta=p$ and with suitable
conditions also on 
$\J^I{}_{\!\!\alpha\dal}$, $\Psi^I{}_{\!\! \alpha}$ to the $SU(4)$
$[0,p-1,1]$ representation by letting 
\eqn\SPsi{\eqalign{
\Psi^{Ii}{}_{\!\!\! \alpha} \to {}&
\Psi^i{}_{\!\! u_1  \dots  u_p \, \alpha} \, ,
\quad  \Psi_{u_1  \dots  u_p \alpha} = \quar p \, \ga_u\bga_{(u_1}\!
\Psi_{u_2  \dots  u_p)u \alpha} \, , \cr
\Rightarrow {}&  \Psi_{u_1  \dots  u_p \alpha} = - 
\ga_{(u_1}\psi_{u_2 \dots u_p)\alpha} \, , \quad
\ga_u \psi_{u\, u_1\dots u_{p-1}\alpha} =0 \, . \cr}
}
For the $[q,0,q]$ representation then similarly 
\eqn\ssPhi{
\Phi^I \to \vphi^{i_1\dots i_q}_{\,j_1\dots j_q} 
= \vphi^{(i_1\dots i_q)}_{\,(j_1\dots j_q)} \, , \qquad
\vphi^{i\, i_2 \dots i_q}_{\,i\, j_2\dots j_q}=0 \, . 
}
For this case 
$(T_{rs})^I{}_{\! J} \Phi^J \to -\half q (\ga_{[r} \bga_{s]})^{(i_1}_{\ \ \, i}
\, \vphi^{i_2\dots i_q)i}_{\, j_1\dots j_q} + \half q \,
\vphi^{i_1\dots i_q}_{j(j_1\dots j_{q-1}}(\ga_{[r} \bga_{s]})^j_{\ j_q)}$
so that, since $(\ga_{[r} \bga_{s]})^i_{\ j} (\ga_{[r} \bga_{s]})^k_{\ \, l} =
- 8\de^i_{\ \, l} \de^k_{\ j} + 2 \de^i_{\ j}\de^k_{\ \, l}$, we have
\eqn\shq{
(\ga_{[r}\bga_{s]})^i_{\ j} \eta^j\, (T_{rs})^I{}_{\! J} \Phi^J \to 
4q \big ( \vphi^{i(i_1\dots i_{q-1}}_{\,j_1\, \dots \, j_q} \eta^{i_q)} -
\de^i_{(j_1} \vphi^{i_1\dots i_q}_{\,j_2\dots j_q)j} \eta^j \big ) \, .
}
It is then easy to see that in \sho\ choosing $\Delta =2q$ removes the
components appropriate  to the $[q{+1},0,q]$ representation. In this
case $\Psi^I{}_{\!\! \alpha}$ is restricted to belong to the
$[q{-1},1,q]$ and $[q,0,q{-1}]$ representations,
\eqn\SSPsi{
\Psi^{Ii}{}_{\!\!\! \alpha}\to \Psi^{i\, i_1\dots i_q}_{\,\,j_1\dots j_q \alpha}
\, , \qquad \Psi^{(i\, i_1\dots i_q)}_{\,\, j_1\dots j_q \, \alpha} = 0 \, .
}
The general case follows
by a combination of the above two examples giving $\Delta =p+2q$ for
a short supermultiplet for lowest weight scalar field belonging to the
self-conjugate $[q,p,q]$ representation.

To illustrate how further constraints, beyond those arising from conformal
invariance arise, we consider then the three point function
\eqn\ppP{
\l \vphi_{rs}(x_1)\, \vphi_{uv}(x_2)\, \Phi^I(x_3)\r = {1\over 
r_{12}^{\,\, 2}} \bigg ({r_{12}\over r_{13}\,r_{23}}\bigg )^{\!\!{1\over 2}\Delta}
D^I{}_{\!\!rs,uv} \, ,
}
for $D^I{}_{\!\!rs,uv}$ a $SU(4)$ invariant tensor. The symmetry condition 
$D^I{}_{\!\!rs,uv}=D^I{}_{\!\!uv,rs}$ here implies that $\Phi^I$
is restricted to belong to the singlet, 20, 84 or 105 dimensional
representations. The conditions for superconformal invariance follow in
a similar fashion to the previous section. We first consider
$\de \langle \psi_{r\alpha}(x_1)\, \vphi_{uv}(x_2)\, \Phi^I(x_3)\rangle=0$
using \defour\ and \deP. For $\Delta \ne 2$ the analysis is simplified
by taking account of the requirement that
$\langle J_{rs\alpha\dal}(x_1)\, \vphi_{uv}(x_2)\, \Phi^I(x_3)\rangle =0$ in
order to comply with the conservation equation $\pr^{\dal\alpha}J_{rs\alpha\dal}
=0$. Using this we then get
\eqn\ppPh{\eqalign{
\l \psi_{r\alpha}(x_1)\, \bpsi{}_{(u\dal}(x_2)\, \Phi^I(x_3)\r \bga_{v)} ={}&
(4-\Delta)\, { i\, \x_{12\alpha\dal} \over r_{12}^{\,\, 3-{1\over 2}\Delta}
r_{13}^{\,\, {1\over 2}\Delta}r_{23}^{\,\, {1\over 2}\Delta}}\,
D^I{}_{\!\!rs,uv} \bga_s \, , \cr
\l \psi_{r\alpha}(x_1)\, \vphi_{uv}(x_2)\, {\bar \Psi}{}^I{}_{\!\dal}(x_3) \r 
={}& \Delta\, { i\, \x_{13\alpha\dal} \over r_{12}^{\,\, 2-{1\over 2}\Delta}
r_{13}^{\,\, 1+ {1\over 2}\Delta}r_{23}^{\,\, {1\over 2}\Delta}}\,
D^I{}_{\!\!rs,uv} \bga_s \, . \cr}
}
The critical conditions follow from $\de \langle \vphi_{rs}(x_1)\, 
\vphi_{uv}(x_2)\, {\bar \Psi}{}^I_{i\dal}(x_3) \rangle =0$. 
Using \clP{b} we may obtain
\eqn\ppdP{\eqalign{
\l \vphi_{rs}(x_1)\, \vphi_{uv}(x_2)\, \de{\bar \Psi}{}^I{}_{\!\dal}(x_3) \r
= {}& \hep^\alpha(x_3)
\l \vphi_{rs}(x_1)\, \vphi_{uv}(x_2)\, \J^I{}_{\!\!\alpha\dal}(x_3)\r\cr
{} - { i \over r_{12}^{\,\,
2-{1\over 2}\Delta} r_{13}^{\,\, {1\over 2}\Delta}r_{23}^{\,\, {1\over 2}\Delta}}&\,\bigg (
\hep^\alpha(x_1)\x_{13\alpha\dal}{1\over r_{13}} +
\hep^\alpha(x_2)\x_{23\alpha\dal}{1\over r_{23}} \bigg )\cr
&{}\times \Big ( \Delta \,
D^I{}_{\!\!rs,uv} + \half (T_{tw})^I{}_{\! J} D^J{}_{\!\!rs,uv}
\ga_{[t}\bga_{w]} \Big ) \, . \cr}
}
Since $D^I{}_{\!\!rs,uv}$ is a $SU(4)$ invariant we have
\eqn\Dinv{
(T_{tw})^I{}_{\! J} D^J{}_{\!\!rs,uv} = -2\de_{r[t}D^I{}_{\!\!w]s,uv}
-2\de_{s[t}D^I{}_{\!\!w]r,uv} + 2D^I{}_{\!\!rs,u[t}\de_{w]v} +
 2D^I{}_{\!\!rs,v[t}\de_{w]u} \, ,
}
so that, with the aid of \ppPh, applying \ppdP\ in the superconformal invariance condition
leads to
\eqn\ppJ{\eqalign{
\hep^\alpha(x_3) &
\l \vphi_{rs}(x_1)\, \vphi_{uv}(x_2)\, \J^I{}_{\!\!\alpha\dal}(x_3)\r \cr
={}& { i \over r_{12}^{\,\,
2-{1\over 2}\Delta} r_{13}^{\,\, {1\over 2}\Delta}r_{23}^{\,\, {1\over 2}\Delta}}\,
\bigg (\hep^\alpha(x_1)\x_{13\alpha\dal}{1\over r_{13}}\Big ( (\Delta-2)
X^I{}_{\!\!rs,uv} -2 Y^I{}_{\!\!rs,uv} \Big ) \cr
\noalign{\vskip -5pt}  
&\qquad \qquad\qquad\qquad{}+ \hep^\alpha(x_2)\x_{23\alpha\dal}{1\over r_{23}} 
\Big ( (\Delta-2) Y^I{}_{\!\!rs,uv} -2 X^I{}_{\!\!rs,uv} \Big ) \bigg ) \, , \cr}
}
for
\eqn\deXY{
X^I{}_{\!\!rs,uv} = \half \ga_{[r}\bga_{t]} D^I{}_{\!\!ts,uv} +
\half \ga_{[s}\bga_{t]} D^I{}_{\!\!tr,uv} \, , \qquad
Y^I{}_{\!\!rs,uv} = \half \ga_{[u}\bga_{t]} D^I{}_{\!\!rs,tv} +
\half \ga_{[v}\bga_{t]} D^I{}_{\!\!rs,tu} \, .
}
For superconformal invariance the right hand side of \ppJ\ must
involve $\hep$ just in the form $\hep^\alpha(x_3) \X_{3[12]\alpha\dal}$ 
and by virtue of \Xep\ this requires
\eqn\cpJ{
(\Delta-4) X^I{}_{\!\!rs,uv}  = - (\Delta-4) Y^I{}_{\!\!rs,uv} \, .
}
This is trivially satisfied if $\Delta=4$ which includes the cases
of short multiplets belonging to the $105$ and $84$ representations
and also the for $\Phi^I$ belonging to the 20 dimensional
representation for which there is no shortening for this $\Delta$.
For $\Phi^I\to \Phi$, a $SU(4)$ singlet, the conditions arising from \ppJ\
are satisfied
for any scale dimension $\Delta$ since then $ X_{rs,uv}=-Y_{rs,uv}$
(for the singlet case the right side of \ppJ\ just involves
$\Delta X_{rs,uv}$ and $\Delta Y_{rs,uv}$).

We now extend this discussion to the case of a self-conjugate operator
of spin $\ell$, 
$\Phi^I{}_{\smash{\! \alpha_1 \dots \alpha_\ell , \dal_1 \dots \dal_\ell}}
= \Phi^I{}_{\smash{\! (\alpha_1 \dots \alpha_\ell) , (\dal_1 \dots \dal_\ell)}}$.
In this case \deP\ becomes
\eqn\dePl{
\de \Phi^I_{\smash{\,\alpha_1 \dots \alpha_\ell,\dal_1 \dots \dal_\ell}} = 
\hep{}_i^\beta
\Psi^{Ii}_{\smash{\,\beta\,\alpha_1 \dots \alpha_\ell,\dal_1 \dots \dal_\ell}} +
{\bar\Psi}{}^I_{\smash {\,i\alpha_1 \dots \alpha_\ell,\dal_1 \dots \dal_\ell\dbe}}
\, \hbep{}^{\, i\dbe} \, .
}
Corresponding to \clP{b} we now have
\eqnn\clPl
$$\eqalignno{
\de_{\hep} 
{\bar\Psi} &{}^I_{\smash {\,\alpha_1 \dots \alpha_\ell,\dal_1 \dots \dal_\ell\dbe}}\cr
= {}& 
-  \hep^\beta i  \pr_{\smash{\beta\dbe}}
\Phi^I{}_{\smash{\! \alpha_1 \dots \alpha_\ell , \dal_1 \dots \dal_\ell}} 
+ 2(\Delta-\ell) \,
\Phi^I{}_{\smash{\! \alpha_1 \dots \alpha_\ell , \dal_1 \dots \dal_\ell}}
\, \bet{}_{\smash{\dbe}} 
+ 4\ell \,
\Phi^I{}_{\smash{\! \alpha_1 \dots \alpha_\ell , \dbe(\dal_1 \dots \dal_{\ell-1}}}
\, \bet{}_{\smash{\dal_\ell})} \cr
&{} + {\ell\over \Delta-1} \,  \hep^\beta i \big ( 
\pr_{\smash{(\alpha_1\dbe}}
\Phi^I{}_{\smash{\! \beta |\alpha_2 \dots \alpha_\ell) , \dal_1 \dots \dal_\ell}}
- \pr_{\smash{\beta(\dal_1}}
\Phi^I{}_{\smash{\! \alpha_1 \dots \alpha_\ell , \dal_2 \dots \dal_\ell)\dbe}}
\big ) \cr
&{} + (T_{rs})^I{}_{\! J} 
\Phi^J{}_{\smash{\! \alpha_1 \dots \alpha_\ell , \dal_1 \dots \dal_\ell}}
\, \bet{}_{\smash{\dbe}} \ga_{[r}\bga_{s]} \cr
{}& - \hep^\beta \ga_{[r}\bga_{s]} \,
(T_{rs})^I{}_{\! J} \, i \Big ( c \,
\pr_{\smash{\beta\dbe}}
\Phi^J{}_{\smash{\! \alpha_1 \dots \alpha_\ell , \dal_1 \dots \dal_\ell}} 
+  e \, \pr_{\smash{(\alpha_1(\dal_1}} \!
\Phi^J{}_{\smash{\! \beta |\alpha_2 \dots \alpha_\ell) , 
\dal_2 \dots \dal_\ell)\dbe }} \cr
& \qquad \qquad\qquad \qquad \ {}
+d \big ( \pr_{\smash{(\alpha_1\dbe}}
\Phi^J{}_{\smash{\! \beta |\alpha_2 \dots \alpha_\ell) , \dal_1 \dots \dal_\ell}}
+ \, \pr_{\smash{\beta(\dal_1}}\!
\Phi^J{}_{\smash{\! \alpha_1 \dots \alpha_\ell , \dal_2 \dots \dal_\ell)\dbe}}
\big ) \Big ) \cr
&{} + \hep^\beta 
\J^I{}_{\!\!\beta\alpha_1 \dots \alpha_\ell,\dal_1 \dots \dal_\ell\dbe} \,  ,
& \clPl \cr}
$$
while the equivalent version of \clP{a} is given by its conjugate. The structure
of \clPl\ is determined as before by the requirement of closure of the
algebra
\eqn\closPl{\eqalign{
[\de_2, \de_1] 
\Phi^I_{\smash{\,\alpha_1 \dots \alpha_\ell,\dal_1 \dots \dal_\ell}}
= {}& - \big ( v{\cdot \pr} + \Delta \, \hla \big ) 
\Phi^I_{\smash{\,\alpha_1 \dots \alpha_\ell,\dal_1 \dots \dal_\ell}} 
+ \half \hht_{rs} (T_{rs})^I{}_{\! J} 
\Phi^J_{\smash{\,\alpha_1 \dots \alpha_\ell,\dal_1 \dots \dal_\ell}} \cr
&{} + \ell \, \hom_{(\alpha_1}{}^{\! \beta}
\Phi^I_{\smash{\,\beta |\alpha_2 \dots \alpha_\ell),\dal_1 \dots \dal_\ell}}
- \ell \, 
\Phi^I_{\smash{\,\alpha_1 \dots \alpha_\ell,\dbe(\dal_1 \dots \dal_{\ell-1}}} 
\, \hbom{}^\dbe{}_{\smash{\dal_\ell)}} \, , \cr}
}
and also that $\de_{\hep}
{\bar \Psi}{}^I_{\smash {\,\alpha_1 \dots \alpha_\ell,\dal_1 \dots \dal_\ell\dbe}}$
has the correct form under conformal transformations which requires
\eqn\cllv{
[\de_\hep, \de_v ]
{\bar \Psi}{}^I_{\smash {\,\alpha_1 \dots \alpha_\ell,\dal_1 \dots \dal_\ell\dbe}}
= \de_{\hep'}
{\bar\Psi}{}^I_{\smash {\,\alpha_1 \dots \alpha_\ell,\dal_1 \dots \dal_\ell\dbe}} \, ,
}
where, with notation as in \vcon\ and \wwh,
\eqn\cvl{\eqalign{
\de_v \Phi^I_{\smash{\,\alpha_1 \dots \alpha_\ell,\dal_1 \dots \dal_\ell}}
= {}& - \big ( v{\cdot \pr} + \Delta \, \hla \big )
\Phi^I_{\smash{\,\alpha_1 \dots \alpha_\ell,\dal_1 \dots \dal_\ell}}\cr
&{} + \ell \, \hom_{(\alpha_1}{}^{\! \beta}
\Phi^I_{\smash{\,\beta |\alpha_2 \dots \alpha_\ell),\dal_1 \dots \dal_\ell}}
- \ell \,
\Phi^I_{\smash{\,\alpha_1 \dots \alpha_\ell,\dbe(\dal_1 \dots \dal_{\ell-1}}}
\, \hbom{}^\dbe{}_{\smash{\dal_\ell)}} \, , \cr
\de_v 
{\bar\Psi}{}^I_{\smash {\,\alpha_1 \dots \alpha_\ell,\dal_1 \dots \dal_\ell\dbe}}
= {}& - \big ( v{\cdot \pr} + (\Delta+\half) \, \hla \big )
{\bar\Psi}{}^I_{\smash {\,\alpha_1 \dots \alpha_\ell,\dal_1 \dots \dal_\ell\dbe}}\cr
&{} + \ell \, \hom_{(\alpha_1}{}^{\! \beta}
{\bar\Psi}{}^I_{\smash{\,\beta |\alpha_2 \dots \alpha_\ell),\dal_1 \dots \dal_\ell\dbe}}
- \ell \,
{\bar\Psi}{}^I_{\smash {\,\alpha_1 \dots \alpha_\ell,\dga(\dal_1 \dots 
\dal_{\ell-1}\dbe}} \, \hbom{}^\dga{}_{\smash{\dal_\ell)}} \cr
&{}- {\bar\Psi}{}^I_{\smash {\,\alpha_1 \dots \alpha_\ell,\dal_1 \dots 
\dal_{\ell}\dga}} \, \hbom{}^\dga{}_{\smash{\dbe}}\, . \cr}
}
The last requirement determines the coefficients $c,d,e$ in \closPl\ through
the relations
\eqn\cde{
(\Delta + \ell-2)e + 2\ell d =0 \, , \quad 
e + (\Delta -1) d + \ell c =0 \, , \quad 
4d + 2(\Delta -\ell ) c =1 \, .
}

To apply these results in the context relevant for this paper we extend \ppP\
to
\eqn\ppPl{
\l \vphi_{rs}(x_1)\, \vphi_{uv}(x_2)\,
\Phi^I_{\smash{\,\alpha_1 \dots \alpha_\ell,\dal_1 \dots \dal_\ell}} \r = 
{1\over r_{12}^{\,\, 2}} 
\bigg ({r_{12}\over r_{13}\,r_{23}}\bigg )^{\!\!{1\over 2}(\Delta-\ell)}
\!\!\!\!\!\!
\X_{\smash{3[12](\alpha_1 \dal_1}} \dots \X_{\smash{3[12]\alpha_\ell)\dal_\ell}}
\, D^I{}_{\!\!rs,uv} \, ,
}
which is uniquely determined by conformal invariance up the $SU(4)$ invariant
$D^I{}_{\!\!rs,uv}$ satisfying
\eqn\symD{
D^I{}_{\!\!rs,uv} = (-1)^\ell D^I{}_{\!\!uv,rs} \, .
}
With the aid of \clPl\ and \cde\ we may then calculate
\eqn\ppdPl{\eqalign{
\l & \vphi_{rs}(x_1)\, \vphi_{uv}(x_2)\, \de
{\bar\Psi}{}^I_{\smash {\,\alpha_1 \dots \alpha_\ell,\dal_1 \dots \dal_\ell\dbe}}
(x_3) \r =  \hep^\beta(x_3)
\l \vphi_{rs}(x_1)\, \vphi_{uv}(x_2)\, 
\J^I{}_{\!\!\beta\alpha_1 \dots \alpha_\ell,\dal_1 \dots \dal_\ell\dbe}(x_3)\r\cr
& {} - 
{i\over r_{12}^{\,\, 2}} 
\bigg ({r_{12}\over r_{13}\,r_{23}}\bigg )^{\!\!{1\over 2}(\Delta-\ell)} \!
\bigg \{ \bigg (
\hep^\beta(x_1)\x_{\smash{13\beta\dbe}}{1\over r_{13}} +
\hep^\beta(x_2)\x_{\smash{23\beta\dbe}}{1\over r_{23}} \bigg )
\X_{\smash{3[12](\alpha_1 \dal_1}} \dots \X_{\smash{3[12]\alpha_\ell)\dal_\ell}}\cr
\noalign{\vskip -5pt}
& \qquad\qquad\qquad\qquad\qquad\qquad{}\times \Big ( (\Delta-\ell) 
D^I{}_{\!\!rs,uv} + \half (T_{tw})^I{}_{\! J} D^J{}_{\!\!rs,uv}
\ga_{[t}\bga_{w]} \Big )  \cr
&\qquad{}+ 2\ell\bigg (
\hep^\beta(x_1)\x_{\smash{13\beta(\dal_1}}{1\over r_{13}} +
\hep^\beta(x_2)\x_{\smash{23\beta(\dal_1}}{1\over r_{23}} \bigg )
\X_{\smash{3[12](\alpha_1 \dbe}} \dots \X_{\smash{3[12]\alpha_\ell)\dal_\ell)}}\,
D^I{}_{\!\!rs,uv} \bigg \} \, . \cr}
}
Assuming
\eqn\ppJ{\eqalign{
\l & \psi_{r\beta}(x_1)\, \vphi_{uv}(x_2)\, 
{\bar\Psi}{}^I_{\smash {\,\alpha_1 \dots \alpha_\ell,\dal_1 \dots 
\dal_\ell\dbe}}(x_3) \r\cr
&{}= - {i\over r_{12}^{\,\, 2}\, r_{13}^{\vphantom g}}
\bigg ({r_{12}\over r_{13}\,r_{23}}\bigg )^{\!\!{1\over 2}(\Delta-\ell)} \!
\Big ( \x_{\smash{13\beta \dbe}}\,
\X_{\smash{3[12](\alpha_1 \dal_1}} \dots \X_{\smash{3[12]\alpha_\ell)\dal_\ell}}
\, P^I{}_{\!\!r,uv} \cr
\noalign{\vskip -6pt}
&\qquad\qquad\qquad\qquad\qquad\qquad {} + \x_{\smash{13\beta(\dal_1}}
\X_{\smash{3[12](\alpha_1 \dbe}} \dots \X_{\smash{3[12]\alpha_\ell)\dal_\ell)}}
\, Q^I{}_{\!\!r,uv} \Big ) \, , \cr}
}
where
\eqn\PQ{
\ga_r P^I{}_{\!\! r,uv} = 0 \, , \qquad
\ga_r Q^I{}_{\!\! r,uv} = 0 \,
}
then since conformal invariance requires $\langle \vphi_{rs}(x_1)\, 
\vphi_{uv}(x_2)\,
\J^I{}_{\!\!\beta\alpha_1\dots\alpha_\ell,\dal_1 \dots \dal_\ell\dbe}(x_3)\rangle$
to be expressed  in terms of $\X_{\smash{3[12]\beta\dbe}}
\X_{\smash{3[12](\alpha_1 \dal_1}} \dots \X_{\smash{3[12]\alpha_\ell)\dal_\ell}}$
and $\X_{\smash{3[12]\beta(\dal}}
\X_{\smash{3[12](\alpha_1 \dbe}} \dots \X_{\smash{3[12]\alpha_\ell)\dal_\ell}}$
we may decompose, using \Xep, 
$\de \langle \vphi_{rs}(x_1)\, \vphi_{uv}(x_2)\, 
{\bar\Psi}{}^I_{\smash {\,\alpha_1 \dots \alpha_\ell,\dal_1 \dots \dal_\ell\dbe}}
(x_3) \rangle =0$ into the following conditions
\eqna\ABJ
$$\eqalignno{
& 2(\Delta-\ell) D^I{}_{\!\!rs,uv} 
+ (T_{tw})^I{}_{\! J} D^J{}_{\!\!rs,uv} \ga_{[t}\bga_{w]} = 
\ga_{(r} P^I{}_{\!\! s),uv} + (-1)^\ell\ga_{(u} P^I{}_{\!\! v),rs}\, , & \ABJ{a}
\cr
& 4 \ell\, D^I{}_{\!\!rs,uv}= \ga_{(r} Q^I{}_{\!\! s),uv} +
(-1)^\ell\ga_{(u} Q^I{}_{\!\! v),rs} \, . & \ABJ{b} \cr}
$$
In general from \Dinv\ and \deXY\ we may write,
\eqn\TXYD{\eqalign{
(T_{tw})^I{}_{\! J} D^J{}_{\!\!rs,uv} = {}& - 4\big ( X^I{}_{\!\!rs,uv} +
Y^I{}_{\!\!rs,uv} \big ) \, \cr  
D^I{}_{\!\!rs,uv}= {}& - \ga_{(r}  D^I{}_{\!\!s)t,uv}
\bga_t +  X^I{}_{\!\!rs,uv} = - \ga_{(u}  D^I{}_{\!\!rs,v)t} \bga_t +
Y^I{}_{\!\!rs,uv} \, . \cr}
}
Using this \ABJ{a} may be simplified to
\eqn\ABJa{
2(\Delta-\ell) D^I{}_{\!\!rs,uv}
= \ga_{(r} {\tilde P}^I{}_{\!\! s),uv} +
(-1)^\ell\ga_{(u} {\tilde P}^I{}_{\!\! v),rs} \, , \qquad
{\tilde P}^I{}_{\!\! r,uv} = 
P^I{}_{\!\!r,uv} + 4 D^I{}_{\!\!rs,uv}\bga_s \, .
}

For further constraints we consider also
$\de \langle \psi_{r\beta}(x_1)\, \vphi_{uv}(x_2)\,
\Phi^I_{\smash{\,\alpha_1 \dots \alpha_\ell,\dal_1 \dots \dal_\ell}} \rangle =0$.
Using \ppPl\ and \ppJ\ this may be decomposed into terms involving $\hbep{}(x_2)$
which give
\eqn\ppPn{\eqalign{
\l \psi_{r\beta}& (x_1)\, \bpsi{}_{\smash{(u \dbe}}(x_2)\,
\Phi^I_{\smash{\,\alpha_1 \dots \alpha_\ell,\dal_1 \dots \dal_\ell}} \r 
\ga_{v)}^{\vphantom g} \cr
= {}&{i\over r_{12}^{\,\, 2}\, r_{13}^{\vphantom g}}
\bigg ({r_{12}\over r_{13}\,r_{23}}\bigg )^{\!\!{1\over 2}(\Delta-\ell)} \!
\bigg \{{1\over r_{12}} \,\x_{\smash{12\beta \dbe}}\,
\X_{\smash{3[12](\alpha_1 \dal_1}} \dots \X_{\smash{3[12]\alpha_\ell)\dal_\ell}}
\, {\tilde P}^I{}_{\!\!r,uv}  \cr
\noalign{\vskip -6pt}
&\qquad\qquad\qquad\qquad\qquad {} -{1\over r_{13}\,r_{23}} \,
\x_{\smash{13\beta(\dal_1}} \x_{\smash{32(\alpha_1 \dbe}}
\X_{\smash{3[12](\alpha_2 \dal_2}} \dots 
\X_{\smash{3[12]\alpha_\ell)\dal_\ell)}}
\, Q^I{}_{\!\!r,uv} \bigg \} \, , \cr}
}
and also if
\eqn\JpP{\eqalign{
\l J_{\smash{rs \beta \dbe }} & (x_1)\, \vphi_{uv}(x_2)\,
\Phi^I_{\smash{\,\alpha_1 \dots \alpha_\ell,\dal_1 \dots \dal_\ell}} \r 
\cr
= {}&{i\over r_{12}^{\,\, 2}}
\bigg ({r_{12}\over r_{13}\,r_{23}}\bigg )^{\!\!{1\over 2}(\Delta-\ell)} \!
\bigg \{\X_{\smash{1[23]\beta \dbe}}\,
\X_{\smash{3[12](\alpha_1 \dal_1}} \dots \X_{\smash{3[12]\alpha_\ell)\dal_\ell}}
\, J^I{}_{\!\! rs,uv} \cr
\noalign{\vskip -6pt}
&\qquad\qquad\qquad\qquad {} +{1\over r_{13}^{\,\, 2}} \,
\x_{\smash{13\beta(\dal_1}} \x_{\smash{31(\alpha_1 \dbe}}
\X_{\smash{3[12](\alpha_2 \dal_2}} \dots \X_{\smash{3[12]\alpha_\ell)\dal_\ell)}}
\, K^I{}_{\!\!rs,uv} \bigg \} \, , \cr}
}
then from  terms involving $\hbep{}(x_1)$
\eqn\JKP{ \eqalign{
J^I{}_{\!\! rs,uv} \bga_s + {\ts {1\over 6}}  J^I{}_{\!\! st,uv} \bga_r
\ga_s \bga_t = {}& - (\Delta - \ell) D^I{}_{\!\!rs,uv} \bga_s - 
P^I{}_{\!\!r,uv} \, , \cr
K^I{}_{\!\! rs,uv} \bga_s + {\ts {1\over 6}}  K^I{}_{\!\! st,uv} \bga_r
\ga_s \bga_t = {}& 2 \ell\,  D^I{}_{\!\!rs,uv} \bga_s +
Q^I{}_{\!\!r,uv} \, . \cr}
}
An additional condition is provided by the conservation of the current
$J_{\smash{rs \beta \dbe }}$ and in \JpP\ this requires
\eqn\Jcon{
2(\Delta -2) J^I{}_{\!\! rs,uv} - ( \Delta - \ell - 4 ) K^I{}_{\!\! rs,uv} = 0 \,.
}

If we combine \ABJ{b}, \ABJa, \JKP\ and \Jcon\ we get
\eqn\all{
(\Delta + \ell -2)(\Delta - \ell -4) \big (D^I{}_{\!\!rs,uv}  
+ \ga_{(r}  D^I{}_{\!\!s)t,uv}\bga_t + (-1)^\ell \ga_{(u}  
D^I{}_{\!\!rs,v)t} \bga_t \big ) =0 \, .
}
For general $\Delta$ this requires, from \deXY,
\eqn\sXY{
X^I{}_{\!\!rs,uv} + Y^I{}_{\!\!rs,uv} = 0 \, ,
}
which, by virtue of \TXYD, is only possible for a singlet operator.
For this case, disregarding an overall constant factor, we have
\eqn\Dsin{
D^I{}_{\!\!rs,uv} \to  D{}_{rs,uv} = \de_{r(u} \de_{v)s} - {\ts {1\over 6}}
\, \de_{rs} \de_{uv} \, ,
}
and then \ABJ{a,b} is easily solved for any $\Delta$ and even $\ell$ by taking
\eqn\PQsin{
P_{\, r,uv} = - (\Delta -\ell) D_{rs,uv} \bga_s =
- (\Delta -\ell) \big ( \de_{r(u} + {\ts {1\over 6}}\, \bga_r \ga_{(u} \big )
\bga_{v)} \, ,  \qquad Q_{\, r,uv} = - 2\ell \,  D_{rs,uv} \bga_s \, .
}
This is compatible with \ppPn\ and \JKP\ ensures that $J^I{}_{\!\! rs,uv}
= K^I{}_{\!\! rs,uv} =0 $ so that the current three point
function \JpP\ is zero as expected by $SU(4)$ symmetry. Apart from
\Dsin\ and \PQsin\ it is also easy to see that for  $J^I{}_{\!\! rs,uv}
= K^I{}_{\!\! rs,uv} =0 $ the only alternative solution, without any
constraint on $D^I{}_{\!\!rs,uv}$, is for $\Delta=4, \, \ell=0$ as was
found earlier (it is easy to derive in this case
$\ell(X^I{}_{\!\!rs,uv} + Y^I{}_{\!\!rs,uv}) = (\Delta-4)
(X^I{}_{\!\!rs,uv} + Y^I{}_{\!\!rs,uv}) =0$).

To find explicit results for  non singlet operators we introduce
\eqn\Cop{
C^I{}_{\!\!rs} = (-1)^\ell C^I{}_{\!\!sr} \, , \qquad C^I{}_{\!\!rr} = 0 \, ,
}
and then define
\eqn\Dnon{
D^I{}_{\!\!rs,uv} = \de_{(r(u} C^I{}_{\!\!s)v)}  - {\ts {1\over 6}}\,
\de_{rs} C^I{}_{\!\!(uv)} - {\ts {1\over 6}}\, \de_{uv} C^I{}_{\!\!(rs)} \, ,
}
satisfying \symD. For $\ell$ even this represents an operator in the
20-representation while $\ell$ odd corresponds to the 15-representation.
If we let
\eqn\Gnon{
G^I{}_{\!\!r,u} = C^I{}_{\!\!ru}1 + {\ts {1\over 6}}\, \bga_r \ga_t C^I{}_{\!\!tu}
+ {\ts {1\over 6}}\, C^I{}_{\!\!rt} \bga_t \ga_u 
+ {\ts {1\over 36}}\, \bga_r \ga_t C^I{}_{\!\!tw} \bga_w \ga_u \, ,
}
then
\eqn\GD{
\ga_{(r} G^I{}_{\!\!s),(u} \bga_{v)} + (-1)^\ell
\ga_{(u} G^I{}_{\!\!v),(r} \bga_{s)} = - 2 D^I{}_{\!\!rs,uv} \, .
}
Hence \ABJ\ with \TXYD\ is solved by taking
\eqn\solPQ{
{\tilde P}^I{}_{\!\!r,uv}  = - (\Delta - \ell -4)
G^I{}_{\!\! r,(u} \bga_{v)}  \, , \qquad Q^I{}_{\!\!r,uv} = - 2\ell \,
G^I{}_{\!\! r,(u} \bga_{v)} \, ,
}
in accord with \PQ\ and \ppPn. Now letting
\eqn\JC{
\J^I{}_{\!\!rs,uv} = \de_{[r(u} C^I{}_{\!\!s]v)}  + {\ts {1\over 6}}\,
\de_{uv} C^I{}_{\!\![rs]} \, ,
}
then
\eqn\JDH{
\J^I{}_{\!\!rs,uv} \bga_s + {\ts {1\over 6}}\, \J^I{}_{\!\! st,uv} \bga_r
\ga_s \bga_t = D^I{}_{\!\!rs,uv} \bga_s - G^I{}_{\!\! r,(u} \bga_{v)} \, ,
}
and using \solPQ\ in \JKP\ gives, with the required symmetries on $rs$ and $uv$,
\eqn\JKJ{
J^I{}_{\!\! rs,uv} = - (\Delta -\ell - 4) \J^I{}_{\!\!rs,uv} \, , \qquad
K^I{}_{\!\! rs,uv} = 2\ell \, \J^I{}_{\!\!rs,uv} \, .
}
Substituting this into \Jcon\ gives finally
\eqn\dcon{
\Delta = 4 + \ell \, , \qquad \Delta = 2 - \ell \, .
}
The second solution is only relevant for $\ell =0$ when $\Phi^I$ may be identified
with the chiral primary operator $\vphi^I$. In general therefore non singlet
15 or 20-representation operators can contribute to the three point function
only for special values of $\Delta$, in agreement with \bpsN.

\newsec{Four Point Functions}

The primary interest in this paper is to consider the constraints arising
from superconformal symmetry
on four point functions for superfields belonging to short representations.
For four points there are two invariants under the conformal group
which we here take as
\eqn\defuv{
u= {r_{12} \, r_{34} \over r_{13} \, r_{24}} \, , \qquad \qquad
v= {r_{14} \, r_{23} \over r_{13} \, r_{24}} \, .
}
In general the four point function depends on arbitrary functions of
$u$ and $v$ which may be analysed using the operator product expansion
in terms of operators with appropriate spins and dimensions. As will be 
shown superconformal invariance provides further conditions which lead
to differential constraints.

For simplicity we consider the $\N=1$ case first although there are no
new constraints in this case. For chiral fields we consider then
\eqn\fourp{
\l \vphi(x_1)\, \bvphi(x_2)\, \vphi(x_3)\, \bvphi(x_4) \r = 
{1\over r_{12}^{\,\, q}\,r_{34}^{\,\, q}} \, a(u,v) \, .
}
Applying the superconformal condition 
$\de \l \psi_\alpha(x_1) \, \bvphi(x_2)\, \vphi(x_3)\, \bvphi(x_4) \r =0$
using \deone\  and its conjugate then leads to
\eqn\pppp{\eqalign{
\l \psi_\alpha(x_1)& \, \bpsi{}_\dal (x_2)\, \vphi(x_3)\, \bvphi(x_4) \r 
\hbep{}^\dal(x_2) + \l \psi_\alpha(x_1) \, \bpsi{}_\dal (x_4) \,
\vphi(x_3) \, \bvphi(x_2) \r  \hbep{}^\dal(x_4) \cr
&{}= 4i {1\over r_{12}^{\,\, q}\,r_{34}^{\,\, q}} \Big ( (qa -u\pr_u a)
{1\over r_{12}} \x_{12\alpha\dal} \hbep{}^\dal(x_2) - v\pr_v a \,
{1\over r_{14}} \x_{14\alpha\dal} \hbep{}^\dal(x_4)\cr
& \qquad \qquad \qquad \quad {}  +(u\pr_u+v\pr_v )a \,
{1\over r_{13}} \x_{13\alpha\dal} \hbep{}^\dal(x_3)  \Big ) \, . \cr}
}
To solve this relation we use
\eqn\relxx{
r_{24}\, \x_{13\alpha\dal} \hbep{}^\dal(x_3) = 
(\x_{13}\tx_{34}\x_{42})_{\alpha\dal}\hbep{}^\dal(x_2) + 
(\x_{13}\tx_{32}\x_{24})_{\alpha\dal}\hbep{}^\dal(x_4) \, ,
}
and hence
\eqn\sppp{\eqalign{
\l \psi_\alpha &(x_1) \, \bpsi{}_\dal (x_2)\, \vphi(x_3)\, \bvphi(x_4) \r \cr
&{}=  4i {1\over r_{12}^{\,\, q}\,r_{34}^{\,\, q}} \bigg \{
{1\over r_{12}} \x_{12\alpha\dal} \, (qa -u\pr_u a) + 
{1\over r_{13}r_{24}}(\x_{13}\tx_{34}\x_{42})_{\alpha\dal}
(u\pr_u+v\pr_v )a \bigg \} \, , \cr
\l \psi_\alpha &(x_1) \, \bpsi{}_\dal (x_4)\, \vphi(x_3)\, \bvphi(x_2) \r \cr
&{}= 4i {1\over r_{12}^{\,\, q}\,r_{34}^{\,\, q}} \bigg \{
- {1\over r_{14}} \x_{14\alpha\dal} \, v\pr_v a  +
{1\over r_{13}r_{24}}(\x_{13}\tx_{32}\x_{24})_{\alpha\dal}
(u\pr_u+v\pr_v )a \bigg \} \, . \cr}
}
Imposing symmetry under $x_2\leftrightarrow x_4$ in \fourp\ requires
$a(v,u)= (v/u)^q a(u,v)$ and this leads to the corresponding relation
between the two expressions in \sppp. The lack of further constraints
in this case is easily understood in terms of $\N=1$ superspace. The
scalar fields $\vphi(x)$ and $\bvphi(x)$ are then the lowest components
of chiral and anti-chiral superfields $\phi(x_+,\theta)$ and ${\bar \phi}
(x_-,\bth)$. The four point function 
$\l \phi(x_{1+},\theta_1)\, {\bar \phi}(x_{2-},\bth_2)\, \phi(x_{3+},\theta_3)
\, {\bar \phi}(x_{4-},\bth_4)\r$ depends on an unconstrained function of two 
variables
since the invariants $u,v$ in \defuv\ may be extended to superconformal 
invariants respecting the required chirality conditions. As shown in \HO\
for two points $(x_i,\theta_i,\bth_i)$ and $(x_j,\theta_j,\bth_j)$ we may
define $\x_{i{\bar\jmath}\alpha\dal}$ which is superconformally covariant
and depends on $(x_{i+},\theta_i)$ and $(x_{j-},\bth_j)$. The superconformal
invariants are the ${\tilde u} = x_{1{\bar 2}}^{\ 2} x_{3{\bar 4}}^{\ 2}/
x_{1{\bar 4}}^{\ 2} x_{3{\bar 2}}^{\ 2}$ and $w=\half \tr \big (
\x_{1{\bar 2}}^{\vphantom g}\, \x_{3{\bar 2}}^{\ -1} 
\x_{3{\bar 4}}^{\vphantom g} \, \x_{1{\bar 4}}^{\ -1} \big )$. 

For $\N=2$ we consider the four point function
\eqnn\fourT
$$\eqalignno{
\l \vphi^{i_1 j_1}(x_1)\, &\bvphi_{i_2 j_2} (x_2)\, \vphi^{i_3 j_3}(x_3)
\,\bvphi_{i_4 j_4} (x_4) \r \cr
= {}& \de^{(i_1}{}_{i_2} \de^{j_1)}{}_{j_2} \, \de^{(i_3}{}_{i_4}
\de^{j_3)}{}_{j_4} \, {1\over r_{12}^{\, \, 2}\,  r_{34}^{\, \, 2} } \, a(u,v)
+ \de^{(i_1}{}_{i_4} \de^{j_1)}{}_{j_4} \, \de^{(i_3}{}_{i_2} \de^{j_3)}{}_{j_2}\, 
{1\over r_{14}^{\, \, 2}\,  r_{23}^{\, \, 2} } \, b(u,v) \cr
&{}+ \de^{(i_1}{}_{(i_2} \de^{j_1)}{}_{(i_4} \, \de^{(i_3}{}_{j_4)}
\de^{j_3)}{}_{j_2)} \,
{1\over r_{12}\, r_{23}\, r_{34}\, r_{14} } \, c(u,v) \, , & \fourT \cr}
$$
and show how the present discussion leads to the conditions of 
Eden {\it et al} \refs{\Eden,\Edent}.
In a similar fashion to \pppp\ we may obtain using \detwo
\eqn\tppp{\eqalign{
\l \psi^{i_1}{}_{\!\!\alpha} &(x_1) \, \bpsi{}_{(i_2\dal} (x_2)\, 
\vphi^{i_3 j_3}(x_3)\, \bvphi_{i_4 j_4}(x_4) \r \,
\hbep{}_{j_2)}{}^{\!\!\dal}(x_2)\cr
&{} + \l \psi^{i_1}{}_{\!\!\alpha} (x_1) \, \bpsi{}_{(i_4\dal} (x_4)\,
\vphi^{i_3 j_3}(x_3)\, \bvphi_{i_2 j_2}(x_2) \r \, 
\hbep{}_{j_4)}{}^{\!\!\dal}(x_4) \cr
&{}+ \vep^{i_3k}\vep^{j_3l}
\l \psi^{i_1}{}_{\!\!\alpha} (x_1) \, \bchi{}_{(i_3\dal} (x_3) \,
\bvphi_{i_2 j_2}(x_2) \, \bvphi_{i_4 j_4}(x_4) \r \,
\hbep{}_{l)}{}^{\!\!\dal}(x_3) \cr
&{}+ \vep^{i_1k} \l J_{\alpha\dal}(x_1) \, \bvphi_{i_2 j_2}(x_2) \,
\vphi^{i_3 j_3}(x_3)\, \bvphi_{i_4 j_4}(x_4) \r \, 
\hbep{}_{k}{}^{\!\!\dal}(x_1) \cr
&{}= 4i {1\over r_{12}^{\,\, 2}\,r_{34}^{\,\, 2}}
\Big ( V^{i_1 k i_3 j_3}_{\, i_2 j_2 i_4 j_4} \,
{1\over r_{12}} \x_{12\alpha\dal} \hbep{}_k{}^{\!\dal}(x_2) +
V'{}^{i_1 k i_3 j_3}_{i_2 j_2 i_4 j_4} \,
{1\over r_{14}} \x_{14\alpha\dal} \hbep{}_k{}^{\!\dal}(x_4)\cr
& \qquad \qquad \qquad \quad {}  + W ^{i_1 k i_3 j_3}_{\, i_2 j_2 i_4 j_4} \,
{1\over r_{13}} \x_{13\alpha\dal} \hbep_k{}^{\!\dal}(x_3)  \Big ) \, , \cr}
}
where the right hand side is determined from \fourT\ giving
\eqnn\derT
$$\eqalignno{\!\!\!\!\!
V^{i_1 k i_3 j_3}_{\, i_2 j_2 i_4 j_4} = {}& 
\de^{(i_1}{}_{i_2} \de^{k)}{}_{j_2} \, \de^{(i_3}{}_{i_4}
\de^{j_3)}{}_{j_4} \, (2a-u\pr_u a) - 
\de^{(i_1}{}_{i_4} \de^{k)}{}_{j_4} \, \de^{(i_3}{}_{i_2} \de^{j_3)}{}_{j_2}
{u^3\over v^2}\pr_u b \cr
&{} + \de^{(i_1}{}_{(i_2} \de^{k)}{}_{(i_4} \, \de^{(i_3}{}_{j_4)}
\de^{j_3)}{}_{j_2)} \, {u\over v}(c - u\pr_u c ) \, , \cr
\!\!\!\!\! V'{}^{i_1 k i_3 j_3}_{i_2 j_2 i_4 j_4} = {}&
- \de^{(i_1}{}_{i_2} \de^{k)}{}_{j_2} \, \de^{(i_3}{}_{i_4}
\de^{j_3)}{}_{j_4} \, {v^3\over u^2}\pr_v a + 
\de^{(i_1}{}_{i_4} \de^{k)}{}_{j_4} \, \de^{(i_3}{}_{i_2} \de^{j_3)}{}_{j_2}
\, (2b-v\pr_v b) \cr
&{} + \de^{(i_1}{}_{(i_2} \de^{k)}{}_{(i_4} \, \de^{(i_3}{}_{j_4)}
\de^{j_3)}{}_{j_2)} \, {v\over u}(c - v\pr_v c ) \, , \cr
\!\!\!\!\! W ^{i_1 k i_3 j_3}_{\, i_2 j_2 i_4 j_4} = {}&
\de^{(i_1}{}_{i_2} \de^{k)}{}_{j_2} \, \de^{(i_3}{}_{i_4}
\de^{j_3)}{}_{j_4} \, {1\over u^2}(u\pr_u {+ v} \pr_v) a +
\de^{(i_1}{}_{i_4} \de^{k)}{}_{j_4} \, \de^{(i_3}{}_{i_2} \de^{j_3)}{}_{j_2}
\, {1\over v^2}(u\pr_u {+ v} \pr_v) b \cr
&{} + \de^{(i_1}{}_{(i_2} \de^{k)}{}_{(i_4} \, \de^{(i_3}{}_{j_4)}
\de^{j_3)}{}_{j_2)} \, {1\over uv} (u\pr_u {+ v} \pr_v) c \, . & \derT \cr}
$$
On the left hand side \tppp\ we may write in general
\eqn\pleft{
\l  \psi^{i_1}{}_{\!\!\alpha} (x_1) \, \bpsi{}_{i_2\dal} (x_2)\,
\vphi^{i_3 j_3}(x_3)\, \bvphi_{i_4 j_4}(x_4) \r 
= 4i \Big (
{\x_{12\alpha\dal}\over r_{12}^{\, \, 3}\,  r_{34}^{\, \, 2}} \, 
R^{i_1i_3j_3}_{i_2i_4j_4} + 
{(\x_{13}\tx_{34}\x_{42})_{\alpha\dal} \over r_{12} r_{13} r_{14}
r_{23} r_{24} r_{34}}  \, S^{i_1i_3j_3}_{i_2i_4j_4} \Big ) \, ,
}
This contains just two independent terms as a consequence of the identity
\eqn\xxx{
\x_{13}\tx_{34}\x_{42} + \x_{14}\tx_{43}\x_{32} = r_{34} \, \x_{12} \, .
}
The $SU(2)$ dependence may be decomposed as
\eqn\RS{\eqalign{
R^{i_1i_3j_3}_{i_2i_4j_4} = {}&
\de^{i_1}{}_{i_2} \, \de^{(i_3}{}_{i_4} \de^{j_3)}{}_{j_4}\, R_1
+ \de^{i_1}{}_{(i_4} \, \de^{(i_3}{}_{j_4)} \de^{j_3)}{}_{i_2}\, R_2 \, , \cr
S^{i_1i_3j_3}_{i_2i_4j_4} = {}&
\de^{i_1}{}_{i_2} \, \de^{(i_3}{}_{i_4} \de^{j_3)}{}_{j_4}\, S_1 + 
\de^{i_1}{}_{(i_4} \, \de^{(i_3}{}_{j_4)} \de^{j_3)}{}_{i_2}\, S_2 \, , \cr}
}
where $R_1,R_2,S_1,S_2$ are functions of $u,v$. A similar expansion to \pleft,
letting $x_2,i_2,j_2 \leftrightarrow x_4,i_4,j_4$, may also be
written for $\l \psi^{i_1}{}_{\!\!\alpha} (x_1)\, \bpsi{}_{i_4\dal} (x_4) \,
\vphi^{i_3 j_3}(x_3)\, \bvphi_{i_2 j_2} (x_2) \r$, defining in this case
$R'_1,R'_2, S'_1,S'_2$. In addition we may write
\eqn\cleft{\eqalign{
\l \psi^{i_1}{}_{\!\!\alpha} (x_1) \, \bchi{}_{i_3\dal} (x_3) \,
\bvphi_{i_2 j_2}& (x_2) \, \bvphi_{i_4 j_4}(x_4) \r  \cr
= {}& 4i \Big (
{\x_{13\alpha\dal}\over r_{13}^{\, \, 3}\,  r_{24}^{\, \, 2}} \, 
T^{i_1}_{i_3i_2j_2i_4j_4} +
{(\x_{12}\tx_{24}\x_{43})_{\alpha\dal} \over r_{12} r_{13} r_{14}
r_{23} r_{24} r_{34}}  \, U^{i_1}_{i_3i_2j_2i_4j_4} \Big ) \, , \cr
T^{i_1}_{i_3i_2j_2i_4j_4} ={}&  \de^{i_1}_{(i_2} \, 
\vep_{j_2)(i_4}^{\vphantom g} \, \vep_{j_4)i_3}^{\vphantom g} \, T_1
+ \de^{i_1}_{(i_4} \vep_{j_4)(i_2}^{\vphantom g} \, 
\vep_{j_2)i_3}^{\vphantom g} T_2 \, ,  \cr
U^{i_1}_{i_3i_2j_2i_4j_4} ={}&  \de^{i_1}_{(i_2} \, 
\vep_{j_2)(i_4}^{\vphantom g} \, \vep_{j_4)i_3}^{\vphantom g} \, U_1
+ \de^{i_1}_{(i_4} \vep_{j_4)(i_2}^{\vphantom g} \, 
\vep_{j_2)i_3}^{\vphantom g} U_2 \, ,  \cr}
}
and, with the definition \defX\ noting that $X_{1[24]} = X_{1[23]}+X_{1[34]}$,
\eqn\Jppp{\eqalign{
\l J_{\alpha\dal}(x_1)&\, \bvphi_{i_2 j_2} (x_2)\, \vphi^{i_3 j_3}(x_3)
\,\bvphi_{i_4 j_4} (x_4) \r \cr 
&{} = 4i\, \de^{(i_3}_{(i_2}\,\vep_{j_2)(i_4}^{\vphantom g}\,\de^{j_3)}_{j_4)} 
\, {1\over r_{13} r_{24}} \Big ({1\over r_{12}r_{34}}\X_{1[23]\alpha\dal}\, A
+ {1\over r_{14}r_{23}}\X_{1[34]\alpha\dal}\, A' \Big ) \, . \cr}
}

Using \pleft, with \RS, \cleft\ and \Jppp\ we may analyse the linear
equations \tppp. The terms involving $S^{i_1i_3j_3}_{i_2i_4j_4}, \,
S'{}^{i_1i_3j_3}_{i_2i_4j_4}$ and $U^{i_1}_{i_3i_2j_2i_4j_4}$ have no
equivalent on the right hand side. Using the relations \relxx\ and also
\eqn\xxep{
(\x_{12}\tx_{24}\x_{43})_{\alpha\dal}\hbep{}^\dal(x_3) -
(\x_{13}\tx_{32}\x_{24})_{\alpha\dal}\hbep{}^\dal(x_4) 
= r_{34}\, \x_{12\alpha\dal} \hbep{}^\dal(x_2) -
r_{23}\, \x_{14\alpha\dal} \hbep{}^\dal(x_4)\, ,
}
this leads to the constraints
\eqn\SSU{
S^{\, i_1i_3j_3}_{(i_2|i_4j_4}\, \de^k_{j_2)} -
S'{}^{i_1i_3j_3}_{\!(i_4|i_2j_2}\, \de^k_{j_4)} = \vep^{(i_3|l}\,
\vep^{j_3)k}\, U^{i_1}_{li_2j_2i_4j_4} \, .
}
With this and \Xep, \tppp\ reduces to
\eqn\VVW{\eqalign{
\!\!\!\! R^{\, i_1i_3j_3}_{(i_2|i_4j_4}\, \de^k_{j_2)}+ {u^2\over v} \,
\vep^{(i_3|l}\, \vep^{j_3)k}\, U^{i_1}_{li_2j_2i_4j_4} + u \, \vep^{i_1 k}\,
\de^{(i_3}_{(i_2}\,\vep_{j_2)(i_4}^{\vphantom g}\,\de^{j_3)}_{j_4)} \, A
= {}& V^{i_1 k i_3 j_3}_{\, i_2 j_2 i_4 j_4} \, , \cr
\!\!\!\! R'{}^{i_1i_3j_3}_{\!(i_2|i_4j_4}\, \de^k_{j_2)}- {v^2\over u} \,
\vep^{(i_3|l}\, \vep^{j_3)k}\, U^{i_1}_{li_2j_2i_4j_4} - v \, \vep^{i_1 k}\,
\de^{(i_3}_{(i_2}\,\vep_{j_2)(i_4}^{\vphantom g}\,\de^{j_3)}_{j_4)} \, A'
= {}& V'{}^{i_1 k i_3 j_3}_{\! i_2 j_2 i_4 j_4} \, , \cr
\!\!\!\! \vep^{(i_3|l}\, \vep^{j_3)k}\, T^{i_1}_{li_2j_2i_4j_4} + 
{1\over uv}\, S^{\, i_1i_3j_3}_{(i_2|i_4j_4}\, \de^k_{j_2)} - \vep^{i_1 k}\,
\de^{(i_3}_{(i_2}\,\vep_{j_2)(i_4}^{\vphantom g}\,\de^{j_3)}_{j_4)}
\Big ( {1\over u}A - {1\over v}A' \Big ) ={}& 
W^{i_1 k i_3 j_3}_{\, i_2 j_2 i_4 j_4} \, .  \cr}
}
The conditions \SSU\ give
\eqn\Scond{
S_1 + U_1 = S_2 -U_2 = S'{}_{\!1} - U_2 = S'{}_{\!2} + U_1 = 0 \, ,
}
and from the terms in \VVW\ antisymmetric in $i_1,k$ we have also
\eqn\AA{\eqalign{
& R_2 - {u^2\over v}(U_1-U_2) = uA \, , \qquad
R'{}_{\!2} - {v^2\over u}(U_1-U_2) = vA' \, , \cr
& T_1 - T_2 -{1\over uv} S_2 = {1\over u}A - {1\over v}A' \, .
\cr}
}
With \derT\ the remaining equations become
\eqn\relabc{\eqalign{
R_1 - {u^2\over v}U_1 = 2a-u\pr_u a \, , \ \ \quad
U_2 = {u\over v}\pr_u b \, , \quad R_2 + {u^2\over v}(U_1+U_2) =
{u\over v} \big (c - u\pr_u c \big ) \, , \cr
R'{}_{\!1} + {v^2\over u}U_2 = 2b-v\pr_v b \, , \quad
U_1 = -{v\over u}\pr_v a \, , \quad R'{}_{\!2} - {v^2\over u}(U_1+U_2) =
{v\over u} \big (c - v\pr_v c \big ) \, , \cr
T_1 - {1\over uv} S_1 = - {1\over u^2}\big ( u\pr_u + v\pr_v \big ) a
\, , \quad \qquad
T_2 = - {1\over v^2}\big ( u\pr_u + v\pr_v \big ) b \, , \cr
T_1 + T_2 +{1\over uv} S_2 = {1\over uv} \big ( u\pr_u + v\pr_v \big )c \,.
\qquad \qquad \cr }
}
In conjunction with \Scond, which implies $S_1+S_2=U_2-U_1$, and \AA,
eliminating $A$ and $A'$, there are two constraints which may be expressed
in the form
\eqn\abcS{\eqalign{
\pr_u c = {}& {v\over u}\pr_v a - \pr_v b
+ {1\over v} ( 1 - u - v ) \pr_u b \, , \cr
\pr_v c = {}& {u\over v} \pr_u b - \pr_u a
+ {1\over u} ( 1 - u - v ) \pr_v a \, . \cr}
}
These equations are invariant under $u\leftrightarrow v, \, 
a \leftrightarrow b$, as is consistent with the crossing symmetry
conditions for the four point function in \fourT, $a(u,v)=b(v,u), \,
c(u,v)=c(v,u)$.

With these relations we may easily find
\eqn\AAp{
A = 2\pr_v a + {1\over v}(c-u\pr_u c ) \, , \qquad
A' = 2\pr_u b + {1\over u}(c-v\pr_v c ) \, ,
}
and the conservation equation $\pr^{\dal\alpha}J_{\alpha\dal}=0$ requires
\eqn\conA{
2uv^2 \pr_u A - (1-u-v)v^2 \pr_v A = 2u^2 v\pr_v A' - (1-u-v)u^2 \pr_u A' \, .
}
As also checked in \Eden\ this is satisfied by virtue of \abcS\foot{Using
\AAp\ for $A$ and \abcS\ for $\pr_u c$ in the resulting term involving 
$\pr_u\pr_u c$ the left hand side of \conA\ becomes 
$(1-u-v)(uvc_{,uv} -uc_{,u} - vc_{,v} +c -2u^2b_{,uu} -2v^2 a_{,vv})
+2uv^2a_{,uv} + 2u^2vb_{,uv} +2v^2a_{,v} +2u^2b_{,u}$. 
This is symmetric under $u\leftrightarrow v,
\ a \leftrightarrow b$ and so is equal to the right hand side.}.

For $\N=4$ superconformal symmetry we again consider the four point
function for the simplest short multiplet whose superconformal transformation
properties were described in section 2. In this case we may write, with
notation defined in appendix A,
\eqn\Fourp{\eqalign{
\l \vphi^{I_1}(x_1) & \, \vphi^{I_2}(x_2)\, \vphi^{I_3}(x_3)\, 
\vphi^{I_4}(x_4)\r \cr
= {}& {\de^{I_1I_2} \de^{I_3I_4} \over 
r_{12}^{\, \, 2}\,  r_{34}^{\, \, 2} }\, a_1 + {\de^{I_1I_3} \de^{I_2I_4} \over
r_{13}^{\, \, 2}\,  r_{24}^{\, \, 2} }\, a_2 + {\de^{I_1I_4} \de^{I_2I_3} \over
r_{14}^{\, \, 2}\,  r_{23}^{\, \, 2} }\, a_3\cr
&{}+ { C^{I_1I_2I_3I_4} \over r_{12}  \, r_{14} \, r_{23}\, r_{34}}\, c_1
+ { C^{I_1I_3I_2I_4} \over r_{13}  \, r_{14} \, r_{23}\, r_{24}}\, c_2
+ { C^{I_1I_2I_4I_3} \over r_{12}  \, r_{13} \, r_{24}\, r_{34}}\, c_3 \, , \cr}
}
involving six functions of $u,v$. This basis is convenient since
for free fields $a_1=a_2=a_3$ and $c_1=c_2=c_3$ are both constants.
The relevant superconformal identity
follows from $\de \l \psi_{r\alpha}(x_1) \, \vphi^{I_2}(x_2)\, 
\vphi^{I_3}(x_3)\, \vphi^{I_4}(x_4)\r =0$  which may be expanded using
\defour\ as
\eqnn\Fourid\hskip -1cm
{$$\eqalignno{
& \l \psi_{r\alpha}(x_1) \, \bpsi{}_{t\dal} (x_2)\, \vphi^{I_3}(x_3)\, 
\vphi^{I_4}(x_4)\r \, C^{I_2}_{ts} \bga_s^{\vphantom g} \hbep{}^{\,\dal}(x_2)  
+ \l \psi_{r\alpha}(x_1) \, \bpsi{}_{t\dal} (x_3)\, \vphi^{I_2}(x_2)\,
\vphi^{I_4}(x_4)\r \, C^{I_3}_{ts} \bga_s^{\vphantom g} \hbep{}^{\,\dal}(x_3) 
\cr
&{}+ \l \psi_{r\alpha}(x_1) \, \bpsi{}_{t\dal} (x_4)\, \vphi^{I_3}(x_3)\,
\vphi^{I_2}(x_2)\r \, C^{I_4}_{ts} \bga_s^{\vphantom g}\hbep{}^{\,\dal}(x_4)\cr
{}& + \l J_{rs\alpha\dal}(x_1) \, \vphi^{I_2}(x_2)\, \vphi^{I_3}(x_3)\, 
\vphi^{I_4}(x_4)\r \bga_s^{\vphantom g} \hbep{}^{\,\dal}(x_1)  + 
{\ts{1\over 6}} \l J_{st\alpha\dal}(x_1) \, \vphi^{I_2}(x_2)\, 
\vphi^{I_3}(x_3)\,\vphi^{I_4}(x_4)\r\bga_r\ga_s\bga_t\hbep{}^{\,\dal}(x_1) \cr
& {}= 2i \bigg ( 
{\x_{12\alpha\dal}\over r_{12}^{\, \, 3}\,  r_{34}^{\, \, 2}} \,
C^I_{rs} \bga_s^{\vphantom g} \hbep{}^{\dal}(x_2) \, Q^{II_2I_3I_4}_{2} +
{\x_{13\alpha\dal}\over r_{13}^{\, \, 3}\,  r_{24}^{\, \, 2}} \,
C^I_{rs} \bga_s^{\vphantom g} \hbep{}^{\dal}(x_3) \, Q^{II_2I_3I_4}_{3} +
{\x_{14\alpha\dal}\over r_{14}^{\, \, 3}\,  r_{23}^{\, \, 2}} \, C^I_{rs} 
\bga_s^{\vphantom g} \hbep{}^{\dal}(x_4)\, Q^{II_2I_3I_4}_{4} \bigg ) \, , \cr
&{}  & \Fourid \cr}
$$}
where, from \Fourp, we have
\eqn\Qfour{\eqalign{
Q^{II_2I_3I_4}_{2} ={}& \de^{II_2} \de^{I_3I_4}(2a_1 - u\pr_u a_1)
- \de^{II_3} \de^{I_2I_4} \, u^3\pr_u a_2 - \de^{II_4} \de^{I_2I_3}\,
{u^3 \over v^2}\pr_u a_3 \cr
&{}+ C^{II_2I_3I_4}{u\over v}(c_1-u\pr_u c_1) - C^{II_3I_2I_4}
{u^3 \over v^2}\pr_u c_2 + C^{II_2I_4I_3} \, u(c_3 - u\pr_u c_3) \, , \cr
Q^{II_2I_3I_4}_{4} ={}& - \de^{II_2} \de^{I_3I_4}\, {v^3 \over u^2}\pr_v a_1
- \de^{II_3} \de^{I_2I_4} \, v^3\pr_v a_2 + \de^{II_4} \de^{I_2I_3}
(2a_3 - v\pr_v a_3) \cr
&{}+ C^{II_2I_3I_4}{v\over u}(c_1-v\pr_v c_1) + C^{II_3I_2I_4}
\, v(c_2 - v\pr_v c_2)  + C^{II_2I_4I_3} \, {v^3 \over u^2}\pr_v c_3 \, ,\cr
Q^{II_2I_3I_4}_{3} ={}&  \de^{II_2} \de^{I_3I_4}\, {1 \over u^2}
(u\pr_u +v\pr_v) a_1
+ \de^{II_3} \de^{I_2I_4} \big (2 a_2 + (u\pr_u +v\pr_v) a_2 \big ) \cr
&{} + \de^{II_4} \de^{I_2I_3} \, {1 \over v^2} (u\pr_u +v\pr_v) a_3 
+ C^{II_2I_3I_4}{1\over uv}(u\pr_u +v\pr_v) c_1 \cr
&{} + C^{II_3I_2I_4}
\, {1\over v} \big ( c_2 + (u\pr_u +v\pr_v) c_2 \big ) +
C^{II_2I_4I_3} \,{1\over u} \big ( c_3+ (u\pr_u +v\pr_v) c_3 \big ) \, .\cr }
}
On the left hand side of \Fourid\ we may write
\eqn\psfour{\eqalign{
\l \psi_{r\alpha}(x_1) \, \bpsi{}_{s\dal} (x_i)\, \vphi^{J}(x_j)\,
\vphi^{K}(x_k)\r = {}& 2i \Big (
{\x_{1i\alpha\dal}\over r_{1i}^{\, \, 3}\,  r_{jk}^{\, \, 2}} \,
R^{JK}_{i,rs} +
{(\x_{1j}\tx_{jk}\x_{ki})_{\alpha\dal} \over r_{12} r_{13} r_{14}
r_{23} r_{24} r_{34}}  \, S^{JK}_{i,rs} \Big ) \, , \cr
\l J_{rs\alpha\dal}(x_1) \, \vphi^{I}(x_2)\,\vphi^{J}(x_3)\,\vphi^{K}(x_4)\r\cr 
= 2i {1\over r_{13} r_{24}} 
\Big ({1\over r_{12}r_{34}}&\X_{1[23]\alpha\dal}\, A^{IJK}_{rs}
+ {1\over r_{14}r_{23}}\X_{1[34]\alpha\dal}\, B^{IJK}_{rs} \Big ) \, .\cr}
}
Using the invariants $T^{(n)JK}_{rs}$ defined in appendix A we
have
\eqn\expRS{
R^{JK}_{i,rs} = \sum_{n=1}^6 R^{(n)}_i T^{(n)JK}_{rs} \, , \qquad
S^{JK}_{i,rs} = \sum_{n=1}^6 S^{(n)}_i T^{(n)JK}_{rs} \, ,
}
and $A^{IJK}_{rs},B^{IJK}_{rs}$ may also be expanded in terms of a basis
formed by $(C^IC^JC^K)_{[rs]}$, $(C^JC^IC^K)_{[rs]}$ and $(C^IC^KC^J)_{[rs]}$.

As before there are conditions necessary to cancel terms involving
$\x_{1j}\tx_{jk}\x_{ki}$ from \psfour. Using \relxx\ and \xxep\ this
requires
\eqn\SSC{
\big ( - (S_2^{I_3I_4}C^{I_2})_{rs}+ (S_3^{I_2I_4}C^{I_3})_{rs} +
(S_4^{I_2I_3}C^{I_4})_{rs} \big ) \bga{}_s = 0 \, .
}
Using \expRS\ and the relations in (A.13,16) this decomposes into 12 linear 
relations for 18 variables which may be written as
\eqn\SSS{\eqalign{
S_2^{(1)} - {\ts{1\over 6}} S_2^{(2)} = {}& - S_3^{(5)} - S_4^{(5)} \, , \cr
S_3^{(1)} - {\ts{1\over 6}} S_3^{(2)} = {}& - S_2^{(5)} + S_4^{(6)} \, , \cr
S_4^{(1)} - {\ts{1\over 6}} S_4^{(2)} = {}& - S_2^{(6)} + S_3^{(6)} \, , \cr
S_2^{(2)} = {}& 2\big ( S_2^{(5)} + S_2^{(6)} + S_3^{(6)} + S_4^{(6)} \big ) 
\, , \cr
S_3^{(2)} = {}& 2\big ( S_2^{(6)} + S_3^{(5)} + S_3^{(6)} - S_4^{(5)} \big ) 
\, , \cr
S_4^{(2)} = {}& 2\big ( S_2^{(5)} - S_3^{(5)} + S_4^{(5)} + S_4^{(6)} \big ) 
\, , \cr
S_2^{(3)} = {} & S_3^{(3)}  = - S_4^{(3)} 
=  2\big ( S_2^{(5)} - S_2^{(6)} + S_3^{(5)} -
S_3^{(6)} - S_4^{(5)} + S_4^{(6)} \big ) \, , \cr
S_2^{(4)} = {}& - S_3^{(5)} + S_4^{(5)} \, , \quad
S_3^{(4)} =  - S_2^{(5)} - S_4^{(6)} \, , \quad
S_4^{(4)} =  - S_2^{(6)} - S_3^{(6)} \, . \cr }
}
There are 9 further relations which are necessary to cancel terms
of the form $E_r^{IJK}$, as defined on (A.14), which cannot occur
elsewhere in \Fourid. This gives 9 relations
\eqn\RRR{\eqalign{
R_2^{(6)} = {}& {u^2\over v} S_3^{(6)}  \, , \qquad
R_4^{(5)} =  - {v^2\over u} S_3^{(5)} \, , \qquad
R_3^{(6)} = {1\over uv} S_2^{(6)} \, , \cr
R_2^{(5)} = {}& {u^2\over v} S_3^{(4)}  \, , \qquad   \,
R_4^{(6)} = {v^2\over u} S_3^{(4)} \, , \qquad \quad 
R_3^{(5)} = {1\over uv} S_2^{(4)} \, , \cr
R_2^{(4)} = {}& {u^2\over v} S_3^{(5)} \, , \qquad \  R_4^{(4)} = {v^2\over u} 
S_3^{(6)} \, , \qquad \quad  R_3^{(4)}  = {1\over uv} S_2^{(6)} \, . \cr }
}
The remaining equations from \Fourid\ which are symmetric in $rs$ correspond
to the terms on the right hand side. With \Qfour, and using \SSS\ and \RRR\
for simplification we have
\eqn\QS{\eqalign{
2\big (&  S_2^{(6)} - S_3^{(5)} + S_3^{(6)} + S_4^{(5)} \big ) = (u\pr_u + v\pr_v)
c_1 \, , \cr
2\big (& - S_2^{(5)} + S_2^{(6)} + S_3^{(6)} - S_4^{(6)} \big ) = -u\pr_u 
c_2 \, , \cr
2\big (&  S_2^{(5)} + S_3^{(5)} - S_4^{(5)} + S_4^{(6)} \big ) =   v\pr_v
c_3 \, , \cr
2S_2^{(5)} = - uv \pr_va_2 \, , & \qquad 2S_2^{(6)} = - {u\over v} (u\pr_u + v\pr_v)
a_3 \, , \qquad 2S_3^{(5)} = - {v\over u} \pr_va_1 \, , \cr
2S_3^{(6)} = {u\over v} \pr_u a_3 \, , & \qquad 
2S_4^{(5)} = -  {v\over u}(u\pr_u + v\pr_v) a_1  \, , \qquad
2S_4^{(6)} = - uv \pr_u a_2 \, , \cr}
}
as well as
\eqn\QR{\eqalign{
R_2^{(2)} - R_2^{(3)} = {}& 2{u\over v}\big (c_1 -u\pr_u c_1\big) 
+ u\pr_v a_1 \, , \cr
R_2^{(2)} + R_2^{(3)} = {}& 2 u \big (c_3 - u(\pr_u+\pr_v) c_3 \big )
- u\pr_v a_1 \, , \cr
R_4^{(2)} + R_4^{(3)} = {}& 2{v\over u}\big (c_1 -v\pr_v c_1\big ) 
+ v\pr_u a_3 \, , \cr
R_4^{(2)} - R_4^{(3)} = {}& 2 v \big (c_2 - v(\pr_u+\pr_v) c_2 \big )
- v\pr_u a_3 \, , \cr
R_3^{(2)} - R_3^{(3)} = {}& 2{1\over v}\big (c_2 + (u\pr_u+v\pr_v) c_1\big) 
+ \pr_v a_2 \, , \cr
R_3^{(2)} + R_3^{(3)} = {}& 2 {1\over u}\big (c_3 + (u\pr_u+v\pr_v) c_3 
- \pr_v c_3 \big ) - \pr_v a_2  \, , \cr
R_2^{(1)} - {\ts {1\over 6}} R_2^{(2)} = {}& 2a_1 - u \pr_u a_1 - 
\half u \pr_v a_1 \, , \cr
R_4^{(1)} - {\ts {1\over 6}} R_4^{(2)} = {}& 2a_3 - v \pr_v a_3 - 
\half v \pr_u a_3 \, , \cr
R_3^{(1)} - {\ts {1\over 6}} R_3^{(2)} = {}& 2a_2 + (u \pr_u + v\pr_v)a_2 - 
\half u \pr_v a_2 \, . \cr}
}
Finally there are three constraints which arise from the requirement that
$\l J_{rs} \, \vphi^{I_2}\, \vphi^{I_3}\, \vphi^{I_4}\r $ has the
conformally covariant form given by \psfour. Using \SSS\ and \RRR\ these
give
\eqn\RRJ{\eqalign{
&{v\over 2u} \big ( R_2^{(2)}- R_2^{(3)} \big ) - 
{u\over 2v} \big ( R_4^{(2)} + R_4^{(3)} \big ) - 3u\, S_3^{(5)}
- 3v\, S_3^{(6)} \cr
&{}+ 2 \big ( S_2^{(6)} + S_3^{(5)} + S_3^{(6)} - S_4^{(5)}\big )
= 0 \, , \cr
&{1\over 2}\big ( R_3^{(2)}- R_3^{(3)} \big ) - 
{1\over 2v^2}\big ( R_4^{(2)} - R_4^{(3)} \big ) - {3\over uv}\, S_2^{(5)}
+ {3\over u}\, S_3^{(6)} \cr
&{}+ 2 \Big ({1\over v}-{1\over u} \Big )
\big ( S_2^{(5)} + S_2^{(6)} + S_3^{(6)} + S_4^{(6)}\big ) = 0 \, , \cr
&{1\over 2}\big ( R_2^{(2)}+ R_2^{(3)} \big ) - 
{u^2\over 2}\big ( R_3^{(2)} + R_3^{(3)} \big ) - {3u\over v}\, S_2^{(5)}
+ {3u^2\over v}\, S_3^{(5)} \cr
&{}+ 2 \Big ({u^2\over v} +{u\over v} - u \Big )
\big ( S_2^{(5)} - S_3^{(5)} + S_4^{(5)} + S_4^{(6)}\big ) = 0 \, . \cr}
}
It is clear that \QS\ provides three constraints and using \QR\ in \RRJ\
gives three more which are essentially integrability conditions. With some
simplification these may be written as
\eqn\ifour{\eqalign{
\pr_u c_1 ={}& {v\over u} \pr_v a_1 + {1\over v}\big ( (1-v) \pr_u a_3
- (u\pr_u + v\pr_v) a_3 \big ) \, , \cr
\pr_v c_1 ={}& {u\over v} \pr_u a_3 + {1\over u}\big ( (1-u) \pr_v a_1
- (u\pr_u + v\pr_v) a_1 \big ) \, , \cr
\pr_u c_2 ={}& -v(\pr_u + \pr_v) a_2 + \pr_v a_3 - {1\over v}(1-u)\pr_u a_3
\, , \cr
\pr_v c_2 ={}& u \pr_u a_2 - (1-u) \pr_v a_2 - {1\over v}
(u\pr_u + v\pr_v) a_3 \, , \cr
\pr_u c_3 ={}& v \pr_v a_2 - (1-v) \pr_u a_2 - {1\over u}
(u\pr_u + v\pr_v) a_1 \, , \cr
\pr_v c_3 ={}& \pr_u a_1 - {1\over u}(1-v)\pr_v a_1 - u(\pr_u + \pr_v) a_2
\, . \cr}
}
These equations are symmetric under $u\leftrightarrow v, \, 
a_1 \leftrightarrow a_3, \, c_2 \leftrightarrow c_3$ and also for
$u\to u/v, \, v\to 1/v, \, a_2 \leftrightarrow a_3, c_1 \leftrightarrow c_3$.
This is consistent with the crossing symmetry relations expected in
\Fourp,
\eqn\crossuv{
a_1(u,v)= a_3(v,u) \, , \quad a_2(u,v)=a_2(v,u) \, , \quad
c_1(u,v)=c_1(v,u) \, , \quad c_2(u,v)= c_3(v,u) \, ,
}
and also
\eqn\cross{\eqalign{
a_1(u,v)={}& a_1(u',v')\, , \quad a_2(u,v)= a_3(u',v') \, , \cr
c_1(u,v)={}&  c_3(u',v') \, ,\quad c_2(u,v)=c_2(u',v') \, , \quad
u'={u\over v} \, , \ v'={1\over v} \, . \cr}
}
The results obtained for $\N=2$ in \abcS\ are obviously a subset of
\ifour\ obtained for $c_1\to c, \, a_1 \to a, \, a_3 \to b$.

\newsec{Solution of Conformal Constraint Equations}

In this section we show how the linear equations \abcS\ and \ifour\
can be simply solved. To this end we introduce new variables $z,x$
defined by\foot{These may be inverted by defining
$\lambda^2 = (z-x)^2= (1-u-v)^2 - 4uv$ and then 
$z={1\over 2}(1-v+u+\lambda)$, $x={1\over 2}(1-v+u-\lambda)$. 
In the Euclidean regime $\sqrt u + \sqrt v \ge 1$ so that
$\lambda^2<0$ and $x=z^*$.}
\eqn\uvxz{
u = xz \, , \qquad v=(1-x)(1-z) \, .
}
With these variables \abcS\ can be rewritten in the form
\eqn\abczx{
{\pr\over \pr x}\Big ( c - {1-z\over z} \, a - {z\over 1-z} \, b \Big ) = 0
 \, , \qquad
{\pr\over \pr z}\Big ( c - {1-x\over x} \, a + {x\over 1-x} \, b \Big ) =0 \, .
}
The solution is then trivial
\eqn\solabc{
c - {1-z\over z} \, a - {z\over 1-z} \, b = f(z) \, , \qquad
c - {1-x\over x} \, a - {x\over 1-x} \, b = f(x) \, ,
}
where we impose symmetry under $z\leftrightarrow x$ as is essential since
$a,b,c$ depend just on $u,v$ which, from \uvxz, are invariant under this
interchange. Eliminating $c$ or $a$ in \solabc\ gives
\eqn\sab{
{1\over u}\, a  - {1\over v}\, b = {f(z)-f(x)\over z-x} \, , \qquad
{1\over v}\, c  - {1-v-u\over v^2}\, b =
{{z\over 1-z}f(z)-{x\over 1-x}f(x)\over z-x} \, ,
}
For later use we define amplitudes corresponding to $SU(2)_R$ quantum
numbers $R=0,1,2$ in the decomposition of  
$\vphi^{i_1 j_1}(x_1)\, \bvphi_{i_2 j_2} (x_2) $  in \fourp,
\eqn\detwo{
A_0 = a + {u^2\over 3v^2}\, b + {u\over 2v}\, c \, , \qquad
A_1 = {u^2\over v^2}\, b + {u\over v}\, c \, , \qquad 
A_2 = {u^2\over v^2} \, b \, .
}
Using \sab\ we then get
\eqn\Detwo{\eqalign{
A_0 = {}& \big ( \half (1+v) - {\ts{1\over 6}}u \big ) \G 
-\half u \, {{2 -  z\over z}\tf(z)- {2 -  x\over x} \tf(x)
\over z-x} \, , \cr   
A_1 = {}& (1-v) \G - u \, { \tf(z)- \tf(x)\over z-x} \, , \cr
A_2 = {}& u \, \G \, ,  \cr}
}
where we define
\eqn\dGf{
\G(u,v)= {u\over v^2} b(u,v) \, , \qquad \tf(z) = {z\over z-1} f(z) \, .
}

For the $\N=4$ case the equations \ifour\ can be similarly simplified
using the variables $z,x$, as given in \uvxz, to three pairs of 
equations which have the form \abczx, involving single partial derivatives
with respect to $z$ and $x$ and which are symmetric under
$z\leftrightarrow x$. The corresponding solutions to \solabc\ 
then involve three initially independent single variable functions $f_{1,2,3}$,
\eqn\solac{\eqalign{
c_1 - {1-z\over z} \, a_1 - {z\over 1-z} \, a_3 = {}& f_1(z) \, , \qquad
c_1 - {1-x\over x} \, a_1 - {x\over 1-x} \, a_3 = f_1(x) \, , \cr
c_2 + (1-z) \, a_2 + {1\over 1-z} \, a_3 = {}& f_2(z) \, , \qquad
c_2 + (1-x) \, a_2 + {1\over 1-x} \, a_3 = f_2(x) \, , \cr
c_3 + z \, a_2 + {1\over z} \, a_1 = {}& f_3(z) \, , \qquad
c_3 + x \, a_2 + {1\over x} \, a_1 = f_3(x) \, . \cr}
}
These may be solved to give relations between any pair of functions 
of $u,v$. For instance eliminating $c_{1,2,3}$
it is easy to see that
\eqn\relac{
{1\over u}\, a_1  - {1\over v}\, a_3 = {f_1(z)-f_1(x)\over z-x} \, , \quad
- a_2 + {1\over v}\, a_3 = {f_2(z)-f_2(x)\over z-x} \, , \quad
a_2 - {1\over u}\, a_1 = {f_3(z)-f_3(x)\over z-x} \, . 
}
Either from \relac\ or directly from \solac\ we must have
\eqn\fff{
f_1(x)+f_2(x)+f_3(x)=k \, ,
}
for $k=\sum_i(a_i+c_i)$, a constant.

In order to discuss the operator product expansion for
$\vphi^{I_1}(x_1) \, \vphi^{I_2}(x_2)$ in terms of operators
belonging to the different possible $SU(4)$ representations, which are here
labelled by their dimensions $R=1,20,84,105,15,175$, we must consider the
corresponding decomposition of the four point function \Fourp,
\eqn\FourR{
\l \vphi^{I_1}(x_1) \, \vphi^{I_2}(x_2)\, \vphi^{I_3}(x_3)\,
\vphi^{I_4}(x_4)\r = {1\over r_{12}^{\, \, 2}\,  r_{34}^{\, \, 2}} 
\sum_R A_R(u,v) P_R^{I_1I_2I_3I_4}\,,
}
where $P_R^{I_1I_2I_3I_4}$ are projection operators which are given
explicitly in appendix A. In terms of \Fourp\ we have
\eqn\Aac{\eqalign{
A_1 = {}& 20a_1 + u^2 a_2 + {u^2\over v^2} \, a_3 + {10\over 3}\Big(
{u\over v}\, c_1 + u c_3 \Big ) + {u^2\over 3v}\, c_2 \, , \cr
A_{20} = {}&  u^2 a_2 + {u^2\over v^2} \, a_3 + {5\over 3}\Big(
{u\over v}\, c_1 + u c_3 \Big ) + {u^2\over 6v}\, c_2 \, , \cr
A_{84} = {}&  u^2 a_2 + {u^2\over v^2} \, a_3 - {u^2\over 2v}\, c_2 \, , \cr
A_{105} = {}&  u^2 a_2 + {u^2\over v^2} \, a_3 + {u^2\over v}\, c_2 \, , \cr
A_{15}= {}&  u^2 a_2 - {u^2\over v^2} \, a_3 -2 \Big( {u\over v}\, c_1 -
u c_3 \Big ) \, , \cr
A_{175} = {}&  u^2 a_2 - {u^2\over v^2} a_3 \, . \cr}
}
Assuming \cross\ we have
\eqn\crossR{
A_R(u,v) = \cases{A_R(u',v')\, , & $R=1,20,84,105$;\cr 
-A_R(u',v')\, , & $R=15,175$.\cr}
}

Taking account of \solac\ we may express $A_R$ just in terms of $A_{105}$ 
in the form
\eqn\AG{\eqalign{
A_1  = {}& {\ts{1\over 3}}\big ( u^2 + 10(1-v)^2 - 8u(1+v) + 60v \big )\G \cr
{}& - {10\over 3}(1-v) \, {\tf_2(z)-\tf_2(x) \over z-x} 
+ {8\over 3}u\, {{2-z\over z}\tf_2(z) - {2-x\over x}\tf_2(x) \over z-x} \cr
{}& + 5u \, {\big ({2-z\over z}\big )^2 \tf(z) - \big ({2-x\over x}\big)^2
\tf(x) \over z-x} - {5\over 3}u\, {\tf(z)-\tf(x)\over z-x} \cr
{}& - 20\big ( f_2(z) + f_2(x) \big ) + 20 k \, , \cr    
A_{20} = {}& {\ts{1\over 6}}\big ( u^2 + 10(1-v)^2 - 5u(1+v) \big ) \G \cr
{}& -{5\over 3}(1-v) \, {\tf_2(z)- \tf_2(x) \over z-x}   
+{5 \over 6}u \, {{2-z\over z} \tf_2(z) - {2-x\over x}\tf_2(x) \over
z-x} \cr     
{}& + {5\over 3}u \, {\tf(z)-\tf(x) \over z-x} \, , \cr    
A_{84} = {}& {\ts{1\over 2}} u \big ( 3(1+v) - u \big ) \G 
- {\ts {3\over 2}} u\, {{2-z\over z}\tf_2(z) - {2-x\over x}\tf_2(x) \over
z-x} \, , \cr      
A_{105} = {}& u^2 \G \, , \cr
A_{15}= {}&  - (1-v) \big ( 2(1+v) - u \big ) \G +
2(1-v) \, {{2-z\over z} \tf_2(z) - {2-x\over x}\tf_2(x)\over z-x} \cr
{}& - u\, {\big ( {2-z\over z} \big )^2 \tf_2(z) - \big ({2-x\over x}\big)^2
\tf_2(x) \over z-x}
- 2u \, {{2-z\over z} \tf(z) - {2-x\over x} \tf(x) \over z-x} \, , \cr
A_{175} = {}& - u (1-v) \G  + u \, {\tf_2(z)-\tf_2(x)\over z-x}\, . \cr}
}      
To achieve the form \AG, which is convenient for application to the
operator product expansion subsequently, we use the definitions,
\eqn\dff{
\tf_2(z)= {z^2 \over 1-z}f_2(z) \, , \qquad
\tf(z) = zf_3(z) - {z\over z-1}f_1(z) \, .
}
The symmetry requirements \crossR\ are satisfied if
\eqn\crossG{
\G(u,v)={1\over v^2} \G(u',v') \, , \qquad \tf_2(z)=\tf_2(z') \, ,
\quad \tf(z) = - \tf(z') \, , \quad z'={z\over z-1} \, .
}     

\newsec{Operator Product Expansion, $\N=2$}

The four point function for quasi-primary fields $\phi_1,\phi_2,\phi_3,\phi_4$
in a conformal field
theory has an expansion in terms of the contributions of all fields
occurring in the operator product expansion of  $\phi_1 \phi_2$ and 
$\phi_3\phi_4$. For simplicity taking $\phi_i = \phi$, a scalar field
of dimension $\Delta_\phi$, this has the form
\eqn\OPEphi{
\l \phi(x_1) \phi(x_2) \phi(x_3) \phi(x_4) \r = 
{1\over (r_{12} \, r_{34})^{\Delta_\phi}} \sum_{\Delta,\ell} 
a_{\Delta,\ell} \, u^{{1\over 2}(\Delta - \ell)} G^{(\ell)}_\Delta (u,v) \, .
}
The functions $ G^{(\ell)}_\Delta (u,v)$ are analytic in $u,1-v$ and
represent the contributions arising from a quasi-primary operator, and
all its derivatives, of dimension $\Delta$ which transforms under the 
rotation group $O(d)$ in $d$ Euclidean dimensions  as a symmetric traceless
rank $\ell$ tensor. For the identity operator $ G^{(0)}_0 =1$. Corresponding 
to $x_1 \leftrightarrow x_2$ or $x_3 \leftrightarrow x_4$ we have
\eqn\Gsym{
G^{(\ell)}_\Delta (u,v) = (-1)^\ell v^{-{1\over 2}(\Delta - \ell)}
G^{(\ell)}_\Delta (u',v') \, , \qquad u'=u/v \, , \ v' = 1/v \, .
}
Recently \Dos\
we obtained compact explicit expressions, using the variables $z,x$ defined
in \uvxz, in four dimensions which are given here in appendix C.

With superconformal symmetry the fields appearing in the operator 
product expansion belong to supermultiplets under the superconformal
group which link the contributions of differing $\ell$. We first discuss
the $\N=2$ case for the four point function shown in \fourT\ and analyse
the contributions to $A_R$, as defined in \detwo, for $R=0,1,2$. For
a long multiplet whose lowest dimension operator, with dimension $\Delta$ 
and spin $\ell$, has $R=0$ the list of operators which may contribute
to the four point function are (only operators in four dimensions with
representation $(j_1,j_2)$ can appear if $j_1=j_2=\half \ell$ and the
$U(1)_R$ charge $r=0$ so this is a subset of the full set
of $2^8(\ell+1)^2$ fields),
\eqnn\listT$$
\eqalignno{
&R=0 \, \qquad \Delta_\ell \qquad (\Delta+1)_{\ell\pm 1} \qquad
(\Delta+2)_{\ell\pm 2}, (\Delta+2)_\ell{}^{\! 2} \qquad
(\Delta+3)_{\ell\pm 1} \qquad (\Delta+4)_\ell \,  \cr
&R=1 \qquad \qquad \quad \ \, (\Delta+1)_{\ell\pm 1} \qquad \qquad  \ 
(\Delta+2)_\ell{}^{\! 3} \qquad \qquad \ \ (\Delta+3)_{\ell\pm 1} & \listT \cr
&R=2 \qquad \qquad \qquad \qquad \qquad \qquad \qquad \ \ (\Delta+2)_\ell 
\quad . \cr}
$$
It is evident that for $R=2$ only a single operator contributes from
this supermultiplet. Consequently  in this case the corresponding
contribution, choosing here the overall coefficient to be one, is just
\eqn\Atwo{
A_2(u,v) = u^{{1\over 2}(\Delta+2 - \ell)} G^{(\ell)}_{\Delta +2}(u,v)\, .
}
According to \Detwo\ this gives trivially $\G(u,v) =
u^{{1\over 2}(\Delta - \ell)} G^{(\ell)}_{\Delta +2}(u,v)$.
Setting $f(z)=0$ this then determines $A_0$ and $A_1$ in
\Detwo. Using the results of appendix C allows $A_0,A_1$ to be expressed in terms
of a sum of contributions $ G^{(\ell)}_{\Delta}$, for suitable $\Delta,\ell$,
with coefficients depending on $\Delta,\ell$. The results are then
\eqn\Aone{\eqalign{
A_1(u,v) = {}& - u^{{1\over 2}(\Delta - \ell)} \bigg (
2\, G^{(\ell+1)}_{\Delta +1}(u,v) + \half u G^{(\ell-1)}_{\Delta +1}(u,v)\cr
& \qquad \qquad \quad {}+ 
{(\Delta + \ell+2)^2 \over 8 (\Delta+\ell+1)(\Delta+\ell+3)} \,
u G^{(\ell+1)}_{\Delta +3}(u,v) \cr
& \qquad \qquad \quad {} +
{(\Delta - \ell)^2 \over 32 (\Delta-\ell-1)(\Delta-\ell+1)}\, 
u^2 G^{(\ell-1)}_{\Delta +3}(u,v)\bigg )  \, , \cr
A_0(u,v) = {}& u^{{1\over 2}(\Delta - \ell)} \bigg (
G^{(\ell)}_{\Delta}(u,v) + {\ts{1\over 12}} u G^{(\ell)}_{\Delta +2}(u,v)\cr
& \qquad \qquad {}+ 
{(\Delta + \ell+2)^2 \over 4 (\Delta+\ell+1)(\Delta+\ell+3)}\,
G^{(\ell+2)}_{\Delta +2}(u,v) \cr
& \qquad \qquad {}+
{(\Delta - \ell)^2 \over 64 (\Delta-\ell-1)(\Delta-\ell+1)}\,  
u^2 G^{(\ell-2)}_{\Delta +2}(u,v)  \cr
& \qquad \qquad {}+
{(\Delta + \ell+2)^2 (\Delta - \ell)^2 \over 256 
(\Delta+\ell+1)(\Delta+\ell+3)(\Delta-\ell-1)(\Delta-\ell+1)}\,  
u^2 G^{(\ell)}_{\Delta +4}(u,v)\bigg )  \, . \cr}
}
These are exactly the contributions expected in an expansion as in
\OPEphi\ corresponding to the operators listed in \listT\ with the constraint
that the operators contributing to $A_1$ have $(-1)^\ell$ differing in sign 
from those appearing in $A_0,A_2$.
The conditions (the overall minus sign 
in $A_1$ arises as a consequence of \Gsym\ and the constraint on $\ell$ just
noticed since the positivity requirement
applies  directly to $\langle \vphi^{i_1 j_1}(x_1)\, \bvphi_{i_2 j_2} (x_2)\, 
\bvphi_{i_3 j_3}(x_3) \,\vphi^{i_4 j_4} (x_4) \rangle$ which is related to
\fourT\ by $x_3 \leftrightarrow x_4$ and hence $u\to u/v, \, v\to 1/v$)
necessary for a unitary theory clearly require only $\Delta > \ell +1$. For
a generic $\N=2$ supermultiplet the unitarity condition for a lowest
weight operator with dimension $\Delta$, belonging to a $SU(2)_R$ 
representation $R$, $U(1)_R$ charge $r$ and with $j_1=j_2=\half \ell$ is
\Short
\eqn\ineq{
\Delta \ge 2 + \ell  + 2R + |r| \, ,
}
and in application to the four point function of interest here $r=0$
and $R=0,1,2$.

In \Aone\ $\ell=0,1$ are special cases but the results may be obtained from
\Aone\ by using, as shown in appendix C,
\eqn\mml{
G_{\Delta}^{(-1)}(u,v)=0 \, , \qquad \quar u \, G_{\Delta}^{(-2)}(u,v)
= - G_{\Delta}^{(0)}(u,v)\, .
}
Thus we may obtain for $\ell=0$,
\eqnn\Aonea
$$\eqalignno{
A_0(u,v) = {}& u^{{1\over 2}\Delta} \bigg ( G^{(0)}_{\Delta}(u,v) 
+ {\Delta^2-4\over 48(\Delta^2-1)} \, u G^{(0)}_{\Delta +2}(u,v)
+ {(\Delta +2)^2 \over 4 (\Delta+1)(\Delta+3)}\,
G^{(2)}_{\Delta +2}(u,v) \cr
& \qquad {}+ { \Delta^2 (\Delta + 2)^2 \over 256
(\Delta+1)^2(\Delta+3)(\Delta-1)} \,
u^2 G^{(0)}_{\Delta +4}(u,v)\bigg )  \, . & \Aonea \cr}
$$
Unitarity here requires $\Delta\ge 2$ in accord with \ineq\ for $R=\ell=0$.

There are also short multiplets in which 
the spectrum of $SU(2)_R$ representations is reduced since some
superconformal transformations annihilate the operators with lowest
dimension. For such supermultiplets $\Delta$ and $\ell$ are related.
The contribution of such operators in the operator product expansion
to the four point function involves considering the function
$\tf(z)$ which arises in the general solution of the superconformal identities 
exhibited in \Detwo. The terms arising from the function $\tf$
may also be written in terms of the operator expansion functions 
$G_\Delta^{(\ell)}(u,v)$ but only in particular cases where  $\Delta$ is given
in terms of $\ell$.  From the results in \Dos, as quoted here in
in appendix C, the essential expression, which matches the form of the
contributions appearing in \Detwo, is
\eqn\Gbps{
G^{(\ell)}_{\ell + 2}(u,v) = {g_{\ell+1}(z) - g_{\ell+1}(x) \over z-x}
\, , \qquad  g_\ell(z) = \big (-\half z\big )^{\ell-1}
z F\big (\ell,\ell;2\ell;z \big ) \, ,
}
with $F$ a hypergeometric function.
Using the formulae in appendix C or standard hypergeometric identities
this satisfies the crucial relations
\eqn\relg{
{2- z \over z} g_\ell (z) = - g_{\ell-1}(z) -
{\ell^2 \over (2\ell-1)(2\ell+1)} \, g_{\ell+1}(z) \, , \qquad
g_\ell(z)=(-1)^\ell g_\ell(z') \, ,
}
with $z'$ as in \crossG.

We first in \Detwo\ set $\G=0$ and therefore
\eqn\Atwo{
A_2(u,v) = 0 \, ,
}
so that there are no contributions from $R=2$ operators.
If we then choose in \Detwo\ $\tf(z)= g_{\ell+2}(z)$ we would have,
by virtue of \Gbps,
\eqn\Aoa{       
A_1(u,v) = - u G^{(\ell+1)}_{\ell + 3}(u,v) \, .
}     
Using \relg\ in \Detwo\ further gives, in conjunction with \Atwo\ and \Aoa,
\eqn\Aoo{      
A_0(u,v) = \half u\, G^{(\ell)}_{\ell+2}(u,v)
+ {(\ell+2)^2 \over 2(2\ell+3)(2\ell+5)}u \, G^{(\ell+2)}_{\ell + 4}(u,v)\, .
}
The results \Atwo, \Aoa\ and \Aoo\ thus represent the contribution of a 
restricted multiplet in which the lowest dimension field has spin $\ell$ and
$R=0, \, \Delta=2+\ell$, saturating the bound \ineq. 
In both \Aoa\ and \Aoo\ the relevant operators
have twist, $\Delta-\ell$, two.  For $\ell=-1$ these results reduce to
\eqn\AO{
A_0 (u,v) = {\ts{1\over 6}}u \, G_3^{(1)}(u,v) \, , \qquad
A_1 (u,v) = - u G_2^{(0)} (u,v) \, , \qquad A_2(u,v) =0 \, ,
}      
in which the $R=1$ operator has the lowest dimension. This corresponds
to the short supermultiplet exhibited in \Ntwo. If $\ell=0$ instead  we get
\eqn\Aem{
A_0(u,v) = \half u\, G^{(0)}_{2}(u,v)
+ {\ts{2 \over 15}}u \, G^{(2)}_{4}(u,v)\, , \quad A_1(u,v) = - u 
G^{(1)}_{3}(u,v) \, , \quad A_2(u,v) = 0 \, ,
}
which represents the contribution of the supercurrent supermultiplet
containing the energy momentum tensor (the $U(1)_R$ current does not
appear in the operator product expansion in this case).

If in \Detwo\ we take $\tf =1$ then
\eqn\Azone{
A_0 =1 \, , \qquad A_1=A_2 =0 \, ,
}
corresponding to the identity operator.\foot{This can also be realised by
\Aoa\ and \Aoo\ if $\ell=-2$ since, by virtue of \mml, $\half u G_0^{(-2)}
(u,v) = - 2 G_0^{(0)}(u,v)=-2$.}  

Besides the case represented by \Atwo, \Aoa\ and \Aoo\ for any $\ell$ there are
also other restricted supermultiplets, with dimensions not involving
a factor $2^8$ as in a generic long multiplets. A further example may be
found by considering the
contributions to the four point function in the operator prodect expansion
obtained by subtracting twice \Aoa\ and \Aoo\ from \Aone\ for $\Delta=\ell+2$. 
This gives, with \Atwo, the results
\eqn\Anew{\eqalign{
A_2(u,v) = {}& u^2 G_{\ell+4}^{(\ell)}(u,v) \, , \cr
A_1(u,v) = {}& - \half u^2  G_{\ell+3}^{(\ell-1)}(u,v)
-{\ts{1\over 24}} u^3  G_{\ell+5}^{(\ell-1)}(u,v) 
- {(\ell+2)^2 \over 2(2\ell+3)(2\ell+5)} \, u^2  G_{\ell+5}^{(\ell+1)}(u,v) \, , \cr
A_0 (u,v) = {}&  {\ts{1\over 12}} u^2 G_{\ell+4}^{(\ell)}(u,v) +
{\ts{1\over 48}} u^3 G_{\ell+4}^{(\ell-2)}(u,v) + 
{(\ell+2)^2 \over 48(2\ell+3)(2\ell+5)} \,
u^3  G_{\ell+6}^{(\ell)}(u,v) \, ,  \cr}
}
which would correspond to a supermultiplet in which the lowest
dimension operator has spin $\ell-1$ and $R=1, \, \Delta=\ell+3$ containing
operators of twist $4,6$. For such multiplets the bound \ineq\ is again 
saturated for the lowest weight operator.

Taking $\ell=0$ in \Anew\ gives
\eqn\Azero{
A_0 (u,v) = {\ts{1\over 180}}u^3 G_6^{(0)}(u,v) \, , \quad
A_1 (u,v) = -{\ts{2\over 15}}u^2 G_5^{(1)}(u,v) \, , \quad
A_2(u,v) =  u^2 G_{4}^{(0)}(u,v) \, ,
}
where the lowest dimension operator has $R=2$. For $\N=2$ the superconformal
group $SU(2,2|2)$ has short multiplets corresponding to lowest weight operators 
which are scalars with $\Delta=2R$, \Short, and can be represented as in
\Ntwo, which is the special case when $R=1$, giving
\eqn\Rtwo{\def\normalbaselines{\baselineskip16pt\lineskip3pt
\lineskiplimit3pt}
\matrix{\Delta&~~~&{~~~}&&{~~~}&&{~~~}&&{~~~}&&{~~~}&&{~~~}&&\cr
2R&&{~~~}&&{~~}&&R_{(0,0)}&&{~~}&&{~~~}&\cr
&&&&&\Bsw&&\Bse&&&&\cr
2R+{\ts{1\over 2}}&&&&\hidewidth~~~~~(R-{\ts{1\over 2}})_{({1\over 2},0)}
\hidewidth&&&&\hidewidth~~(R-{\ts{1\over 2}})_{(0,{1\over 2})}~~\hidewidth&&&\cr&&&\Bsw&&\Bse&&\Bsw&&\Bse&&\cr
2R+1&&\hidewidth~~~~~~(R-1)_{(0,0)}\hidewidth&&&&
\hidewidth(R-1)_{({1\over 2},{1\over 2})}\hidewidth&&&&
\hidewidth(R-1)_{(0,0)}\hidewidth&\cr
&&&\Bse&&\Bsw&&\Bse&&\Bsw&&\cr
2R+{\ts {3\over 2}}&&&
&\hidewidth~~(R-{\ts{3\over 2}})_{(0,{1\over 2})}\hidewidth&&&
&\hidewidth(R-{\ts{3\over 2}})_{({1\over 2},0)}\hidewidth&&&\cr
&&&&&\Bse&&\Bsw&&&&\cr
2R+2&&&&&
&\hidewidth(R-2)_{(0,0)}\hidewidth&&&&&\cr
r&&1&&{\ts{1\over 2}}&&0&&-{\ts{1\over 2}}&&-1&\cr}
}
where the representations are denoted by $R_{(j_1,j_2)}$ with $(j_1,j_2)$
determining the spin, $R$ the $SU(2)_R$ representation and for $r$
the $U(1)_R$ charge. The total dimension is $16(2R-1)$. 
For $R=2$ the operators listed in \Rtwo\ with zero $r$ are just those 
required to give the operator product contributions in \Azero.

For a general four point function, 
represented as in \Detwo, the superconformal operator product 
expansion can thus be written simply as
\eqn\OPEtwo{
\G(u,v) = \sum_{\Delta,\ell} a_{\Delta,\ell}(-1)^\ell \, 
u^{{1\over 2}(\Delta - \ell)} G^{(\ell)}_{\Delta+2} (u,v) \, , \qquad
\tf(z) = A + \sum_{\ell\ge -1} a_\ell (-1)^\ell \, g_{\ell+2}(z) \, .
}
For generic $\Delta$ unitarity requires $a_{\Delta,\ell}>0$. From
the above it is sufficient if $\Delta=\ell+2$ to impose
$ a_\ell + 2a_{\ell+2,\ell} \ge 0$, for $\ell=0,1,\dots$,  and also $A>0$.

As an illustration we may consider the result for a free theory when
in \fourT\ we have a trivial solution of the superconformal constraints,
$a=b$ and $c$ constants. In this case, from \solabc\ and \dGf, we have
\eqn\solGf{
\G(u,v) = a\, {u\over v^2} \, , \qquad
\tf(z) = a \bigg ( 1 + {z^2 \over (1-z)^2} \bigg ) - c \, {z \over 1-z} \, .
}
Using our previous results \Dos\ we have
\eqn\aDe{
a_{\Delta,\ell}  = a\, 2^{\ell-1}{(\ell+t-1)!(\ell+t)!\big((t-1)!\big )^2
\over (2\ell+2t-1)!(2t-2)!}\,(\ell+1)(\ell+2t) \, \de_{\Delta,\ell+2t} \, ,
\quad {\ell=0,1,2\dots \atop t=1,2,\dots} \, .
}
Furthermore, by analysing the expansion of $\tf(z)$ in powers of $z$,
\eqn\aexp{
A=a \, , \qquad
a_\ell + 2a_{\ell+2,\ell} = c \, 2^{\ell+1}{\big ((\ell+1)!\big )^2 
\over (2\ell+2)! } \, , \quad \ell=-1,0, \dots \, .
}
Clearly positivity conditions are satisfied for $a,c>0$. From \aexp\
it is evident that the short multiplets represented by \Atwo, \Aoa, 
\Aoo\ and also \Anew, including the special cases \AO\ and
\Azero, are relevant in the operator product expansion for this
free theory.

\newsec{Operator Product Expansion, $\N=4$}

For the analysis of the operator product expansion for the
four point function for the simplest chiral primary operators 
as in \Fourp, the implications of $\N=4$ superconformal symmetry can
be derived following a similar procedure as in the $\N=2$ case. This
depends essentially on the representation \AG\ for the contributions
of operators belonging to different representations of the $SU(4)_R$
symmetry.

We first consider the contributions of long supermultiplets, without
any constraints,  in which the lowest dimension operator is a $SU(4)_R$
singlet of dimension $\Delta$. According to \bpsN\ and also section 4
this is the only possibility for generic $\Delta$. For the case when this 
operator also
has spin zero the decomposition of all $2^{16}$ operators into representations
of $SU(4)_R$ and also $SO(3,1)$, which are labelled $(j_1,j_2)$, was
listed by Andrianopoli and Ferrara \AF.  Here we are interested in just
those operators contributing in the operator product expansion
to the scalar four point function, which
have $j_1=j_2=\half \ell$, and we also consider only those operators
with $Y=0$, corresponding to those occurring in a superfield expansion
with equal numbers of $\theta$'s and $\bth$'s. With a similar notation
as in \listT\ these operators are
\eqn\listN{\def\normalbaselines{\baselineskip16pt\lineskip3pt
 \lineskiplimit3pt}
\!\!\!\matrix{R~&1&15&20&84&175&105\cr
&\Delta_0\cr
&(\Delta+1)_1&(\Delta+1)_1\cr
&~(\Delta+2)_{0,2}&~(\Delta+2)_{0,2}&(\Delta+2)_2&(\Delta+2)_0\cr
&~(\Delta+3)_{1,3}&\,~~(\Delta+3)_{1,1,3}&(\Delta+3)_1&(\Delta+3)_1&(\Delta+3)_1\cr
&\,~~(\Delta+4)_{0,2,4}&\,~~(\Delta+4)_{0,2,2}&~(\Delta+4)_{0,2}&~(\Delta+4)_{0,2}&
(\Delta+4)_0&(\Delta+4)_0\cr
&~(\Delta+5)_{1,3}&\,~~(\Delta+5)_{1,1,3}&(\Delta+5)_1&(\Delta+5)_1&(\Delta+5)_1\cr
&~(\Delta+6)_{0,2}&~(\Delta+6)_{0,2}&(\Delta+6)_2&(\Delta+6)_0\cr
&(\Delta+7)_1&(\Delta+7)_1\cr
&(\Delta+8)_0&. \cr}
}
For the case of the lowest dimension operator having spin $\ell$ we may
easily obtain, since there are no constraints, the corresponding table
by tensoring with $(j,j), \ \ell=2j$. Thus in \listN, for any $\Delta$, we
have $\Delta_0 \to \Delta_\ell$, $\Delta_1 \to \Delta_{\ell\pm 1}$,
$\Delta_2 \to \Delta_{\ell\pm 2,\ell}$, $\Delta_3 \to 
\Delta_{\ell\pm 3,\ell\pm 1}$ and $\Delta_4 \to \Delta_{\ell\pm 4,\ell\pm 2,\ell}$.
This representation is unitary if $\Delta \ge \ell+2$.

The crucial observation for our purposes
arising from \listN\ is that there is only
one operator belonging to the 105-representation. We therefore consider the
contribution of this operator in the operator product expansion by writing
\eqn\Afive{
A_{105}(u,v)=  u^{{1\over 2}(\Delta+4 - \ell)} G^{(\ell)}_{\Delta +4}(u,v)\, .
}
According to \AG\ this now gives $\G(u,v) =
u^{{1\over 2}(\Delta - \ell)} G^{(\ell)}_{\Delta +4}(u,v)$. Discarding
$\tf(z)$ and $\tf_2(z)$ we may now determine the remaining $A_R$ from
\AG, in a similar fashion to \Aone, by using the results of appendix C.
We give the results in order of increasing complexity. For $R=175$,
\eqn\Aseven{\eqalign{
A_{175}(u,v) = {}& u^{{1\over 2}(\Delta +2 - \ell)} \bigg (
2 G^{(\ell+1)}_{\Delta +3}(u,v) + \half u G^{(\ell-1)}_{\Delta +3}(u,v)\cr
& \qquad \qquad \quad {}+
{(\Delta + \ell+4)^2 \over 8 (\Delta+\ell+3)(\Delta+\ell+5)} \,
u G^{(\ell+1)}_{\Delta +5}(u,v) \cr
& \qquad \qquad \quad {} +
{(\Delta - \ell+2)^2 \over 32 (\Delta-\ell+1)(\Delta-\ell+3)} \,
u^2 G^{(\ell-1)}_{\Delta +5}(u,v)\bigg )  \, . \cr}
}
For $R=84$,
\eqn\Aeight{\eqalign{
A_{84}(u,v) = {}& 3u^{{1\over 2}(\Delta +2 - \ell)} \bigg (
G^{(\ell)}_{\Delta+2}(u,v) + {\ts{1\over 12}} u G^{(\ell)}_{\Delta +4}(u,v)\cr
& \qquad \quad {}+
{(\Delta + \ell+4)^2 \over 4 (\Delta+\ell+3)(\Delta+\ell+5)} \,
G^{(\ell+2)}_{\Delta +4}(u,v) \cr
& \qquad \quad {}+
{(\Delta - \ell+2)^2 \over 64 (\Delta-\ell+1)(\Delta-\ell+3)} \,
u^2 G^{(\ell-2)}_{\Delta +4}(u,v)  \cr
& \ {}+
{(\Delta + \ell+4)^2 (\Delta - \ell+2)^2 \over 256
(\Delta+\ell+3)(\Delta+\ell+5)(\Delta-\ell+1)(\Delta-\ell+3)} \,
u^2 G^{(\ell)}_{\Delta +6}(u,v)\bigg )  \, . \cr}
}
For $R=20$,
\eqnn\Atwen
$$\eqalignno{\!\!\!
A_{20}(u,v) = {}& {\ts{5\over 3}}u^{{1\over 2}(\Delta - \ell)} \bigg (
4G^{(\ell+2)}_{\Delta+2}(u,v) + u G^{(\ell)}_{\Delta+2}(u,v) +
{\ts{1\over 4}} u^2 G^{(\ell-2)}_{\Delta +2}(u,v)\cr
& \quad {}+
{(\Delta + \ell+4)^2 \over 4 (\Delta+\ell+3)(\Delta+\ell+5)} \,
u G^{(\ell+2)}_{\Delta +4}(u,v) \cr
& \quad {}+
{(\Delta - \ell+2)^2 \over 64 (\Delta-\ell+1)(\Delta-\ell+3)} \,
u^3 G^{(\ell-2)}_{\Delta +4}(u,v)  +{1\over 10}
u^2 G^{(\ell)}_{\Delta +4}(u,v) \cr
& \ {}+ {1\over 8}\Big ({1\over(\Delta+\ell+1)(\Delta+\ell+5)}
+ {1\over (\Delta-\ell-1)(\Delta-\ell+3)} \Big ) 
u^2 G^{(\ell)}_{\Delta +4}(u,v) \cr
& \ {}+
{(\Delta + \ell+4)^2 (\Delta + \ell+6)^2 \over 64
(\Delta+\ell+3)(\Delta+\ell+5)^2(\Delta+\ell+7)} \,
u^2 G^{(\ell+2)}_{\Delta +6}(u,v)  \cr
& \ {}+
{(\Delta + \ell+4)^2 (\Delta - \ell+2)^2 \over 256
(\Delta+\ell+3)(\Delta+\ell+5)(\Delta-\ell+1)(\Delta-\ell+3)} \,
u^3 G^{(\ell)}_{\Delta +6}(u,v) \cr 
& \ {}+
{(\Delta - \ell+2)^2 (\Delta - \ell+4)^2 \over 2^{10}
(\Delta-\ell+1)(\Delta-\ell+3)^2(\Delta-\ell+5)} \,
u^4 G^{(\ell-2)}_{\Delta +6}(u,v)\bigg ) \, . & \Atwen \cr}
$$
For $R=15$,
\eqnn\Afif
$$\eqalignno{\!\!\!\!\!\!\!\!\!\!\!\!
A_{15}(u,v) = {}& u^{{1\over 2}(\Delta - \ell)} \bigg (
8G^{(\ell+1)}_{\Delta+1}(u,v) + 2u G^{(\ell-1)}_{\Delta+1}(u,v) \cr
&  {}+
{2(\Delta + \ell+4)^2 \over (\Delta+\ell+3)(\Delta+\ell+5)} \,
G^{(\ell+3)}_{\Delta +3}(u,v) 
+ {(\Delta + \ell+ 3)^2 -3 \over (\Delta+\ell+1)(\Delta+\ell+5)} \,
u G^{(\ell+1)}_{\Delta +3}(u,v) \cr
&  {}+ {(\Delta - \ell +1 )^2 -3 \over 4(\Delta-\ell-1)(\Delta-\ell+3)} \,
u^2 G^{(\ell-1)}_{\Delta +3}(u,v) \cr
& {}+  {(\Delta - \ell + 2 )^2  \over 32(\Delta-\ell+1)(\Delta-\ell+3)} \,
u^3 G^{(\ell-3)}_{\Delta +3}(u,v) \cr
&  {}+
{(\Delta + \ell+4)^2(\Delta + \ell+6)^2  \over 8(\Delta+\ell+3)
(\Delta+\ell+5)^2(\Delta+\ell+7)} \, u G^{(\ell+3)}_{\Delta + 5}(u,v)\cr
&  {} + {(\Delta + \ell+4)^2 \big((\Delta - \ell+ 1)^2 -3\big) \over 
16(\Delta-\ell-1)(\Delta-\ell+3)(\Delta+\ell+3)(\Delta+\ell+5)} \,
u^2 G^{(\ell+1)}_{\Delta +5}(u,v) \cr
&  {}+ {(\Delta - \ell + 2 )^2\big ((\Delta + \ell +3 )^2 -3\big) 
\over 64(\Delta-\ell+1)(\Delta-\ell+3)(\Delta+\ell+1)(\Delta+\ell+5)} \,
u^3 G^{(\ell-1)}_{\Delta +5}(u,v) \cr
& {}+ {(\Delta - \ell+2)^2 (\Delta - \ell+4)^2 \over 2^{9}
(\Delta-\ell+1)(\Delta-\ell+3)^2(\Delta-\ell+5)} \,
u^4 G^{(\ell-3)}_{\Delta +5}(u,v) \cr
&  {}+
{(\Delta - \ell+2)^2(\Delta + \ell+4)^2 (\Delta + \ell+6)^2 \over 2^9
(\Delta-\ell+1)(\Delta-\ell+3)(\Delta+\ell+3)(\Delta+\ell+5)^2(\Delta+\ell+7)} \,
u^3 G^{(\ell+1)}_{\Delta +7}(u,v)  \cr
& {}+
{(\Delta - \ell+2)^2 (\Delta - \ell+4)^2 (\Delta + \ell+4)^2 \over 2^{11}
(\Delta-\ell+1)(\Delta-\ell+3)^2(\Delta-\ell+5)(\Delta+\ell+3)(\Delta+\ell+5)} \,
u^4 G^{(\ell-1)}_{\Delta +7}(u,v)\bigg )
\, . \cr
&& \Afif \cr}
$$
Finally the singlet contribution is
\eqnn\Aon
\hskip-1cm{$$\eqalignno{
A_{1}&(u,v) \cr
= {}& {\ts{1\over 3}}u^{{1\over 2}(\Delta - \ell)} \bigg (
60  G^{(\ell)}_{\Delta}(u,v) +
{10(\Delta + \ell+2)(\Delta + \ell+4) \over (\Delta+\ell+1)(\Delta+\ell+5)} \,
G^{(\ell+2)}_{\Delta +2}(u,v) \cr
& {}+ 
{4} \, uG^{(\ell)}_{\Delta+2}(u,v) + 
{5(\Delta - \ell)(\Delta - \ell+2) \over 8 (\Delta-\ell-1)(\Delta-\ell+3)} \,
u^2 G^{(\ell-2)}_{\Delta + 2}(u,v) \cr
& {}+
{15(\Delta + \ell+4)^2(\Delta + \ell+6)^2 \over 4 (\Delta+\ell+3)
(\Delta+\ell+5)^2(\Delta+\ell+7)} \,
G^{(\ell+4)}_{\Delta +4}(u,v) \cr
& {}+
{(\Delta + \ell+4)^2 \over (\Delta+\ell+3)(\Delta+\ell+5)} \,
u G^{(\ell+2)}_{\Delta +4}(u,v) \cr
& {}+ {5(\Delta-\ell)(\Delta-\ell+2)(\Delta+\ell+2)(\Delta+\ell+4)
\over 48(\Delta-\ell-1)(\Delta-\ell+3)(\Delta+\ell+1)(\Delta+\ell+5)}\,
u^2 G^{(\ell)}_{\Delta +4}(u,v) 
+{1\over 3} u^2 G^{(\ell)}_{\Delta +4}(u,v)\cr
& {}+
{(\Delta - \ell+2)^2 \over 16 (\Delta-\ell+1)(\Delta-\ell+3)} \,
u^3 G^{(\ell-2)}_{\Delta +4}(u,v)  \cr
& {}+
{15(\Delta - \ell+2)^2 (\Delta - \ell+4)^2  \over 2^{10} 
(\Delta-\ell+1)(\Delta-\ell+3)^2(\Delta-\ell+5)} \,
u^4 G^{(\ell-4)}_{\Delta +4}(u,v)  \cr
& {}+
{5(\Delta - \ell)(\Delta - \ell+2)(\Delta + \ell+4)^2 
(\Delta + \ell+6)^2 \over 128 (\Delta - \ell-1)(\Delta - \ell+3)
(\Delta+\ell+3)(\Delta+\ell+5)^2(\Delta+\ell+7)} \,
u^2 G^{(\ell+2)}_{\Delta +6}(u,v)  \cr
& {}+
{(\Delta - \ell+2)^2(\Delta + \ell+4)^2 \over 64
(\Delta-\ell+1)(\Delta-\ell+3)(\Delta+\ell+3)(\Delta+\ell+5)} \,
u^3 G^{(\ell)}_{\Delta +6}(u,v) \cr
& {}+
{5(\Delta - \ell+2)^2 (\Delta - \ell+4)^2(\Delta + \ell+2)(\Delta + \ell+4) 
\over 2^{11} (\Delta-\ell+1)(\Delta-\ell+3)^2(\Delta-\ell+5)(\Delta+\ell+1)
(\Delta+\ell+5)} \,
u^4 G^{(\ell-2)}_{\Delta +6}(u,v)\cr
& {}+
{15(\Delta - \ell+2)^2 (\Delta - \ell+4)^2(\Delta + \ell+4)^2(\Delta + \ell+6)^2 
\over 2^{14} (\Delta-\ell+1)(\Delta-\ell+3)^2(\Delta-\ell+5)(\Delta+\ell+3)
(\Delta+\ell+5)^2(\Delta+\ell+7)} \,
u^4 G^{(\ell)}_{\Delta +8}(u,v) \bigg ) . \cr
&& \Aon \cr}
$$
It is evident that these results correspond exactly to what would be expected
for a long multiplet whose lowest dimension operator is a singlet of
spin $\ell$. The symmetry conditions \crossR, using \Gsym, require $\ell$
to be even. 

For later use a more succinct notation is useful so the $A_R$ are assembled
as a vector,
\eqn\Avec{
\A = (A_1,A_{15},A_{20},A_{84},A_{175},A_{105}) \, ,
}
and then the equations \Afive, \Aseven, \Aeight, \Atwen, \Afif\ and
\Aon\ are expressed as
\eqn\AGvec{
\A(u,v) = \G_{\Delta,\ell}(u,v) \, ,
}
defining implicitly the vector $ \G_{\Delta,\ell}$.

If $\ell=0$ then the above results are valid if we use  \mml\ and also
\eqn\mmll{
u^3  G^{(-4)}_\Delta(u,v) = -64  G^{(2)}_\Delta(u,v) \, , \qquad
u^2  G^{(-3)}_\Delta(u,v) = -16  G^{(1)}_\Delta(u,v) \, .
}
The corresponding results are then
\eqnn\Azero\hskip-1cm{
$$\eqalignno{
A_{105}(u,v)=  {}& u^{{1\over 2}(\Delta+4)} G^{(0)}_{\Delta +4}(u,v)\,, \cr
A_{175}(u,v) = {}& u^{{1\over 2}(\Delta +2)} \bigg (
2 G^{(1)}_{\Delta +3}(u,v) 
+ {(\Delta +4)^2 \over 8 (\Delta+3)(\Delta+5)} \,
u G^{(1)}_{\Delta +5}(u,v) \bigg ) \, , \cr
A_{84}(u,v) = {}& u^{{1\over 2}(\Delta +2)} \bigg (
3G^{(0)}_{\Delta+2}(u,v)  +
{\Delta (\Delta +4) \over 16 (\Delta+1)(\Delta+3)} \,
u G^{(0)}_{\Delta +4}(u,v)  \cr
& {} + {3(\Delta +4)^2 \over 4 (\Delta+3)(\Delta+5)} \,
G^{(2)}_{\Delta +4}(u,v) +
{3(\Delta +2)^2 (\Delta +4)^2 \over 256
(\Delta+3)^2(\Delta+5)(\Delta+1)} \,
u^2 G^{(0)}_{\Delta +6}(u,v)\bigg )  ,  \cr
A_{20}(u,v) = {}& u^{{1\over 2}\Delta} \bigg (
{20\over 3}G^{(2)}_{\Delta+2}(u,v) +
{5(\Delta +4)^2 \over 12 (\Delta+3)(\Delta+5)} \,
u G^{(2)}_{\Delta +4}(u,v) \cr
& {}+ {\Delta(\Delta+4)\over 16 (\Delta-1)(\Delta+5)} \,
u^2 G^{(0)}_{\Delta +4}(u,v)  +
{5(\Delta +4)^2 (\Delta +6)^2 \over 192
(\Delta+3)(\Delta+5)^2(\Delta+7)} \,
u^2 G^{(2)}_{\Delta +6}(u,v)\bigg ) , \cr
A_{15}(u,v) = {}& u^{{1\over 2}\Delta} \bigg (
8 G^{(1)}_{\Delta+1}(u,v) +
{2(\Delta +4)^2 \over (\Delta+3)(\Delta+5)} \,
G^{(3)}_{\Delta +3}(u,v)  + {(\Delta +4)^2 \over 2(\Delta+3)(\Delta+5)} \,
u G^{(1)}_{\Delta +3}(u,v) \cr
& {}+ {(\Delta+4)^2(\Delta+6)^2\over 
8 (\Delta+3)(\Delta+5)} \, u G^{(3)}_{\Delta +5}(u,v) 
+ {\Delta^2(\Delta+4)^2\over 
32(\Delta-1)(\Delta+3)(\Delta+5)}\, 
u^2 G^{(1)}_{\Delta +5}(u,v) \cr
&{}+ {(\Delta+2)^2(\Delta +4)^2 (\Delta +6)^2 \over 2^9
(\Delta+1)(\Delta+3)^2(\Delta+5)^2(\Delta+7)} \,
u^3 G^{(1)}_{\Delta +7   }(u,v)\bigg ) ,  \cr
A_{1}(u,v) = {}& u^{{1\over 2}\Delta} \bigg (
{20}G^{(0)}_{\Delta}(u,v) + 
{10(\Delta+2)(\Delta +4) \over 3(\Delta+1)(\Delta+5)} \,
G^{(2)}_{\Delta + 2}(u,v) +
{(\Delta-2)(\Delta +4) \over 2(\Delta-1)(\Delta+3)} \,
u G^{(0)}_{\Delta + 2}(u,v) \cr
&{}+ {5(\Delta+4)^2(\Delta +6)^2 \over 
4 (\Delta+3)(\Delta+5)^2(\Delta+7)} \,
G^{(4)}_{\Delta + 4}(u,v) +
{(\Delta-2)(\Delta +6) \over 
48 (\Delta+1)(\Delta+3)^2(\Delta+5)} \,
u G^{(2)}_{\Delta + 4}(u,v) \cr
& {}+ {\Delta^2(\Delta+2)^2(\Delta+4)\over 16
(\Delta-1)(\Delta+1)(\Delta+3)(\Delta+5)} \,
u^2 G^{(0)}_{\Delta +4}(u,v) \cr
&{}+ {5\Delta(\Delta+2)(\Delta +4)^2 (\Delta +6)^2 \over 384
(\Delta-1)(\Delta+3)^2(\Delta+5)^2(\Delta+7)} \,
u^2 G^{(2)}_{\Delta +6}(u,v) \cr
&{}+ {\Delta(\Delta +6) \over 2^9
(\Delta+1)^2(\Delta+3)^2(\Delta+5)^2} \,
u^3 G^{(0)}_{\Delta +6}(u,v) \cr
&{}+ {5(\Delta+2)^2(\Delta +4)^2 (\Delta +6)^2 \over 2^{14}
(\Delta+1)(\Delta+3)^3(\Delta+5)^3(\Delta+7)} \,
u^4 G^{(0)}_{\Delta +8}(u,v) \bigg ) \, . & \Azero \cr}
$$}
The different contributions correspond exactly with those expected
according to \listN\ with $\ell$ restricted to be even or odd
for the $1,20,84,105$ or $15,175$-representations.

We now turn to an analysis of the of the contributions corresponding to
the functions $\tf(z), \, \tf_2(z)$ in \AG. As in the $\N=2$ case
these are associated with operators in which $\Delta$ is related to
$\ell$. First if we restrict to only the contribution of the term $k$
in \AG\ we have, setting $k=1$,
\eqn\Aid{
A_1 =20 \, , \ A_{105}=A_{175}=A_{84} = A_{20}=A_{15} = 0 \quad
\Rightarrow \quad \A(u,v) = \I \, ,
}
where  $\I=(20,0,0,0,0,0)$.
This is obviously the contribution corresponding to the identity operator.

If we consider now in \AG\ just the function $\tf$ we must have
\eqn\AAAz{
A_{105}=A_{175}=A_{84} =0 \, .
}
Assuming $\tf(z)= g_{\ell+1}(z)$, in the notation of \Gbps, we have
using \relg,
\eqn\AAA{\eqalign{       
A_{20}(u,v) = {}&{\ts{5\over 3}}\, u G^{(\ell)}_{\ell + 2}(u,v) \, , \cr
A_{15}(u,v) = {}& 2\, u G^{(\ell-1)}_{\ell+1}(u,v) + {2(\ell+1)^2 
\over (2\ell+1)(2\ell+3)}\, u G^{(\ell+1)}_{\ell + 3}(u,v)\, , \cr
A_{1}(u,v) = {}&5\,u G^{(\ell-2)}_{\ell}(u,v) +
{10\ell(\ell+1) \over 3(2\ell-1)(2\ell+3)}\, u G^{(\ell)}_{\ell + 2}(u,v)\cr
&{}+ {5(\ell+1)^2(\ell+2)^2 \over (2\ell+1)(2\ell+3)^2(2\ell+5)}\, 
u G^{(\ell+2)}_{\ell + 4}(u,v) \, . \cr}
}
The results given by \AAAz\ and \AAA, with the notation in \Avec,
may now be written as
\eqn\ABvec{
\A(u,v) = \B_\ell(u,v) \, ,
}
defining the vector $\B_\ell$.  The operators 
which give \AAA\ in the operator product expansion have twist two and
to comply with \crossG\ $\ell$ must be even.

If we consider $\ell=0$ in \AAA\ this gives
\eqn\AAzero{\eqalign{
A_{20}(u,v) = {}& {\ts{5\over 3}}\, u G^{(0)}_{2}(u,v) \, , \qquad
A_{15}(u,v) = {\ts{2\over 3}}\, u G^{(1)}_{3}(u,v) \, , \cr
A_{1}(u,v) = {}& {\ts{4\over 9}}\, u G^{(2)}_{4}(u,v) -20 \, . \cr}
}
The term $-20$ in $A_1$ corresponds to the identity operator so that
from \ABvec\ and \Aid\ this may be written as
\eqn\ABz{
\B_0(u,v) = {\hat \B}_0 (u,v) - \I \, .
}
The result given by ${\hat \B}_0$
then represents precisely the contribution in the 
operator product expansion of the basic short multiplet exhibited
in \Nfour\ where the lowest dimension operator is belongs to a
$SU(4)_R$ $20$-representation with $\Delta=2$. The expression for $A_1$
arises from the energy momentum tensor in the operator product
expansion and for $A_{15}$ from the $SU(4)_R$ conserved current.

For $\tf_2$ the results are more involved. We first take 
$\tf_2(z) = g_{\ell+2}(z)$ in \AG\ with $\ell$  even to ensure
\crossG\ holds. By virtue of \relg\ this leads
to $f_2$ being a linear combination of $g_\ell, \, g_{\ell+2}, \,
g_{\ell+4}$. Using \relg\ and also
\eqn\ggrel{
g_\ell(z) + g_\ell(x) = -2 \, G_\ell^{(\ell)}(u,v) + \half \, u
G_\ell^{(\ell-2)}(u,v) + {\ell^2\over 2(2\ell-1)(2\ell+1)}\, u
G_{\ell+2}^{(\ell)}(u,v) \, ,
}
which is derived in appendix C, we may then obtain
\eqn\Aftwo{\eqalign{
A_{105}(u,v)={}& 0 \, , \qquad A_{175}(u,v) = u G_{\ell+3}^{(\ell+1)}(u,v)\, ,
\qquad A_{20}(u,v) = {\ts{10\over 3}}\,  G_{\ell+2}^{(\ell+2)}(u,v) \, , \cr
A_{84}(u,v) = {}& {\ts{3\over 2}} \, u G_{\ell+2}^{(\ell)}(u,v)
+ {3(\ell+2)^2\over 2(2\ell+3)(2\ell+5)}\, u G_{\ell+4}^{(\ell+2)}(u,v) \, ,\cr
A_{15}(u,v) = {}& 4 \, G_{\ell+1}^{(\ell+1)}(u,v)
+ {4(\ell+2)^2\over (2\ell+3)(2\ell+5)}\, G_{\ell+3}^{(\ell+3)}(u,v) \, ,\cr
A_{1}(u,v) = {}& 10\, G_{\ell}^{(\ell)}(u,v) +
{20(\ell+1)(\ell+2)\over 3(2\ell+1)(2\ell+5)}\, G_{\ell+2}^{(\ell+2)}(u,v)\cr
&{}+ {10(\ell+2)^2(\ell+3)^2\over (2\ell+3)(2\ell+5)^2(2\ell+7)}\,
G_{\ell+4}^{(\ell+4)}(u,v)\cr
&{}- {\ts{5\over 2}} \, u G_{\ell}^{(\ell-2)}(u,v) - 
{3\ell(\ell+2)\over 2(2\ell-1)(2\ell+5)}\, u G_{\ell+2}^{(\ell)}(u,v)\cr
&{}-{3(\ell+1)(\ell+2)^2(\ell+3)\over 2(2\ell+1)(2\ell+3)(2\ell+5)(2\ell+7)}\,
u G_{\ell+4}^{(\ell+2)}(u,v)\cr
&{}-{5(\ell+2)^2(\ell+3)^2(\ell+4)^2\over 
2(2\ell+3)(2\ell+5)^2(2\ell+7)^2(2\ell+9)}\,
u G_{\ell+6}^{(\ell+4)}(u,v) \, .  \cr}
}
For this case these results define  the vector $\C_\ell$
so that  \Aftwo\ becomes
\eqn\ACvec{
\A(u,v) = \C_\ell(u,v) \, .
}

The results \Aftwo\ are not acceptable in isolation since the 
corresponding operators do not satisfy the necessary unitarity conditions.
This precludes any operators with twist $\Delta-\ell=0$. To eliminate
such terms, and also to obtain positive coefficients, we consider the 
combination 
\eqn\CBA{
\D_\ell= \G_{\ell,\ell} - 2\C_\ell - \B_\ell - {(\ell+2)^2\over 
(2\ell+3)(2\ell+5)} \, \B_{\ell+2} \, ,
}
which removes all twist 0 and twist 2 terms. The detailed results 
for $\D_\ell$ are then
\eqnn\Adn{\hskip -0.5cm$$\eqalignno{
A_{105}(u,v)={}& u^2  G_{\ell+4}^{(\ell)}(u,v) \, , \cr
A_{175}(u,v) = {}& \half u^2 G_{\ell+3}^{(\ell-1)}(u,v) +
{(\ell+2)^2\over 2(2\ell+3)(2\ell+5)}\, 
u^2 G_{\ell+5}^{(\ell+1)}(u,v) + {\ts {1\over 24}}\,
u^3 G_{\ell+5}^{(\ell-1)}(u,v) \, , \cr
A_{84}(u,v) = {}& {\ts{1\over 4}} \, u^2 G_{\ell+4}^{(\ell)}(u,v) +
{\ts{1\over 16}} \, u^3 G_{\ell+4}^{(\ell-2)}(u,v) 
+ {(\ell+2)^2\over 16(2\ell+3)(2\ell+5)}\, u^3 G_{\ell+6}^{(\ell)}(u,v) \, ,\cr
A_{20}(u,v) = {}& {\ts{5\over 12}}\, u^2 G_{\ell+2}^{(\ell-2)}(u,v) +
{\ts{5\over 144}}\, u^3 G_{\ell+4}^{(\ell-2)}(u,v) +
{\ts{7\over 72}}\, u^2 G_{\ell+4}^{(\ell)}(u,v)  \cr
&{}+ {5\over 24 (2\ell+1)(2\ell+5)}\,  u^2 G_{\ell+4}^{(\ell)}(u,v) +
{5(\ell+2)^2(\ell+3)^2\over 12 (2\ell+3)(2\ell+5)^2(2\ell+7)}\,  
u^2 G_{\ell+6}^{(\ell+2)}(u,v)\cr
&{}+ {5(\ell+2)^2\over 144 (2\ell+3)(2\ell+5)}\, 
u^3 G_{\ell+6}^{(\ell)}(u,v) +{\ts{1\over 432}} \, u^4 
G_{\ell+6}^{(\ell-2)}(u,v) \, , \cr
A_{15}(u,v) = {}& {\ts {1\over 6}} \,u^2 G_{\ell+3}^{(\ell-1)}(u,v)
+ {\ts {1\over 24}} \,u^3 G_{\ell+3}^{(\ell-3)}(u,v) +
{(\ell+2)^2\over 6(2\ell+3)(2\ell+5)}\, u^2G_{\ell+5}^{(\ell+1)}(u,v) \cr
&{}+ {(2\ell+3)^2-3\over 48(2\ell+1)(2\ell+5)}\, u^3G_{\ell+5}^{(\ell-1)}(u,v)
+ {\ts {1\over 360}} \,u^4 G_{\ell+5}^{(\ell-3)}(u,v)\cr
&{}+ {(\ell+2)^2(\ell+3)^2\over 24(2\ell+3)(2\ell+5)^2(2\ell+7)}\, 
u^3G_{\ell+7}^{(\ell+1)}(u,v) \cr
&{} + 
{(\ell+2)^2\over 360(2\ell+3)(2\ell+5)}\, u^4G_{\ell+7}^{(\ell-1)}(u,v)\, , \cr
A_{1}(u,v) = {}& {\ts{1\over 9}}\, u^2 G_{\ell+4}^{(\ell)}(u,v) +
{\ts{1\over 36}}\, u^3 G_{\ell+4}^{(\ell-2)}(u,v) +
{\ts{1\over 144}}\, u^4 G_{\ell+4}^{(\ell-4)}(u,v) \cr
&{}+ {(\ell+2)^2\over 36(2\ell+3)(2\ell+5)}\, u^3 G_{\ell+6}^{(\ell)}(u,v)
+ {(\ell+1)(\ell+2)\over 216(2\ell+1)(2\ell+5)}\, u^4 
G_{\ell+6}^{(\ell-2)}(u,v) \cr
&{}+ {(\ell+2)^2(\ell+3)^2\over 144(2\ell+3)(2\ell+5)^2(2\ell+7)}\, u^4
G_{\ell+8}^{(\ell)}(u,v) \, . & \Adn \cr}
$$}
For $\ell=2,4\dots$ these results correspond to the lowest dimension operator 
having $\Delta=\ell+2$, belonging to the $20$-representation, with
maximum dimension $\Delta=\ell+8$.
$\ell=0$ is again a special case. Using \mml\ and \mmll\ we obtain
\eqnn\Adnz$$\eqalignno{
A_{105}(u,v)={}& u^2  G_{4}^{(0)}(u,v) \, , \quad
A_{175}(u,v) = {\ts{2\over 15}}\, u^2 G_{5}^{(1)}(u,v)\, , \quad 
A_{84}(u,v) = {\ts{1\over 60}} \, u^3 G_{6}^{(0)}(u,v) \, , \cr
& A_{20}(u,v) =  - {\ts{5\over 3}}\, u G_{2}^{(0)}(u,v) +
{\ts{1\over 35}}\, u^2 G_{6}^{(2)}(u,v) \, , \cr
& A_{15}(u,v) =  - {\ts {2\over 3}} \,u G_{3}^{(1)}(u,v)
+ {\ts {1\over 350}} \,u^3 G_{7}^{(1)}(u,v) \, , \cr
& \ A_{1}(u,v) =  - {\ts{4\over 9}}\, u G_{4}^{(2)}(u,v) +
{\ts{1\over 2100}}\, u^4 G_{8}^{(0)}(u,v) \, . & \Adnz \cr}
$$
The negative terms present in the results for
$A_1, A_{15}, A_{20}$  in \Adnz\ are just those
corresponding to the short supermultiplet built on the scalar with
$\Delta=2$ belonging to the 20-representation, as given in \AAzero.
Thus we may write
\eqn\ADz{
\D_0 = - {\hat \B}_0 + {\hat \D}_0 \, ,
}
where the terms in ${\hat \D}_0$ then represent
the contribution for a lowest dimension scalar operator with $\Delta=4$
belonging to the 105-representation. This is a chiral primary
operator and the contributions to ${\hat \D}_0$  correspond exactly to the
short supermultiplet operators with $Y=0$ for this case, 
as listed in appendix B for $p=4$.

When $\Delta=\ell+2$ the contributions given by the operator product
expansion in $\G_{\Delta,\ell}$ can also be decomposed since we may write
\eqn\GBE{
\G_{\ell+2,\ell} = 4\, \B_{\ell+2} + \E_\ell \, ,
}
where $\E_\ell$ corresponds to operators with twist $\ge 4$. The detailed
results for  $\E_\ell$ are then given by
\eqnn\AE{\hskip -0.5cm$$\eqalignno{
A_{105}(u,v)={}& u^3  G_{\ell+6}^{(\ell)}(u,v) \, , \cr
A_{175}(u,v) = {}& 2\, u^2 G_{\ell+5}^{(\ell+1)}(u,v) +
\half \,  u^3 G_{\ell+5}^{(\ell-1)}(u,v) \cr
{}& + {(\ell+3)^2\over 2(2\ell+5)(2\ell+5)}\, 
u^3 G_{\ell+7}^{(\ell+1)}(u,v) + {\ts {1\over 30}}\,
u^4 G_{\ell+7}^{(\ell-1)}(u,v) \, , \cr
A_{84}(u,v) = {}& 3 \, u^2 G_{\ell+4}^{(\ell)}(u,v) +
\quar \, u^3 G_{\ell+6}^{(\ell)}(u,v) 
+ {3(\ell+3)^2\over (2\ell+5)(2\ell+7)}\, u^3 G_{\ell+6}^{(\ell+2)}(u,v) \cr
&{}+ {\ts{1\over 20}} \, u^4 G_{\ell+6}^{(\ell-2)}(u,v) +
{(\ell+3)^2\over 20(2\ell+5)(2\ell+7)}\, u^4 G_{\ell+8}^{(\ell)}(u,v) \, ,\cr
A_{20}(u,v) = {}& {\ts{5\over 3}}\, u^2 G_{\ell+4}^{(\ell)}(u,v) +
{\ts{5\over 12}}\, u^3 G_{\ell+4}^{(\ell-2)}(u,v) +
{5(\ell+3)^2\over 3(2\ell+5)(2\ell+7)}\, u^2 G_{\ell+6}^{(\ell+2)}(u,v) \cr
&{}+ {5\big ((2\ell+5)^2 -3 \big ) \over 24 (2\ell+3)(2\ell+7)}\,  
u^3 G_{\ell+6}^{(\ell)}(u,v)+ 
{\ts{1\over 36}}\, u^4 G_{\ell+6}^{(\ell-2)}(u,v) \cr
&{}+ {5(\ell+3)^2(\ell+4)^2\over 12 (2\ell+5)(2\ell+7)^2(2\ell+9)}\,  
u^3 G_{\ell+8}^{(\ell+2)}(u,v)
+ {(\ell+3)^2\over 36(2\ell+5)(2\ell+7)}\, 
u^4 G_{\ell+8}^{(\ell)}(u,v) \cr 
&{} +{\ts{1\over 560}} \, u^5 G_{\ell+8}^{(\ell-2)}(u,v) \, , \cr
A_{15}(u,v) = {}& 2 \,u^2 G_{\ell+3}^{(\ell-1)}(u,v) \cr
&{} + {(2\ell+5)^2 -3  \over (2\ell+3)(2\ell+7)}\,  
u^2 G_{\ell+5}^{(\ell+1)}(u,v) 
+ {\ts {3\over 10}} \,u^3 G_{\ell+5}^{(\ell-1)}(u,v) +
{\ts {1\over 30}} \,u^4 G_{\ell+5}^{(\ell-3)}(u,v) \cr
&{}+ {2(\ell+3)^2(\ell+4)^2\over (2\ell+5)(2\ell+7)^2(2\ell+9)}\,
u^2 G_{\ell+7}^{(\ell+3)}(u,v) +
{3(\ell+3)^2\over 10(2\ell+5)(2\ell+7)}\, u^3G_{\ell+7}^{(\ell+1)}(u,v) \cr
&{}+ {(2\ell+5)^2-3\over 60(2\ell+3)(2\ell+7)}\, u^4
G_{\ell+7}^{(\ell-1)}(u,v)
+ {\ts {3\over 1400}} \,u^5 G_{\ell+7}^{(\ell-3)}(u,v)\cr
&{}+ {(\ell+3)^2(\ell+4)^2\over 30(2\ell+5)(2\ell+7)^2(2\ell+9)}\, 
u^4 G_{\ell+9}^{(\ell+1)}(u,v) + 
{3(\ell+2)^2\over 1400(2\ell+5)(2\ell+7)}\, u^5 
G_{\ell+9}^{(\ell-1)}(u,v)\, , \cr
A_{1}(u,v) = {}& {\ts{4\over 3}}\, u^2 G_{\ell+4}^{(\ell)}(u,v) +
{\ts{1\over 3}}\, u^3 G_{\ell+4}^{(\ell-2)}(u,v)\cr 
&{}+ {4(\ell+3)^2\over 3(2\ell+5)(2\ell+7)}\, u^2 G_{\ell+6}^{(\ell+2)}(u,v)
+ {(2\ell+5)^2-3\over 6 (2\ell+3)(2\ell+7)}\, u^3
G_{\ell+6}^{(\ell)}(u,v) \cr
&{} + {\ts{1\over 45}}\, u^4 G_{\ell+6}^{(\ell-2)}(u,v)  +
{\ts{3\over 560}}\, u^5 G_{\ell+6}^{(\ell-4)}(u,v) \cr
&{}+ {(\ell+3)^2\over 45(2\ell+5)(2\ell+7)}\, u^4
G_{\ell+8}^{(\ell)}(u,v)  
+ {(\ell+2)(\ell+3)\over 280(2\ell+3)(2\ell+7)}\, u^5 
G_{\ell+8}^{(\ell-2)}(u,v) \cr
&{}+ {3(\ell+3)^2(\ell+4)^2\over 560(2\ell+5)(2\ell+7)^2(2\ell+9)}\, u^5
G_{\ell+10}^{(\ell)}(u,v) \, . 
& \AE \cr}
$$}
In this the lowest dimension operator occurs in the 15-representation
with $\Delta=\ell+3$. $\ell=0$ is again a special case. In this case \AE\
reduces to 
\eqnn\AEz$$\eqalignno{
A_{105}(u,v)={}& u^3  G_{6}^{(0)}(u,v) \, , \qquad
A_{175}(u,v) = 2 \, u^2 G_{5}^{(1)}(u,v) + 
{\ts {9\over 70}} \, u^3 G_{7}^{(1)}(u,v) \, , \cr
A_{84}(u,v) = {}& 3\, u^2  G_{4}^{(0)}(u,v) +
{\ts{1\over 20}} \, u^3 G_{6}^{(0)}(u,v) +
{\ts{27\over 35}} \, u^2 G_{6}^{(2)}(u,v) +
{\ts{9\over 700}} \, u^4 G_{8}^{(0)}(u,v)\, , \cr
A_{20}(u,v) = {}&  {\ts{3\over 7}}\, u^2 G_{6}^{(2)}(u,v) +
{\ts{3\over 28}}\, u^3 G_{6}^{(0)}(u,v) +
{\ts{4\over 147}}\, u^3 G_{8}^{(2)}(u,v) \, , \cr
A_{15}(u,v) = {}& {\ts {18\over 35}} \,u^2 G_{5}^{(1)}(u,v)
+ {\ts {32\over 245}} \,u^2 G_{7}^{(3)}(u,v) 
+ {\ts {3\over 70}} \,u^3 G_{7}^{(1)}(u,v) 
+ {\ts {8\over 3.5^2.7^2}} \,u^4 G_{9}^{(1)}(u,v) \, , \cr
A_{1}(u,v) =  {}& {\ts{3\over 35}}\, u^3 G_{6}^{(0)}(u,v) +
{\ts{16\over 15.7^2}}\, u^2 G_{8}^{(2)}(u,v) 
+ {\ts{2\over 5^2.7^2}}\, u^4 G_{8}^{(0)}(u,v)
+ {\ts{3\over 5^2.7^3}}\, u^5 G_{10}^{(0)}(u,v) \, . \cr
&& \AEz \cr}
$$
Thus for $\E_0$ the lowest dimension relevant operator is a scalar
belonging to the 84-representation with $\Delta=4$, corresponding 
to a well known short multiplet.

As a result of the above considerations we expect in general expansions
of the form
\eqn\OPEfour{\eqalign{
& \G(u,v) = \sum_{\Delta,\ell} a_{\Delta,\ell} \,
u^{{1\over 2}(\Delta - \ell)} G^{(\ell)}_{\Delta+4} (u,v) \, , \cr
\tf(z) = {}& \sum_{\ell} b_\ell  \, g_{\ell+1}(z) \, , \quad
\tf_2(z) = \sum_{\ell} c_\ell  \, g_{\ell+2}(z) \, , \qquad \ell=0,2,4
\dots \, , \cr}
}
where, along with $k$, $a_{\Delta,\ell}, \,  b_\ell, \,  c_\ell$
determine the operator product expansion coefficients  for all $A_R$
since we may write
\eqn\OPER{
\A(u,v) = k\I + 
\sum_\ell \big ( b_\ell \, \B_\ell(u,v) + c_\ell \, \C_\ell (u,v) \big ) 
+ \sum_{\Delta,\ell} a_{\Delta,\ell}\, \G_{\Delta,\ell}(u,v) \, . 
}
If we use \ABz, \CBA, \ADz, assuming $c_\ell + 2a_{\ell,\ell}=0$ as
is necessary for unitarity, this may be rewritten in the following form
\eqn\opeN{\eqalign{
\A(u,v) = {}& C \,\I + b_0 \, {\hat \B}_0(u,v) + 
d_0 \, {\hat \D}_0(u,v) + e_0 \, {\E}_0(u,v) \cr
&{} + \sum_{\ell\ge 2} \big (  d_\ell \, \D_\ell (u,v) +
e_\ell \, \E_\ell(u,v) \big ) +
\sum_{\Delta,\ell}  
{\hat a}_{\Delta,\ell}\, \G_{\Delta,\ell}(u,v) \, , \cr}
}
where
\eqn\Cde{ 
C = k-b_0+\half c_0 \, , \quad  d_\ell= - \half c_\ell \, ,
\quad e_\ell = - \quar b_{\ell+2} + {\ts {1\over 8}} c_{\ell+2} 
+ {(\ell+2)^2 \over 8(2\ell+1)(2\ell+3)}\, c_{\ell} \, , 
}
and ${\hat a}_{\Delta,\ell}$ is identical with $a_{\Delta,\ell}$
save that ${\hat a}_{\ell,\ell}=0$ and ${\hat a}_{\ell+2,\ell} =
a_{\ell+2,\ell} - e_\ell$. In the first line of \opeN\ the
contribution of various short multiplets is explicit. The
range of scale dimensions and spins of the sets of 
operators which contribute to the different terms in \opeN\ are 
summarised in the following table,

\hskip1cm
\vbox{\tabskip=0pt \offinterlineskip
\hrule
\halign{&\vrule# &\strut \ \hfil#\  \cr
height0pt&\omit&&\omit &&\omit &&\omit&&\omit&&\omit&&\omit&\cr
&\hfil &&\multispan5 \hfil lowest weight operator \hfil&& && && &\cr
height0pt&\omit&&\omit &&\omit &&\omit&&\omit&&\omit&&\omit&\cr
\noalign{\vskip-8pt}
&\hfil &&\multispan5\hrulefill&& && && & \cr
\noalign{\vskip-4pt}
&\hfil &&\ $R$ \ \  && $\Delta$ \quad && spin \ \ 
&& $\Delta_{\rm max}$ && $\ell_{\rm min}\quad $&&$\ell_{\rm max}$\quad &\cr 
height2pt&\omit&&\omit&&\omit&&\omit&&\omit&&\omit&&\omit&\cr
\noalign{\hrule}
&\ $\G_{\Delta,\ell}$ \quad && \ 1 \ && \ $\Delta$ \quad &
& \ $\ell \, $  \quad && \ $\Delta+8$ \ &&\ $\ell-4$ \ &&\ $\ell+4$ \ &\cr
&\ $\D_\ell$ \quad && \ 20 \ && \ $\ell+2$ \ &
& \ $\ell-2$ \ &&\ $\ell+8$ \ && \ $\ell-4$ \  &&\ $\ell+2$ \ & \cr
&\ $\E_\ell$ \quad && \ 15 \ && \ $\ell+3$ \ &
&\ $\ell-1$ \ &&\ $\ell+10$ \ && \ $\ell-4$ \  &&\ $\ell+3$ \ & \cr
&\ ${\hat \B}_0$ \quad && \ 20 \ && \ 2 \quad &
&\ 0 \quad  &&\ 4 \ \quad &&\ 0 \quad &&\ 2 \quad & \cr
&\ ${\hat \D}_0$ \quad && \ 105 \ && \ 4 \quad &
&\ 0 \quad &&\ 8 \ \quad && \ 0 \quad  &&\ 2 \quad & \cr
&\ $\E_0$ \quad && \ 84 \ && \ 4 \quad &
&\ 0 \quad &&\ 10  \quad && \ 0 \quad  &&\ 3 \quad &\cr
height0pt&\omit&&\omit&&\omit&&\omit&&\omit&&\omit&\cr}
\hrule}

As an illustration of these results we consider the free case when 
$a_1=a_2=a_3=a$ and $c_1=c_2=c_3=c$ are both constants. From \Aac\ 
and \AG\ we then have
\eqn\Gfree{
\G(u,v) = a \Big ( 1 + {1\over v^2} \Big ) + c\, {1\over v} \, ,
}
and from \solac, \fff\ and \dff,
\eqn\kff{\eqalign{
& k=3(a+c) \, , \cr
\tf(z) =  a (z^2 - z'^2) +  c(z-z')\, , \quad  &
\tf_2(z) = a (z^2 + z'^2) - c(z+z') \, , \quad z'={z\over z-1} \, . \cr}
}
The expansion coefficients in \OPEfour\ for this case may be determined 
in terms of our previous results \Dos,
\eqn\al{\hskip-0.3cm
a_{\Delta,\ell}  =  2^{\ell}{(\ell+t)!(\ell+t+1)!(t!)^2
\over (2\ell+2t+1)!(2t)!}\,\Big ( a (\ell+1)(\ell+2t+2) + c(-1)^t \Big ) 
\de_{\Delta,\ell+2t} \, ,
\quad {\ell=0,2\dots \atop \, t=0,1,\dots} \, , 
}
and also
\eqn\bcl{\hskip -0.5cm
b_\ell = 2^{\ell+1}\,{(\ell!)^2 \over (2\ell)!} \big ( {- \ell}(\ell+1)a
+c \big ) \, , \  \
c_\ell = -2^{\ell+1}{\ell!(\ell+1)! \over (2\ell+1)!} 
\big ( (\ell+1)(\ell+2)a +c \big ) \, , \quad \ell=0,2 \dots \, . }
}
Clearly the condition $2a_{\ell,\ell} + c_\ell =0$ is satisfied
and in \opeN\ we have by using \Cde
\eqn\kb{\eqalign{
C = a \, , \qquad b_0 = {}& 2c \,  , \qquad d_\ell = 
2^{\ell}{\ell!(\ell+1)! \over (2\ell+1)!}
\big ( (\ell+1)(\ell+2)a +c \big ) \, , \cr
e_\ell ={}&  2^{\ell}{\big((\ell+2)!\big)^2 \over (2\ell+4)!} \,
\big ( (\ell+1)(\ell+4)a - 3c \big )\, ,\cr}
}
and furthermore
\eqn\ahat{
{\hat a}_{\ell+2,\ell} = 2^{\ell+1}{\big((\ell+2)!\big)^2 \over (2\ell+4)!} 
\, c \, .
}

\newsec{Results for Weak Coupling and Large $N$}

We show here how the results just exhibited in general in section 7
for the operator product expansion may be used to recover and extend
previous results \OPEW\ obtained in an analysis of perturbative corrections
to the four point correlation function in \Fourp. It is convenient
first to rewrite the solution of the superconformal identities in the
form
\eqn\Solac{\eqalign{
a_1 = {}& \quar \hN^2 + \half\hN \, u\F \, , \qquad
a_2 = \quar \hN^2 + \half\hN \, \F \, , \qquad
a_3 = \quar \hN^2 + \half\hN \, v\F \, , \cr
c_1 = {}& \hN \big ( 1 + \half (1-v-u)\F \big ) \, , \qquad
c_2 = \hN \big ( 1 - \half (1+v-u)\F \big ) \, , \cr
&\qquad \quad c_3 = \hN \big ( 1 - \half (1-v+u)\F \big ) \, . \cr}
}
Here we have used the normalisation convention of \twofourP\ and also,
following \Edent, assumed that the contributions corresponding to the
functions $f_{1,2,3}$ are given just by free field results. The
non-trivial dynamical results are therefore contained in the function
$\F(u,v)$ which for \cross\ and \crossuv\ must satisfy
$\F(u,v)=\F(v,u)=\F(u',v')/v$. From \Aac\ and \AG\ we have
\eqn\GF{
\G(u,v) = \quar \hN^2 \Big ( 1 + {1\over v^2} \Big ) + \hN \, {1\over v}
\Big ( 1 + \half \, u \F(u,v) \Big ) \, .
}

The contributions of short supermultiplets to the operator product
expansion are therefore given just by the free results and we use 
the formulae obtained in section 7 with $a=\quar \hN^2 , \ c=\hN$.
Denoting the operator product coefficients for operators in 
representation $R$ by $a^R_{\Delta,\ell}$
the identity operator and the three short multiplets are given, from \kb,
by
\eqn\ashort{
a^1_{0,0} = \quar \hN^2 \, , \qquad a^{20}_{2,0} = {\ts{10\over 3}}\hN \, 
\qquad a^{84}_{4,0} = \half \hN^2 \Big ( 1 - {3\over \hN} \Big ) \, ,
\qquad a^{105}_{4,0} = \half \hN^2 \Big ( 1 + {2\over \hN} \Big ) \, ,
}
where we have absorbed the relevant coefficients from \AAzero, \AEz\
for $R=20,\, 84$. There are also infinite sequences of unrenormalised
operators associated with the coefficients $d_\ell, \, e_\ell$ \opeN\
which correspond to
\eqn\als{\eqalign{
a^{20}_{\ell+4,\ell} = {}& \hN^2 2^\ell {5(\ell+2)!(\ell+3)!\over 3(2\ell+5)!}
\bigg ( \quar (\ell+3)(\ell+4) +{1\over \hN} \bigg ) \, , \cr
a^{15}_{\ell+5,\ell+1} = {}& \hN^2 2^{\ell+3} {\big((\ell+4)!\big)^2\over 
(2\ell+8)!} \bigg ( \quar (\ell+3)(\ell+6) +{3\over \hN} \bigg ) \, . \cr}
}
The formulae \ashort\ and \als\ are in agreement with the results in \OPEW.

The ${\rm O}(g^2)$ result for the four point function in \Fourp\ 
in a $SU(N)$ $\N=4$ superconformal theory, when  $\hN=N^2-1$ is the
dimension of the adjoint representation,  has been calculated by
various authors \refs{\Gonz,\Eloop,\Bia}.  It is reducible to a standard 
conformal integral giving for $\F$ 
\eqn\FP{
\F(u,v) = - {\tilde \lambda} \, \Phi^{(1)}(u,v) \, , \qquad
{\tilde \lambda} = {g^2 N \over 4\pi^2} \, ,
}
where, using the variables $z,x$ defined in \uvxz,
\eqn\Pone{
\Phi^{(1)}(u,v) = {1\over z-x} \bigg ( \ln zx \, \ln {1-z \over 1-x}  +
2{\rm Li}_2(z) - 2{\rm Li}_2 (x) \bigg ) \, ,
}
with ${\rm Li}_2$ the dilogarithm function.  The $\ln u = \ln zx$
term in \Pone\ generates a shift in the dimensions of long 
supermultiplets, the dilogarithm terms generate a series which is
analytic in $u,1-v$. To analyse the consequences of the $\ln u$ term
we expand
\eqn\expln{
- {u\over v} \, {1\over z-x}  \ln {1-z \over 1-x} = \sum_{t,\ell}
\ep_{t,\ell} \, u^t G^{(\ell)}_{\ell+2t+4}(u,v) \, .
}
By matching terms in a power series expansion we find
\eqn\atl{\eqalign{
\ep_{t,\ell} = {}& 2^{\ell+3}{\big( (\ell+t+1)!\, t!\big )^2
\over (2\ell+2t+2)!\,(2t)!}\,\sum_{j=1}^{\ell+t+1}{1\over j} \, ,
\qquad \ {\ell=0,2\dots \atop \, t=1,3,\dots} \, , \cr
\ep_{t,\ell} = {}& - 2^{\ell+2}{\big( (\ell+t)!\big )^2 (t+1)!\, t!
\over (2\ell+2t+1)!\, (2t)!}\,\sum_{j=1}^{t}{1\over j} \, ,
\qquad {\ell=0,2\dots \atop \, t=2,4,\dots} \, .\cr} 
}

For long multiplets the dimensions of the lowest weight operators
in each supermultiplet may be written as
\eqn\Dd{
\Delta_{t,\ell} = 2t + \ell + \eta_{t,\ell} \, , \qquad
t=1,2, \dots \, , \quad \ell = 0,2,\dots \, .
}
We first consider $t=1$. Using \ahat, with $c=\hN$, then the perturbative
results give
\eqn\deone{
\eta_{1,\ell} = {\ep_{1,\ell} \over {\hat a}_{\ell+2,2}} \, 
{\tilde \lambda} \hN = 2 \sum_{j=1}^{\ell+2} {1\over j} \, {\tilde \lambda}\, ,
}
which is the result given in the introduction in \Konl.
Note that $\eta_{1,0}=3{\tilde \lambda}, \, 
\eta_{1,2}={25\over 6}{\tilde \lambda}, \, 
\eta_{1,4} ={49\over 10}{\tilde \lambda}$ which coincide with the results
of Anselmi \Ans. For $\ell=0$ this corresponds to the Konishi
supermultiplet. This and also $\eta_{1,2}$ were obtained in \OPEW\ by
analysis of the perturbative four point function, but here there is no need to 
separate carefully the contributions of
the three free field theory $\Delta=4, \, \ell=2$ operators, which include 
the energy momentum tensor and a descendant of the Konishi scalar operator.
For $t>1$ we just quote the large $N$ result based on the tacit
assumption that there is just one long supermultiplet contributing
to the operator product expansion in each case,
\eqn\etatl{\eqalign{
\eta_{t,\ell} ={} & {16\over (\ell+1)(\ell+2t+2)} \!
\sum_{j=1}^{\ell+t+1}\! {1\over j} \, {{\tilde \lambda} \over \, N^2} \, ,
\qquad\qquad\qquad\quad \ {\ell=0,2\dots \atop \, t=3,5,\dots} \, , \cr
\eta_{t,\ell} ={} & -{16(t+1)\over (\ell+1)(\ell+t+1)(\ell+2t+2)} 
\sum_{j=1}^{t} {1\over j} \, {{\tilde \lambda} \over \, N^2} \, ,
\qquad \ {\ell=0,2\dots \atop \, t=2,4,\dots} \, . \cr}
}
These results correspond to `double trace' operators.

We now consider the analogous results for large $N$ and also large
$\tilde \lambda$ which may be obtained from supergravity calculations
on $AdS_5 \times S^5$ using the AdS/CFT correspondence. Arutunov and
Frolov \Arut\ calculated the leading large $N$ behaviour of the the
four point function for chiral primary operators in the 20-representation
which is expressed here in the form \Fourp. As shown in appendix D
their result may be simplified to the form given in \Solac\ with
\eqn\FDs{
\F(u,v)=-2 \oD_{2224}(u,v) \, ,
}
where $\oD_{\Delta_1\Delta_2\Delta_3\Delta_4}(u,v)$ are functions of
the conformal invariants $u,v$ which arise from Feynman type integrals
on AdS space. For our purposes we make use of the analysis in \Dos\ 
to write
\eqn\Dexp{
\oD_{2224}(u,v) = {\ts{1\over 3}} F(2,1;4;1-v) +  
G(u,v) \, u \ln u + {\rm O}(u) \, ,
}
where the discarded terms are analytic in $u,1-v$. From \Dos\foot{In terms
of the results in \Dos\ we have $\oD_{2224}(u,v)=H(2,1,0,4;u,v) =
uH(3,2,2,6;u,v)$ and $G(u,v) = {1\over 5}G(3,2,2,6;u,1-v)$. 
We may also express $G$ more explicitly by
$$\eqalign{
G(u,v) = {}& \bigg ( {120uv^2\over(z-x)^7} + 
{12v\over (z-x)^5}(1+u+v) \bigg ) \ln{1-z\over 1-x} \cr
& {} + {60uv\over (z-x)^6}(1+v-u) +
{2 \over (z-x)^4} \Big ( (1+v)(1+v-u)+16v  \Big  ) \, . \cr}
$$}
the function $G$ has the expansion
\eqn\Gexp{
G(u,v) = {1\over 5}\sum_{m,n=0} {(3)_m(4)_m \over m!\, (2)_m} \, 
{(3)_{m+n}(2)_{m+n}\over n!\, (6)_{2m+n}} u^m(1-v)^n \, , \qquad
(\alpha)_m = {\Gamma(\alpha+m) \over \Gamma(\alpha)} \, .
}
The $\ln u$ terms in \Dexp\ are absorbed by a shift in the scale dimensions of 
the operators appearing in the operator product expansion. To determine
these we first expand
\eqn\Gop{
{u^2\over v}G(u,v) = \sum_{t,\ell}
{\tilde \ep}_{t,\ell} \, u^t G^{(\ell)}_{\ell+2t+4}(u,v) \, .
}
By using the expansion \Gexp\ the coefficients ${\tilde \ep}_{t,\ell}$
may be determined iteratively and can then be fitted to the formula
\eqn\epN{
{\tilde \ep}_{t,\ell} = 2^{\ell-2}(t-1){t!\, (t+2)!\, (\ell+t)!\, (\ell+t+1)!
\over (2t-1)!\, (2\ell+2t+1)!} \, , \qquad \ 
{\ell=0,2\dots \atop \, t=2,3,\dots} \,.
}
Assuming a single lowest weight operator for a long representation for
any $\Delta,\ell$ the order $1/N^2$ corrections to the dimensions in \Dd\
may be determined from $\eta_{t,\ell}= -2N^2 {\tilde \ep}_{t,\ell}/
a_{\Delta,\ell}$ where $a_{\Delta,\ell}$ is given by \al\ for $t=2,3\dots$
and $a \to \quar N^4$,
\eqn\etN{
\eta_{t,\ell} = - {4(t-1)t(t+1)(t+2)\over (\ell+1)(\ell+2t+2)}\, 
{1\over N^2} \, .
}
For $t=2$ this coincides with the result of Hoffman, Mesref and R\"uhl \HMR\
who considered the four point function involving, in our notation, the
descendant fields $\Phi$ and  $\bPhi$. For the simplest case,
$t=2,\ell=0$, $\eta_{2,0}= -16/N^2$  was derived
earlier in \Hokone\ and also from the results of \Arut\ in \OPEN.

We may also take account of the first term in \Dexp\ by expanding
\eqn\Fexp{
{u\over 3v} F(2,1;4;1-v) = \sum_{\ell=0} b_\ell \, u G_{\ell+6}^{(\ell)}(u,v)
+ \hbox{higher twist} \, ,
}
or equivalently
\eqn\Fexpt{
{1\over 1-z}\, {\ts{1\over 3}}  F(2,1;4;z) = \sum_{\ell=0} b_\ell
(-\half z)^\ell F(\ell+3,\ell+3;2\ell+6;z) \, ,
}
which gives
\eqn\bell{
b_\ell = 2^{\ell+1}{\big((\ell+2)!\big)^2 \over (2\ell+4)!} \, , \qquad
\ell =0,2, \dots \, .
}
The contribution to $a_{\Delta,\ell}$ in \OPEfour\ is then $-b_\ell N^2$ for 
$\Delta=\ell+2$. This just cancels ${\hat a}_{\ell+2,\ell}$ in \ahat\ 
which is a reflection of the fact that the supermultiplet containing the
Konishi scalar and its partners disappear from the spectrum in the large
$N$ limit \OPEN.

\newsec{Discussion}

We have endeavoured to show in this paper the exact compatibility with
superconformal identities for the four point function of chiral primary
operators belonging to the simplest short representations of $\N=2$ and
$\N=4$ superconformal symmetry with the representation content of the
various possible supermultiplets of operators which may contribute in the
operator product expansion. In both cases the superconformal identities
are solved in terms of a single function $\G(u,v)$ of the two conformal
invariants which may be expanded in terms of operators belonging to
long supermultiplets whose lowest dimension operator is a singlet under
the appropriate $R$ symmetry group with arbitrary $\Delta$. This is in
agreement with the analysis of three point functions in \refs{\Non,\bpsN,\Hes}.
In our case the analysis depended on the result that for such supermultiplets
there was just one operator contributing to the operator product expansion
belonging to the $R=2$ representation for $\N=2$ and the 105-dimensional
representation for $\N=4$. Although we have not shown in detail the
impossibility of incorporating alternative long multiplets in the operator
product expansion for the four point functions here it is clear,
from the structure of \Detwo\ and \AG, that for contributions for general
$\Delta$ alone it is impossible to have operators with a finite range of
dimensions and spins if any other representation is assumed to contain
just one operator.

In addition the contribution of possible short supermultiplets involve
the single variable functions which are present in the solution of the
superconformal identities. The arguments of Eden {\it et al} \Edent\ 
show that such contributions have no perturbative corrections and are 
essentially given by the results for free field theory. Thus for $\N=2$
we have exhibited the contributions for lowest weight operators scalars
with $\Delta = 2R$, for $R=1,2$, and for $\N=4$ for chiral primary 
operators belonging to the $[q,p,q]$ representation, $\Delta=p+2q$, for
the relevant cases here of $q=0, \, p=1,2$ and $q=2, \, p=0$. The terms
appearing in the operator product expansion tie in with the expected
representation content.\foot{For a recent discussion of short multiplets
with $q\ne 0$ in $\N=4$ theories see \Hok.} For these short representations
the crucial restrictions arise at the first level.

There are also contributions for arbitrary spin $\ell$ which arise for
$\N=2$ when the basic inequality \ineq\ is saturated for $R=0,1, \, r=0$.
For $\N=4$ the corresponding unitarity inequality for a lowest weight operator 
with spin $\ell$ belonging to the $[q,p,q]$-representation is \Short,
\eqn\InN{
\Delta \ge 2 + \ell + p + 2q \,
}
In the results in section 8 we have found contributions to the operator
product expansion when this is an equality for
$\ell$ even for the $20$-dimensional representation, $p=2, \, q=0$,
represented by $\D_{\ell+2}$, and for $\ell$ odd for the $15$-dimensional
representation, $p=0, \, q=1$, represented by $\E_{\ell+1}$. The
necessity for a protected operator in the $20$-representation with
$\Delta=4$ was first shown in \OPEN. The full
supermultiplet structure does not seem to have been exhibited in these
cases but our results seem to imply that they do not correspond to a 
full long supermultiplet with dimensions proportional to $2^{16}$. 
The constraint on the dimensions is perhaps an extension of the
condition that the conservation condition on a current is only compatible
with conformal invariance in $d$-dimensions if $\Delta=d-1$.  A related
discussion $\N=4$ supersymmetric theories is given in \HH. 
The simplest example in this context is the supercurrent multiplet in $\N=2$
superconformal symmetry \Soh. With a similar notation to \Rtwo, where the
representations are denoted by $R_{(j_1,j_2)}$, this has the structure
\eqn\NJT{\def\normalbaselines{\baselineskip16pt\lineskip3pt
\lineskiplimit3pt}
\matrix{\Delta&~~~&{~~~}&&{~~~}&&{~~~}&&{~~~}&&{~~~}&&{~~~}&&\cr
2&&{~~~}&&{~~}&&0_{(0,0)}&&{~~}&&{~~~}&\cr
&&&&&\Bsw&&\Bse&&&&\cr
{\ts{5\over 2}}&&&&\hidewidth~~~~~\half_{({1\over 2},0)}
\hidewidth&&&&\hidewidth~~\half_{(0,{1\over 2})}~~\hidewidth&&&\cr&&&\Bsw&&
\Bse&&\Bsw&&\Bse&&\cr
3&&\hidewidth~~~~~~0_{(1,0)}\hidewidth&&&&
\hidewidth 0_{({1\over 2},{1\over 2})},1_{({1\over 2},{1\over 2})}\hidewidth&&&&
\hidewidth0_{(0,1)}\hidewidth&\cr
&&&\Bse&&\Bsw&&\Bse&&\Bsw&&\cr
{\ts{7\over 2}}&&&
&\hidewidth~~\half_{(1,{1\over 2})}\hidewidth&&&
&\hidewidth\half_{({1\over 2},1)}\hidewidth&&&\cr
&&&&&\Bse&&\Bsw&&&&\cr
4&&&&&
&\hidewidth 0_{(1,1)}\hidewidth&&&&&\cr
r&&~1&&{\ts{1\over 2}}&&0&&-{\ts{1\over 2}}&&-1&\cr}
}
This contains the conserved currents for $U(2)_R$ symmetry and the conserved
energy momentum tensor as well as the conserved spinorial supercurrents
so that there are 24 bosonic and 24 fermionic degrees of freedom. There
is however no shortening at the first level, so there is no BPS-like
condition in this case, although shortening appears at level 2. 
The operators appearing in \NJT\ for $r=0$  match those necessary for the 
operator product expansion results for the four point function in \Aem.
In general when \InN\ becomes an equality there are possible constraints on 
superfields \HH\ which may imply shortening at level 2.

A remaining issue is whether the conditions derived in section 5, and
solved in section 6, exhaust all the implications of superconformal 
symmetry in these cases. This seems highly probable in that the
conditions derived reflect directly the shortening conditions at the first
level in the supermultiplets to which the chiral primary operators 
belong but although they are necessary there is no guarantee of
sufficiency as yet. Of course in the future it would be very nice to
investigate more general correlation functions than those considered in
this paper.

\bigskip
\noindent{\bf Acknowledgements}

One of us (FAD) would like to thank the EPSRC, the National University of
Ireland and Trinity College, Cambridge for support and is also very
grateful to  David Grellscheid for help with Mathematica. HO would like to
thank Massimo Bianchi, Anastasios Petkou and Johanna Erdmenger for useful 
conversations.

\vfill\eject
\appendix{A}{$SU(4)$ formulae}

The link between the 4-dimensional indices $i,j,\dots $ and the
6-dimensional indices $r,s,\dots$ is given by the antisymmetric gamma
matrices,
\eqn\defga{
\ga_r{}^{\! ij} = - \ga_r{}^{\! ji} \, , \qquad
\bga_{rij} = \half \vep_{ijkl} \ga_r{}^{\! kl} \, ,
}
where we impose the completeness/orthogonality relations
\eqn\gao{
\ga_r{}^{\! ij} \bga_{sij} = 4 \de_{rs} \, , \qquad
\ga_r{}^{\! ij} \bga_{r kl} = 4\, \de^{[i}{}_{\!k} \, \de^{j]}{}_{\!l} \quad
\Rightarrow \quad \bga_{rij} \bga_{rkl} = 2\, \vep_{ijkl} \, .
}
It is then easy to see that we have the usual $\ga$-matrix algebra
\eqn\mat{
\ga_r \bga_s + \ga_s \bga_r = - 2 \de_{rs} 1\, , \qquad 
\bga_r \ga_s + \bga_s \ga_r = - 2 \de_{rs} 1 \, ,
}
with $1$ the $4\times 4$ unit matrix.  These $\ga$-matrices also satisfy
\eqn\gaep{
\ga_{[r}\bga_s\ga_t\bga_u\ga_v\bga_{w]}  = i \, \vep_{rstuvw} \,1 \, ,
}
involving the six dimensional antisymmetric symbol. This
leads to the useful relations
\eqn\garel{\eqalign{
\ga_{[r}\bga_s\ga_t\bga_u\ga_{v]} = {}& -i \, \vep_{rstuvw} \ga_w \, , \qquad
\ga_{[r}\bga_s\ga_t\bga_{u]} = -\half i \, \vep_{rstuvw} \ga_v\bga_w \, ,\cr
\ga_{[r}\bga_s\ga_{t]} = {}& {\ts{1\over 6}}i \, \vep_{rstuvw}
\ga_u\bga_v\ga_w \, . \cr}
}

For discussion of the four point function for the 20-dimensional
representation formed by $\vphi_{rs}$ it is also convenient to define a
basis by $C^I_{rs}= C^I_{sr}, \, C^I_{rr}=0$ satisfying 
\eqn\Cao{
C^I_{rs}C^J_{rs}= \de^{IJ} \, , \qquad C^I_{rs}C^I_{uv}=
\half\big ( \de_{ru}\de_{sv} + \de_{rv}\de_{su} \big ) -
{\ts {1\over 6}}\, \de_{rs}\de_{uv} \, .
}

For the correlation functions of 
$\vphi^I\equiv C^I_{rs}\vphi_{rs}^{\vphantom g}$
the relevant invariant tensors are then
\eqn\defCn{
\tr\big (C^{I_1}C^{I_2} \dots C^{I_n} \big ) = C^{I_1I_2\dots I_n} =
C^{I_n I_{n-1} \dots I_1} = C^{I_nI_1 \dots I_{n-1}} \, ,
}
and we also define
\eqn\CC{
C^{I_1I_2\dots I_n}_{rs} = C^I_{rs} C^{II_1I_2\dots I_n} =
(C^{I_1} \dots C^{I_n})_{(rs)} - {\ts{1\over 6}} \de_{rs}
C^{I_1I_2\dots I_n} \, .
}

Since $[0,2,0]\otimes [0,2,0] = [1,2,1]\oplus
[0,4,0] \oplus [2,0,2] \oplus [0,2,0] \oplus [1,0,1] \oplus [0,0,0]$
$(20\times 20 = 175 + 105 + 84 + 20 + 15 + 1)$, the four point function
$\langle \vphi^{I_1} \vphi^{I_2}\vphi^{I_3}\vphi^{I_4}\rangle$ may be
decomposed into six irreducible representations for which the associated
projection operators\foot{With a different normalisation these were
given in \OPEN.} are
\eqn\defPP{\eqalign{
P_1^{I_1I_2I_3I_4} = {}& {\ts{1\over 20}} \de^{I_1I_2} \de^{I_3I_4} \, , \cr
P_{15}^{I_1I_2I_3I_4} = {}& - \quar \big (C^{I_1I_2I_3I_4} - C^{I_2I_1I_3I_4}
\big ) \, , \cr
P_{20}^{I_1I_2I_3I_4} = {}& {\ts{3\over 10}} \big (C^{I_1I_2I_3I_4} +
C^{I_2I_1I_3I_4} \big ) - {\ts{1\over 10}} \de^{I_1I_2} \de^{I_3I_4} \, , \cr
P_{84}^{I_1I_2I_3I_4} = {}& {\ts{1\over 3}} \big ( \de^{I_1I_3}\de^{I_2I_4}
+ \de^{I_2I_3} \de^{I_1I_4} \big ) + 
{\ts{1\over 30}} \de^{I_1I_2} \de^{I_3I_4} \cr
{}& - {\ts{2\over 3}} C^{I_1I_3I_2I_4} - {\ts{1\over 6}}
\big (C^{I_1I_2I_3I_4} + C^{I_2I_1I_3I_4} \big ) \, , \cr
P_{105}^{I_1I_2I_3I_4} = {}& {\ts{1\over 6}} \big ( \de^{I_1I_3}\de^{I_2I_4}
+ \de^{I_2I_3} \de^{I_1I_4} \big ) +  
{\ts{1\over 60}} \de^{I_1I_2} \de^{I_3I_4} \cr
{}& + {\ts{2\over 3}} C^{I_1I_3I_2I_4} - {\ts{2\over 15}}
\big (C^{I_1I_2I_3I_4} + C^{I_2I_1I_3I_4} \big ) \, , \cr
P_{175}^{I_1I_2I_3I_4} = {}& {\ts{1\over 2}} \big ( \de^{I_1I_3}\de^{I_2I_4}
- \de^{I_2I_3} \de^{I_1I_4} \big ) + \quar \big (C^{I_1I_2I_3I_4} - 
C^{I_2I_1I_3I_4} \big ) \, , \cr}
}
which satisfy
\eqn\PPP{
P_R^{I_1I_2IJ} P_{R'}^{IJI_3I_4} = \de_{RR'}^{\vphantom g}
P_R^{I_1I_2I_3I_4} \, , \qquad
\sum_R P_R^{I_1I_2I_3I_4} = \de^{I_1I_3}\de^{I_2I_4} \, , \qquad
P_R^{IJIJ} = R \, .
}

For the product of the fermion field $\psi_{ri}$ and its conjugate 
$\bpsi{}^{\,j}_s$ we have $[0,1,1]\otimes [1,1,0] = [1,2,1]\oplus
[2,0,2] \oplus [2,1,0] \oplus [0,1,2] \oplus [0,2,0] \oplus [1,0,1]
\oplus [1,0,1] \oplus [0,0,0]$
$(20\times 20 = 175 + 84 + 45 + 45 + 20 + 15 + 15 + 1)$. Hence in
the four point function $\langle \psi_{r} \bpsi{}_s\vphi^I \vphi^J\rangle$ 
we may construct six invariant tensors,
\eqn\defTT{\eqalign{
T^{(1)IJ}_{rs} = {}& \big ( \de_{rs} 1 + {\ts{1\over 6}} \bga_r \ga_s 
\big ) \de^{IJ} \, , \cr
T^{(2)IJ}_{rs} = {}& C^{IJ}_{rs} + {\ts{1\over 6}} \bga_r \ga_u
C^{IJ}_{us} + {\ts{1\over 6}} C^{IJ}_{rv} \bga_v \ga_s \, , \cr
\tT^{(3)IJ}_{rs} = {}& (C^IC^J)_{[rs]} + {\ts{1\over 8}} \bga_r \ga_u
(C^IC^J)_{[us]} + {\ts{1\over 8}} (C^IC^J)_{[rv]} \bga_v \ga_s \cr
{}& + {\ts{1\over 48}}i\, \vep_{rsuvtw} (C^IC^J)_{[uv]} \bga_t\ga_w \, ,\cr
\tT^{(4)IJ}_{rs} = {}&  \de_{rs} (C^IC^J)_{[uv]} \bga_u \ga_v + \quar
\bga_r \ga_u (C^IC^J)_{[us]} + \quar (C^IC^J)_{[rv]} \bga_v \ga_s \cr
{} & + {\ts{5\over 24}}i\, \vep_{rsuvtw} (C^IC^J)_{[uv]} \bga_t\ga_w \, ,\cr 
T^{(5)IJ}_{rs} = {}& C^I_{ru} \bga_u^{\vphantom g} \, \ga_v^{\vphantom g} 
C^J_{vs} \, , \cr
T^{(6)IJ}_{rs} = {}& C^J_{ru} \bga_u^{\vphantom g} \, \ga_v^{\vphantom g} 
C^I_{vs} \, . \cr}
}
For each term in \defTT\ we have
\eqn\Tga{
\ga_r^{\vphantom g} T^{(n)IJ}_{rs} = 0 \, , \qquad \quad 
T^{(n)IJ}_{rs} \bga_s^{\vphantom g} = 0 \, .
}
$T^{(1)}$, $T^{(2)}$  correspond to the contribution of the singlet, 
20- representation while $\tT^{(3)},\tT^{(4)}$ arise from the 
two 15-representations.
For the analysis of the four point function we need
\eqn\TCg{\eqalign{
T^{(1)IJ}_{rt}C^K_{ts}\bga_s^{\vphantom g} = {}& \de^{IJ} 
C^K_{rs}\bga_s^{\vphantom g} \, , \cr
T^{(2)IJ}_{rt}C^K_{ts}\bga_s^{\vphantom g} = {}& - {\ts{1\over 6}} \de^{IJ} 
C^K_{rs}\bga_s^{\vphantom g} + \half \big ( C^{IJK}_{rs} + C^{JIK}_{rs} 
\big ) \bga_s^{\vphantom g} + \half \big ( D^{IJK}_r + D^{JIK}_r \big )\, , \cr
\tT^{(3)IJ}_{rt}C^K_{ts}\bga_s^{\vphantom g} = {}& 
\half \big ( C^{IJK}_{rs} - C^{JIK}_{rs} \big ) \bga_s^{\vphantom g}
+{\ts {7\over 12}} \big ( D^{IJK}_r - D^{JIK}_r \big )
+ {\ts {1\over 24}} E^{KIJ}_r \, ,  \cr
\tT^{(4)IJ}_{rt}C^K_{ts}\bga_s^{\vphantom g} = {}& 
- \big ( C^{IJK}_{rs} - C^{JIK}_{rs} \big ) \bga_s^{\vphantom g}
+{\ts {11\over 6}} \big ( D^{IJK}_r - D^{JIK}_r \big )
+ {\ts {17\over 12}} E^{KIJ}_r \, ,  \cr
T^{(5)IJ}_{rt}C^K_{ts}\bga_s^{\vphantom g} = {}& - \de^{JK}
C^I_{rs}\bga_s^{\vphantom g}
- \big ( C^{IJK}_{rs} - C^{IKJ}_{rs} \big ) \bga_s^{\vphantom g}
-  D^{IJK}_r + D^{IKJ}_r  +  E^{IJK}_r \, ,  \cr
T^{(6)IJ}_{rt}C^K_{ts}\bga_s^{\vphantom g} = {}& - \de^{IK}
C^J_{rs}\bga_s^{\vphantom g}
- \big ( C^{JIK}_{rs} - C^{IKJ}_{rs} \big ) \bga_s^{\vphantom g}
-  D^{JIK}_r - D^{IKJ}_r  +  E^{JIK}_r \, ,  \cr}
}
where
\eqn\DE{\eqalign{
D^{IJK}_r = {}& (C^IC^JC^K)_{[rs]} + {\ts{1\over 6}}
(C^IC^JC^K)_{[ts]}\bga_r \ga_t \bga_s \, , \cr
E^{IJK}_r = {}& C^I_{r[t} (C^JC^K)_{uv]}^{\vphantom g}\bga_t^{\vphantom g}
\ga_u^{\vphantom g}\bga_v^{\vphantom g}+ {\ts{1\over 6}}
(C^IC^JC^K - C^IC^KC^J)_{[ts]}^{\vphantom g} \bga_r^{\vphantom g}
\ga_t^{\vphantom g}\bga_s^{\vphantom g} \, , \cr}
}
which necessarily satisfy, from \Tga, $\ga_r^{\vphantom g} D^{IJK}_r =
\ga_r^{\vphantom g} E^{IJK}_r =0$. As a consequence of \TCg\ it is more
convenient to use a basis where
\eqn\TTn{
T^{(3)IJ}_{rs} = {\ts {1\over 36}}\big ( 34 \tT^{(3)IJ}_{rs} - \tT^{(4)IJ}_{rs} 
\big ) \, , \qquad 
T^{(4)IJ}_{rs} = {\ts {1\over 3}}\big ( 2 \tT^{(3)IJ}_{rs} + \tT^{(4)IJ}_{rs} 
\big )  \, ,
}
which from \TCg\ satisfy
\eqn\TCgg{\eqalign{
T^{(3)IJ}_{rt}C^K_{ts}\bga_s^{\vphantom g} = {}& 
\half \big ( C^{IJK}_{rs} - C^{JIK}_{rs} \big ) \bga_s^{\vphantom g}
+ \half \big ( D^{IJK}_r - D^{JIK}_r \big )  \, ,  \cr
\tT^{(4)IJ}_{rt}C^K_{ts}\bga_s^{\vphantom g} = {}& 
D^{IJK}_r - D^{JIK}_r  + \half E^{KIJ}_r \, .  \cr}
}

\vfill\eject
\appendix{B}{Superconformal Short Multiplets}
The component fields in four dimensional $\N=4$ supersymmetry may be denoted
by $[k,p,q]_{(j_1,j_2)}$ where $[k,p,q]$ are the Dynkin labels for the
$SU(4)$ $R$-symmetry representation and $(j_1,j_2)$ label $(2j_1+1)(2j_2+1)$
dimensional representations of the Lorentz group. For a short multiplet
of the superconformal group starting from a scalar $[0,p,0]$ field with
dimension $\Delta=p$ the field content is obtained by superconformal 
transformations $\ep$ $(\swarrow)$ and $\bep$ $(\searrow)$, following
\Class, as follows
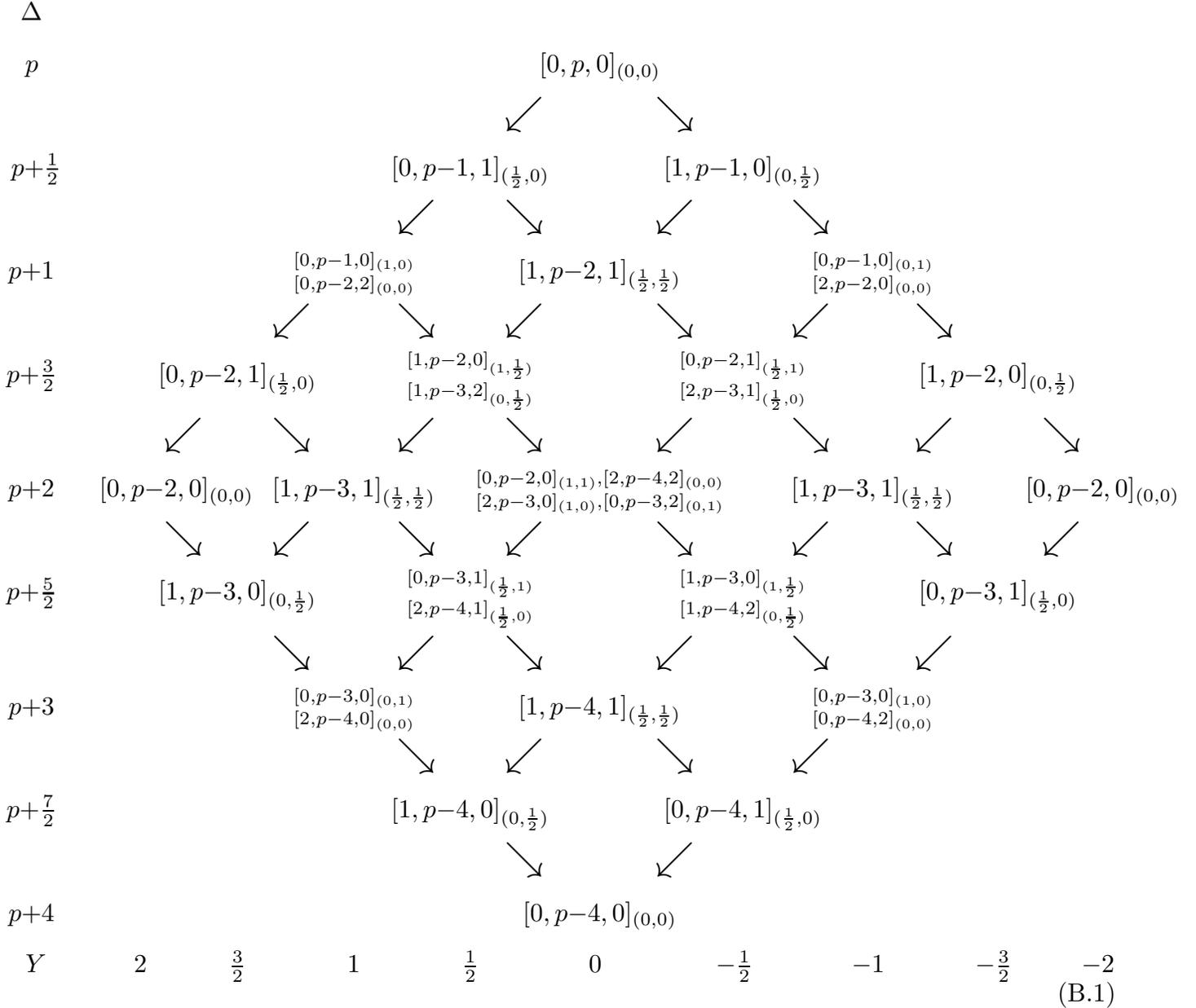
\eqn\Nall{\def\normalbaselines{\baselineskip20pt\lineskip3pt
 \lineskiplimit3pt}\hskip-1.5cm
\matrix{\Delta~~\cr
p~~&~~~~~&&{~~~}&&{~~~~~}&&{~~~}&&\hidewidth[0,p,0]_{(0,0)}\hidewidth&
&{~}&&{~}&&{~}&&\cr
&&&&&&&&\Bsw&{~~~~~~~~}&\Bse&&&&&&&\cr
p{+\half}~&&&&&&&\hidewidth[0,p{-1},1]_{({1\over 2},0)}\hidewidth&&&&
\hidewidth[1,p{-1},0]_{(0,{1\over 2})}\hidewidth&&&&&&\cr
&&&&&&\Bsw&&\Bse&&\Bsw&&\Bse&&&&&\cr
p{+1}~~&&&&&\hidewidth {[0,p-1,0]_{(1,0)}\atop[0,p-2,2]_{(0,0)}}\hidewidth&&&&
\hidewidth [1,p{-2},1]_{({1\over 2},{1\over 2})}\hidewidth&&&&
\hidewidth{[0,p-1,0]_{(0,1)}\atop[2,p-2,0]_{(0,0)}}\hidewidth&&&&\cr
&&&&\Bsw&&\Bse&&\Bsw&&\Bse&&\Bsw&&\Bse&&&\cr
p{+{\ts{3\over 2}}}~~&&&\hidewidth[0,p{-2},1]_{({1\over 2},0)}\hidewidth&&&
&\hidewidth{[1,p-2,0]_{(1,{1\over 2})}\atop[1,p-3,2]_{(0,{1\over 2})}}\hidewidth&&&&
\hidewidth{[0,p-2,1]_{({1\over 2},1)}\atop[2,p-3,1]_{({1\over 2},0)}}
\hidewidth&&&&\hidewidth[1,p{-2},0]_{(0,{1\over 2})}\hidewidth&&\cr 
&&\Bsw&&\Bse&&\Bsw&&\Bse&&\Bsw&&\Bse&&\Bsw&&\Bse&\cr
p{+2}~~&[0,p{-2},0]_{(0,0)}\hidewidth&&&&
\hidewidth[1,p{-3},1]_{({1\over 2},{1\over 2})}\hidewidth&&&&
\hidewidth {[0,p-2,0]_{(1,1)},[2,p-4,2]_{(0,0)}\atop
[2,p-3,0]_{(1,0)},[0,p-3,2]_{(0,1)}} \hidewidth 
&&&&\hidewidth[1,p{-3},1]_{({1\over 2},{1\over 2})}\hidewidth
&&&&\hidewidth[0,p{-2},0]_{(0,0)}\hidewidth\cr
&&\Bse&&\Bsw&&\Bse&&\Bsw&&\Bse&&\Bsw&&\Bse&&\Bsw&\cr
p{+{\ts{5\over 2}}}~~&&&\hidewidth[1,p{-3},0]_{(0,{1\over 2})}\hidewidth&&&&\hidewidth
{[0,p-3,1]_{({1\over 2},1)}\atop[2,p-4,1]_{({1\over 2},0)}}\hidewidth&&&&
\hidewidth{[1,p-3,0]_{(1,{1\over 2})}\atop[1,p-4,2]_{(0,{1\over 2})}}
\hidewidth&&&&\hidewidth[0,p{-3},1]_{({1\over 2},0)}\hidewidth&&\cr
&&&&\Bse&&\Bsw&&\Bse&&\Bsw&&\Bse&&\Bsw&&&\cr
p{+3}~~&&&&&\hidewidth {[0,p-3,0]_{(0,1)}\atop[2,p-4,0]_{(0,0)}}\hidewidth&&&&
\hidewidth [1,p{-4},1]_{({1\over 2},{1\over 2})}\hidewidth&&&&
\hidewidth{[0,p-3,0]_{(1,0)}\atop[0,p-4,2]_{(0,0)}}\hidewidth&&&&\cr
&&&&&&\Bse&&\Bsw&&\Bse&&\Bsw&&&&&\cr
p{+{\ts{7\over 2}}}~~&&&&&&&\hidewidth[1,p{-4},0]_{(0,{1\over 2})}\hidewidth&&&&
\hidewidth[0,p{-4},1]_{({1\over 2},0)}\hidewidth&&&&&&\cr
&&&&&&&&\Bse&&\Bsw&&&&&&&\cr
p{+4}~~&&&{~~}&&{~~~}&&{~~}&&\hidewidth[0,p{-4},0]_{(0,0)}\hidewidth&&{~~}&&{~~~}&&{~~}&&\cr
Y~&~~~~2\hidewidth&&{\ts{3\over 2}}&&1&&\half&&0~&&-\half~~&&-1~&
&-{\ts{3\over 2}}~&&\hidewidth-2~\hidewidth\cr}
}
This representation has dimension $2^8\times{1\over 12}p^2(p^2-1)$ for
$p=2,3,\dots$. We may note that ${1\over 12}p^2(p^2-1)={\rm dim}[0,p-2,0]$
which may be understood since the whole supermultiplet may be generated
from the representation with $Y=2$ using $\bep{}^{\, i \dal}$ supersymmetry
transformations, the dimensions of the representations in each column add
to ${8\choose 8-4Y}{1\over 12}p^2(p^2-1)$.

The lowest weight operator, belonging to the $[0,p,0]$-representation, 
can be represented by  symmetric, traceless tensor, $\vphi_{r_1\dots r_p}$.
The closure of the superconformal algebra follows directly from the
considerations of section 4, from \deP, \sPhi\ and \SPsi, and  if
\eqn\ptph{
\de \vphi_{r_1\dots r_p} = - \hep \ga_{(r_p} \psi_{r_1\dots r_{p-1})} +
\bpsi{}_{(r_1\dots r_{p-1}} \bga_{r_p)} \hbep \, , 
}
where $\psi_{i r_1\dots r_{p-1}\alpha}$ and
$\bpsi{}^i{}_{\!r_1\dots r_{p-1}\dal}$ belong to the $[0,p-1,1]$ and
$[1,p-1,0]$-representations respectively. From \clP{a,b}
\eqn\ptps{\eqalign{
\de \psi_{r_1\dots r_{p-1}\alpha} ={}&
i \pr_{\alpha\dal} \vphi_{r_1\dots r_{p-1}s} \, \bga_s \hbep{}^{\, \dal}
+ 2p \, \vphi_{r_1\dots r_{p-1}s} \, \bga_s \eta_\alpha + \dots \, ,\cr
\de \bpsi{}_{r_1\dots r_{p-1}\dal} = {}& \hep^\alpha \ga_s \, i  
\pr_{\alpha\dal} \vphi_{r_1\dots r_{p-1}s} - 
2p\, \vphi_{r_1\dots r_{p-1}s}\, \bta{}^\dal \ga_s + \dots \, , \cr}
}
so that we may then obtain
\eqn\pcp{
[\de_2 , \de_1] \vphi_{r_1\dots r_p} = - v{\cdot \pr} \vphi_{r_1\dots r_p} 
- p\, \hla\, \vphi_{r_1\dots r_p} + ip\, \hht_{(r_p|s}\, 
\vphi_{r_1\dots r_{p-1})s} \, ,
}
with the notation of \lfour\ and \defla.

Following the discussion in section 4 we may also for completeness
consider supermultiplets with the lowest weight operators belonging to 
$(j,0)$ or $(0,j)$ spin representations
and rederive the usual shortening conditions on $\Delta$ for these cases. 
Thus for a field $\Phi^I{}_{\! \alpha_1 \dots \alpha_{2j}}
= \Phi^I{}_{\! (\alpha_1 \dots \alpha_{2j})}$ transforming according to
a $(j,0)$ representation \deP\ is extended to
\eqn\dePP{
\de \Phi^I_{\, \alpha_1 \dots \alpha_{2j}} = \hep{}_i^\beta
\Psi^{Ii}_{\, \beta \, \alpha_1 \dots \alpha_{2j}} + 
{\Lambda}^I_{\smash {\, i \alpha_1 \dots \alpha_{2j}  \dbe}}
\, \hbep{}^{\, i\dbe} \, .
}
The relevant superconformal variations of 
$\Psi^{I}_{\,\beta\,\alpha_1\dots \alpha_{2j}}$
and ${\Lambda}^I_{\smash{\, \alpha_1 \dots \alpha_{2j} \dbe}}$ are then 
\eqna\clPP{\hskip-0.8cm
$$\eqalignno{
\de_{\hbep} \Psi^{I}_{\,\beta\,\alpha_1 \dots \alpha_{2j}} = {} & (2-\nu)\, 
i \pr_{\smash{\beta\dbe}} \Phi^I_{\, \smash{\alpha_1 \dots \alpha_{2j}}}
 \, \hbep{}^{\,\dbe} + 2(\Delta-2j)\,  
\Phi^I_{\smash{\, \alpha_1 \dots \alpha_{2j}}} \eta_\beta^{\vphantom g} + 8j \,
\Phi^I_{\, \beta (\alpha_1 \dots \alpha_{2j-1}}\, 
\eta_{\alpha_{2j})}^{\vphantom g} \cr
&\!\!{}  - (T_{rs})^I{}_{\! J} \Phi^J_{\, \alpha_1 \dots \alpha_{2j}} 
\, \ga_{[r}\bga_{s]} \eta_\beta^{\vphantom g}  + a \, i
\pr_{\smash{(\alpha_1\dbe}}\Phi^I_{\smash{\,\beta |\alpha_2 \dots \alpha_{2j})}}
\, \hbep{}^{\,\dbe} \cr
&{}\!\! - {1\over 2\Delta}\, (T_{rs})^I{}_{\! J} \big ( \nu \, 
i\pr_{\smash{\beta\dbe}} \Phi^J_{\smash{\, \alpha_1 \dots \alpha_{2j}}}
- a \, i\pr_{\smash{(\alpha_1\dbe}} 
\Phi^J_{\smash{\, \beta |\alpha_2 \dots \alpha_{2j})}} \big )
\, \ga_{[r}\bga_{s]} \hbep{}^{\,\dbe} + 
\dots \, , & \clPP{a} \cr
\de_{\hep} {\Lambda}^I_{\smash {\,  \alpha_1 \dots \alpha_{2j}  \dbe}} =  {}& 
- \nu \, \hep^\beta i  \pr_{\smash{\beta\dbe}} 
\Phi^I_{\, \alpha_1 \dots \alpha_{2j}} + 2\Delta \, 
\Phi^I_{\, \alpha_1 \dots \alpha_{2j}} \, \bet{}_\dbe 
+ (T_{rs})^I{}_{\! J} \Phi^J_{\, \alpha_1 \dots \alpha_{2j}}
\, \bet{}_\dbe \ga_{[r}\bga_{s]} \cr
&{} + a \,  \hep^\beta i 
\pr_{\smash{(\alpha_1\dbe}}\Phi^I_{\smash{\,\beta |\alpha_2 \dots \alpha_{2j})}}
\cr
{}& - {1\over 2\Delta} \,   \hep^\beta \ga_{[r}\bga_{s]} \,
(T_{rs})^I{}_{\! J}  \big ( \nu \,
i\pr_{\smash{\beta\dbe}} \Phi^J_{\smash{\, \alpha_1 \dots \alpha_{2j}}}
- a \, i\pr_{\smash{(\alpha_1\dbe}} 
\Phi^J_{\smash{\, \beta |\alpha_2 \dots \alpha_{2j})}} \big )
+  \dots \, , & \clPP{b} \cr }
$$}
with $T_{rs}$ the appropriate $SU(4)$ generators and the coefficients
$\nu,a$ are given by
\eqn\nua{
\nu = {\Delta (\Delta +j -1) \over (\Delta +j )(\Delta -j -1)} \, , \qquad
a = {2\Delta j  \over (\Delta +j )(\Delta -j -1)} \, .
}
The form of \clPP{a,b}, with \nua, is determined by the requirement of
closure of the algebra
\eqn\closPP{
[\de_2, \de_1] \Phi^I_{\, \alpha_1 \dots \alpha_{2j}} = - \big ( v{\cdot \pr} 
+ \Delta \, \hla \big ) \Phi^I_{\, \alpha_1 \dots \alpha_{2j}}
+ 2j \, \hom_{(\alpha_1}{}^{\! \beta}
\Phi^I_{\, \beta |\alpha_2\dots \alpha_{2j})}
+ \half \hht_{rs} (T_{rs})^I{}_{\! J} \Phi^J_{\, \alpha_1 \dots \alpha_{2j}} \, ,
}
with notation as in \lfour\ and \defla, and also by requiring
\eqn\clvP{
[\de_\hep, \de_v ]{\Lambda}^I_{\smash {\, \alpha_1 \dots \alpha_{2j}\dbe}}
= \de_{\hep'}{\Lambda}^I_{\smash {\, \alpha_1 \dots \alpha_{2j}\dbe}} \, ,
\qquad 
[\de_{\hbep},\de_v ] \Psi^{I}_{\,\beta\,\alpha_1 \dots \alpha_{2j}} =
\de_{\hbep{}^\prime} \Psi^{I}_{\,\beta\,\alpha_1 \dots \alpha_{2j}} \, ,
}
for $\hep'$ defined in \epp\ and, as in \dO\ with \vcon\ and \wwh,
\eqn\dPP{\eqalign{
\de_v  \Phi^I_{\, \alpha_1 \dots \alpha_{2j}} = {}& - \big ( v{\cdot \pr}
+ \Delta \, \hla \big ) \Phi^I_{\, \alpha_1 \dots \alpha_{2j}}
+ 2j \, \hom_{(\alpha_1}{}^{\! \beta}
\Phi^I_{\, \beta |\alpha_2\dots \alpha_{2j})}\, , \cr
\de_v  \Psi^{I}_{\,\beta\,\alpha_1 \dots \alpha_{2j}} = {}&- \big ( v{\cdot \pr}
+ (\Delta + \half) \hla \big ) \Psi^{I}_{\,\beta\,\alpha_1 \dots \alpha_{2j}}
+ 2j \, \hom_{(\alpha_1}{}^{\!\! \gamma}
\Psi^{I}_{\,\beta\gamma|\alpha_1 \dots \alpha_{2j})} +
\hom_{\beta}{}^{ \gamma}
\Psi^{I}_{\,\gamma\,\alpha_1 \dots \alpha_{2j}}\, , \cr
\de_v  {\Lambda}^I_{\smash {\, \alpha_1 \dots \alpha_{2j}\dbe}}  = {}& -
\big ( v{\cdot \pr} + (\Delta + \half) \hla \big ) 
{\Lambda}^I_{\smash {\, \alpha_1 \dots \alpha_{2j}\dbe}} 
+ 2j \, \hom_{(\alpha_1}{}^{\!\! \gamma}
{\Lambda}^I_{\smash {\, \gamma|\alpha_1 \dots \alpha_{2j})\dbe}}
- {\Lambda}^I_{\smash {\, \alpha_1 \dots \alpha_{2j}\dga}}  \,
\hbom{}^\dga{}_{\smash\dbe} \, .
\cr}
}

As a consequence of \clPP{b} it is easy to see that multiplet shortening
occurs if, suppressing the now irrelevant spinorial indices,
\eqn\shoP{
2\Delta\, \Phi^I \, \bet + 
(T_{rs})^I{}_{\! J} \Phi^J \, \bet \ga_{[r}\bga_{s]} \, ,
}
does not span the full representation space. As an illustration we
consider the $[k,0,q]$-representation so that, similarly to \ssPhi,
$\Phi^I \to \vphi^{i_1\dots i_k}_{\,j_1\dots j_q}$ and
\eqn\shP{
(T_{rs})^I{}_{\! J} \Phi^J \, \bet_i (\ga_{[r}\bga_{s]})^i_{\ j} \to 
4k \, \de^{(i_1}_{\,j}\! \vphi^{i_2\dots i_{k})i}_{\,j_1\, \dots \, j_q} \,
\bet_i^{\vphantom g} - 4q \, 
\vphi^{i_1\dots i_k}_{\,j(j_1\dots j_{q-1}} \bet_{j_q)}^{\vphantom g}
-(k-q)\, \vphi^{i_1\dots i_k}_{\,j_1\dots j_q} \bet_j^{\vphantom g} \, .
}
It is then easy to see that in \shoP\ that if
\eqn\Dkq{
2\Delta = 3q +k \, ,
}
then the $[k,0,q+1]$-representation does not appear on the right hand
side of  \clPP{b} allowing 
${\Lambda}^I_{\smash {\,i\alpha_1 \dots\alpha_{2j}\dbe}}$ to be 
restricted to the $[k,1,q-1]$ and $[k-1,0,q]$-representations.

\appendix{C}{Recurrence Relations for Conformal Partial Waves}

In four dimensions we recently \Dos\ derived an explicit formula for the
contribution of a quasi-primary operator of dimension $\Delta$ and belonging
to a $(j,j)$ representation under the Lorentz group including all its
derivative descendents to a four point function for scalar fields
$\langle \phi_1(x_1) \phi_2(x_2) \phi_3(x_3) \phi_4(x_4) \rangle$. Assuming
for simplicity  the dimensions of $\phi_i$ satisfy $\Delta_1 = \Delta_2, \,
\Delta_3 = \Delta_4$ and letting $\ell=2j$ the contribution of such a conformal
block is given by $u^{{1\over 2}(\Delta-\ell)} G_\Delta^{(\ell)}(u,v)$
where
\eqn\OPEs{\eqalign{
G_\Delta^{(\ell)}(u,v)
&{}= {1\over z-x} \Big (\big (-\half z\big )^\ell
z F\big (\half(\De+\ell),\half(\De+\ell);\De+\ell;z \big )  \cr
&\qquad \qquad \qquad
{}\times F\big (\half(\De-\ell-2),\half(\De-\ell-2);\De-\ell-2;x\big )
- z \leftrightarrow x \Big ) \, , \cr }
}
where the conformal invariants are as in \defuv\ and $z,x$ are defined
in \uvxz. 
Formally the definition \OPEs\ extends to $\ell <0$ and it
is easy to see that then
\eqn\ml{
\big ( \quar u \big )^{\ell-1} G_{\Delta}^{(-\ell)}(u,v) = -
G_{\Delta}^{(\ell-2)}(u,v) \, .
}
Using \ml\ it is straightforward to obtain \mml\ and \mmll.

The crucial results obtained here depend entirely on the following relation
for hypergeometric functions of the form which appear in \OPEs,
\eqn\hyper{\eqalign{
( 1-\half z) & F(\half \de , \half \de ; \de ; z ) \cr
&{}= F( \half \de -1 , \half \de -1 ; \de -2 ; z ) 
+ {\de^2 \over 16(\de-1)(\de+1)} \, z^2
F(\half \de+1 , \half \de+1 ; \de +2 ; z ) \, . \cr}
}
This is easily verified by considering the power series expansion of both 
sides, or using combinations of standard hypergeometric identities.

For $\Delta=\ell+2$ \OPEs\ gives directly \Gbps. For $\Delta=\ell$
we have, with the definition of $g_\ell$ in  \Gbps,
\eqn\Gell{\eqalign{
G_\ell^{(\ell)}(u,v)
={}& {1\over z-x} \Big (-\half  z(1-\half x) g_\ell(z) 
- z \leftrightarrow x \Big ) \, , \cr
= {}& - \half \big ( g_\ell(z) + g_\ell(x)\big )  -\quar u\,
{{2-z\over z}g_\ell(z) - {2-x\over x}g_\ell(x) \over z-x} \, .\cr}
}
With the aid of \relg, which follows from \hyper, we may
obtain \ggrel.

Other identities also follow from \hyper.
To derive a relation for $-\half(1-v)G_{\Delta}^{(\ell)}(u,v)$. 
we first obtain
\eqn\relone{\eqalign{
& {(\Delta+\ell)^2 \over 8 (\Delta+\ell-1)(\Delta+\ell+1)} \, u \,
G_{\Delta+1}^{(\ell+1)}(u,v)\cr
&{}= -{zx\over z-x} \Big (\big (-\half z\big )^\ell (1-\half z)
F\big (\half(\De+\ell),\half(\De+\ell);\De+\ell;z \big )  \cr
&\qquad \qquad \qquad
{}\times F\big (\half(\De-\ell-2),\half(\De-\ell-2);\De-\ell-2;x\big )
- z \leftrightarrow x \Big ) \cr
& \quad {} - \half u \, G_{\Delta-1}^{(\ell-1)} (u,v)\, , \cr}
}
and hence, since $1-v = z+x -zx$,
\eqn\reltwo{\eqalign{
& -\half(1-v)G_{\Delta}^{(\ell)}(u,v) - 
{(\Delta+\ell)^2 \over 16 (\Delta+\ell-1)(\Delta+\ell+1)} \, u \, 
G_{\Delta+1}^{(\ell+1)}(u,v) - \quar u \, G_{\Delta-1}^{(\ell-1)} (u,v) \cr
&{}= -{1\over z-x} \Big (z^2\big (-\half z\big )^\ell (1-\half x)
F\big (\half(\De+\ell),\half(\De+\ell);\De+\ell;z \big )  \cr
&\qquad \qquad \qquad
{}\times F\big (\half(\De-\ell-2),\half(\De-\ell-2);\De-\ell-2;x\big )
- z \leftrightarrow x \Big ) \, .\cr}
}
Using \hyper\ once again then gives
\eqn\Gvm{\eqalign{
-\half(1-v)G_{\Delta}^{(\ell)}(u,v) = {}& G_{\Delta-1}^{(\ell+1)} (u,v)
+ {(\Delta+\ell)^2 \over 16 (\Delta+\ell-1)(\Delta+\ell+1)} \, u \,
G_{\Delta+1}^{(\ell+1)}(u,v) \cr
{}& + \quar u \, G_{\Delta-1}^{(\ell-1)} (u,v) 
+ {(\Delta-\ell-2)^2 \over 64 (\Delta-\ell-3)(\Delta-\ell-1)} \, u^2 \,
G_{\Delta+1}^{(\ell-1)}(u,v) \, . \cr}
}

We also consider similarly $\half(1+v)G_{\Delta}^{(\ell)}(u,v)$. We first
obtain
\eqnn\relthre
$$\eqalignno{
G_{\Delta}^{(\ell+2)}(u,v) 
= {}& {1\over z-x} \Big (z^2\big (-\half z\big )^{\ell+2} (1-\half x)
F\big (\half(\De+\ell),\half(\De+\ell);\De+\ell;z \big )  \cr
&\qquad \qquad \qquad
{}\times F\big (\half(\De-\ell-2),\half(\De-\ell-2);\De-\ell-2;x\big )
- z \leftrightarrow x \Big ) \cr
{}& - {(\Delta-\ell-2)^2 \over 64 (\Delta-\ell-3)(\Delta-\ell-1)} \, u^2 \,
G_{\Delta+2}^{(\ell)}(u,v) \, , & \relthre \cr}
$$
and hence, with $\half(1+v) - (1-\half x)(1-\half z) = \quar u$,
\eqnn\relfour
$$\eqalignno{
& \half(1+v) G_{\Delta}^{(\ell)}(u,v) \cr
& {} - {(\Delta+\ell)^2 \over 4 (\Delta+\ell-1)(\Delta+\ell+1)} \bigg (
G_{\Delta}^{(\ell+2)}(u,v) + 
{(\Delta-\ell-2)^2 \over 64 (\Delta-\ell-3)(\Delta-\ell-1)} \, u^2 \,
G_{\Delta+2}^{(\ell)}(u,v) \bigg ) \cr
& {}= \quar u \, G_{\Delta}^{(\ell)}(u,v) \cr
& \quad  {} - {1\over z-x} \Big (z\big (-\half z\big )^{\ell} (1-\half x)
F\big (\half(\De+\ell-2),\half(\De+\ell-2);\De+\ell-2;z \big )  \cr
&\qquad \qquad \qquad
{}\times F\big (\half(\De-\ell-2),\half(\De-\ell-2);\De-\ell-2;x\big )
- z \leftrightarrow x \Big ) \, . & \relfour \cr}
$$
With the aid of \hyper\ once more we may then derive
\eqnn\Gvp
$$\eqalignno{
\half(1+v)G_{\Delta}^{(\ell)}(u,v) = {}&  \quar u \, G_{\Delta}^{(\ell)}(u,v)
+  {(\Delta+\ell)^2 \over 4 (\Delta+\ell-1)(\Delta+\ell+1)} \, 
G_{\Delta}^{(\ell+2)}(u,v) \cr
&{} + {(\Delta+\ell)^2 (\Delta-\ell-2)^2\over 256 (\Delta+\ell-1)(\Delta+\ell+1)
(\Delta-\ell-3)(\Delta-\ell-1)} \, u^2\, G_{\Delta+2}^{(\ell)}(u,v) \cr
&{}+ G_{\Delta-2}^{(\ell)} (u,v)
+ {(\Delta-\ell-2)^2 \over 64 (\Delta-\ell-3)(\Delta-\ell-1)} \, u^2 \,
G_{\Delta}^{(\ell-2)}(u,v) \, . & \Gvp \cr}
$$
As a consistency check it is easy to verify that \Gvm\ and \Gvp\ are invariant
under \ml\ and we may also check compatibility with \Gsym.

The results \Gvm\ and \Gvp\ are also true for $\ell=0,1$ by using \mml.
Thus \Gvp\ implies
\eqnn\Gvpp
$$\eqalignno{
\half(1+v)G_{\Delta}^{(0)}(u,v) = {}& 
{\Delta^2 \over 4 (\Delta-1)(\Delta+1)} \,
G_{\Delta}^{(2)}(u,v) \cr
&{} + {\Delta^2 (\Delta-2)^2\over 256 (\Delta-1)^2(\Delta+1)
(\Delta-3)} \, u^2\, G_{\Delta+2}^{(0)}(u,v) \cr
&{}+ G_{\Delta-2}^{(0)} (u,v)
+ {3(\Delta-2)^2 -4 \over 16 (\Delta-3)(\Delta-1)} \, u \,
G_{\Delta}^{(0)}(u,v) \, . & \Gvpp \cr}
$$

For further application in the text we also use \Gvp\ and \Gvm\  twice
to calculate $\quar (1+v)^2 G_{\Delta}^{(\ell)}(u,v) -
\quar (1-v)^2 G_{\Delta}^{(\ell)}(u,v)$ giving 
\eqnn\Gv
\hskip -0.5cm{$$
\eqalignno{
v G_{\Delta}^{(\ell)}(u,v) {}& =G_{\Delta-4}^{(\ell)}(u,v)\cr
&\ {} - {(\Delta+\ell-1)^2 - 5 \over 2(\Delta+\ell-3)(\Delta+\ell+1)}
G_{\Delta-2}^{(\ell+2)}(u,v)
- {(\Delta-\ell-3)^2 - 5 \over 32(\Delta-\ell-5)(\Delta+\ell-1)}
u^2 G_{\Delta-2}^{(\ell-2)}(u,v) \cr
&\ {}+ {(\Delta+\ell)^2(\Delta+\ell+2)^2  \over 16(\Delta+\ell-1)
(\Delta+\ell+1)^2(\Delta+\ell+3)} G_{\Delta}^{(\ell+4)}(u,v)\cr
&\ {}+ {(\Delta-\ell)^2(\Delta-\ell-2)^2  \over 2^{12}(\Delta-\ell-3)
(\Delta-\ell-1)^2(\Delta-\ell+1)} u^4 G_{\Delta}^{(\ell-4)}(u,v)\cr
&\ {}+ {\big ((\Delta+\ell-1)^2 - 5\big )\big ((\Delta-\ell-3)^2 - 5\big ) 
\over 64 (\Delta+\ell-3)(\Delta+\ell+1)(\Delta-\ell-5)(\Delta+\ell-1)}
u^2 G_{\Delta}^{(\ell)}(u,v) \cr
& \ {}- {(\Delta+\ell)^2(\Delta+\ell+2)^2 \big ((\Delta-\ell-3)^2 - 5 \big )
\over 2^9 (\Delta+\ell-1)(\Delta+\ell+1)^2(\Delta+\ell+3)(\Delta-\ell-5)
(\Delta-\ell-1)} u^2 G_{\Delta+2}^{(\ell+2)}(u,v) \cr
& \ {}- {(\Delta-\ell)^2(\Delta-\ell-2)^2 \big ((\Delta+\ell-1)^2 - 5 \big )
\over 2^{13} (\Delta-\ell-3)(\Delta-\ell-1)^2(\Delta-\ell+1)(\Delta+\ell-3)
(\Delta+\ell+1)} u^4 G_{\Delta+2}^{(\ell-2)}(u,v)  \cr
& \ {}+ {(\Delta+\ell)^2(\Delta+\ell+2)^2(\Delta-\ell)^2(\Delta-\ell-2)^2 
\over 2^{16} (\Delta+\ell-1)(\Delta+\ell+1)^2(\Delta+\ell+3)
(\Delta-\ell-3)(\Delta-\ell-1)^2(\Delta-\ell+1)}
u^4 G_{\Delta+4}^{(\ell)}(u,v) \, .  \cr
& & \Gv \cr}
$$}

\appendix{D} {Simplification of strong coupling result}

By virtue of the AdS/CFT correspondence Arutyunov and Frolov \Arut\
calculated the four point function \Fourp\ in the leading large $N$
limit for large $g^2N$. The result is rather complicated and
expressed in terms of conformal integrals over AdS with four 
bulk/boundary propagators. We here define the corresponding functions
of $u,v$ $\oD_{\Delta_1\Delta_2\Delta_3\Delta_4}$ by,
\eqn\Cint{\eqalign{
{\prod_{i=1}^4 \Gamma(\Delta_i) \over \Gamma(\Sigma-\half d)} \, 
{2\over \pi^{{1\over 2}d}} \int_0^\infty \!\!\!\!\! \d z & \int \! \d^d x
\, \prod_{i=1}^4 \bigg ({z \over z^{\,2} + (x-x_i)^2} \bigg )^{\! \Delta_i}\cr
&{} = {r_{14}{}^{\raise 2pt\hbox{$\scriptstyle \!\! \Sigma-\Delta_1-\Delta_4$}}
r_{34}{}^{\raise 2pt\hbox{$\scriptstyle \!\! \Sigma-\Delta_3-\Delta_4$}}
\over r_{13}{}^{\raise 2pt\hbox{$\scriptstyle \!\! \Sigma-\Delta_4$}}
\, r_{24}{}^{\raise 2pt\hbox{$\scriptstyle \!\! \Delta_2$}}} \, 
\oD_{\Delta_1\Delta_2\Delta_3\Delta_4}(u,v) \, ,  \cr}
}
where
\eqn\defSig{
\Sigma=\half \sum_{i=1}^N \Delta_i \, .
}
In \Cint\ $\oD_{\Delta_1\Delta_2\Delta_3\Delta_4}$ is independent of the
dimension $d$. The $\oD$-functions satisfy the identities
\eqna\Dsym
$$\eqalignno{
\oD_{\Delta_1\, \Delta_2\, \Delta_3\, \Delta_4}(u,v) = {}&
\oD_{\Sigma{-\Delta_3}\,\Sigma{-\Delta_4}\,\Sigma{-\Delta_1}\,
\Sigma{-\Delta_2}}(u,v) & \Dsym a \cr
={}& v^{-\Delta_2}\oD_{\Delta_1\, \Delta_2\, \Delta_4\, \Delta_3}(u/v,1/v)
& \Dsym b \cr
={}& v^{\Delta_4-\Sigma} \,
\oD_{\Delta_2\, \Delta_1\, \Delta_3\, \Delta_4}(u/v,1/v) & \Dsym c \cr
={}& v^{\Delta_1+\Delta_4-\Sigma} \,
\oD_{\Delta_2\, \Delta_1\, \Delta_4\, \Delta_3}(u,v) & \Dsym d \cr
={}& \oD_{\Delta_3\, \Delta_2\, \Delta_1\, \Delta_4}(v,u) & \Dsym e \cr
={}& u^{\Delta_3+\Delta_4-\Sigma} \,
\oD_{\Delta_4\, \Delta_3\, \Delta_2\, \Delta_1}(u,v) \, , & \Dsym f \cr}
$$
which reflect permutation symmetries of the basic integral in \Cint.
We also have the relations
\eqn\relD{\eqalign{
(\Delta_2 + \Delta_4 - \Sigma)
\oD_{\Delta_1\, \Delta_2\, \Delta_3\, \Delta_4}(u,v) ={}&
\oD_{\Delta_1\, \Delta_2{+1}\, \Delta_3\, \Delta_4{+1}}(u,v) -
\oD_{\Delta_1{+1}\, \Delta_2\, \Delta_3{+1}\, \Delta_4}(u,v) \, , \cr
(\Delta_1 + \Delta_4 - \Sigma)
\oD_{\Delta_1\, \Delta_2\, \Delta_3\, \Delta_4}(u,v) ={}&
\oD_{\Delta_1{+1}\, \Delta_2\, \Delta_3\, \Delta_4{+1}}(u,v) -
v \oD_{\Delta_1\, \Delta_2{+1}\, \Delta_3{+1}\, \Delta_4}(u,v) \, , \cr
(\Delta_3 + \Delta_4 - \Sigma)
\oD_{\Delta_1\, \Delta_2\, \Delta_3\, \Delta_4}(u,v) ={}&
\oD_{\Delta_1\, \Delta_2\, \Delta_3{+1}\, \Delta_4{+1}}(u,v) - u
\oD_{\Delta_1{+1}\, \Delta_2{+1}\, \Delta_3\, \Delta_4}(u,v) \, , \cr}
}
and
\eqn\relDa{\eqalign{
\Delta_4 \oD_{\Delta_1\, \Delta_2\, \Delta_3\, \Delta_4}(u,v) ={}&
\oD_{\Delta_1\, \Delta_2\, \Delta_3{+1}\, \Delta_4{+1}}(u,v)
+ \oD_{\Delta_1\, \Delta_2{+1}\, \Delta_3\, \Delta_4{+1}}(u,v) \cr
&{}+ \oD_{\Delta_1{+1}\, \Delta_2\, \Delta_3\, \Delta_4{+1}}(u,v) \, .\cr }
}
There are also relations which arise since if $\Delta_i=0$ the
integral \Cint\ reduces to a three point function. Thus
\eqn\Din{
\Delta_2 \oD_{\Delta_1\, \Delta_2\, \Delta_3\, \Delta_4}
\big |_{\Delta_2=0} = \Gamma(\Sigma-\Delta_1) \Gamma(\Sigma-\Delta_3)
\Gamma(\Sigma-\Delta_4) \, ,
}
and by using \Dsym{a} we have
\eqn\Dz{
(\Sigma-\Delta_4) \oD_{\Delta_1\, \Delta_2\, \Delta_3\, \Delta_4}
\big |_{\Sigma-\Delta_4=0} = \Gamma(\Delta_1) \Gamma(\Delta_2) 
\Gamma(\Delta_3) \, .
}
{}From \relDa\ and the sum of eqs.\relD\ this may be rewritten as
\eqn\DDz{\eqalign{
(\oD_{\Delta_1{+1}\, \Delta_2\, \Delta_3{+1}\, \Delta_4} {}&
+ u\oD_{\Delta_1{+1}\, \Delta_2{+1}\, \Delta_3\, \Delta_4}
+ v\oD_{\Delta_1\, \Delta_2{+1} \, \Delta_3{+1}\, \Delta_4})
\big |_{\Delta_4=\Delta_1+\Delta_2+\Delta_3} \cr
&{} = \Gamma(\Delta_1) \Gamma(\Delta_2) \Gamma(\Delta_3) \, . \cr}
}

For compatibility with superconformal symmetry the results obtained
in \Arut\ by using the AdS/CFT correspondence must be expressible
in the form \Solac\ in terms of a single function $\F$. This was
demonstrated by Eden {\it et al} \Edent\ by direct calculation
starting from
\eqn\Done{
\oD_{1111}(u,v) = \Phi^{(1)}(u,v) \, ,
}
where $ \Phi^{(1)}$ is given here by  \Pone.  We show here that this
requirement also follows from the above $\oD$ identities and that the 
resulting expression for $\F$ may be simplified to a single $\oD$
function. 

For $a_1$ in \Fourp, with the normalisation
conventions as in \Solac, expression obtained in \Arut\
may be written in terms of $\oD$ functions in the form
\eqn\Naone{
a_1 =  \quar \hN^2 + \half\hN u^2 \big ( -\oD_{2211} + 3\oD_{2222}
+ (1+v-u) \oD_{3322} \big ) \, .
}
Using \Dsym{f} this gives from \Solac,
\eqn\FDDD{
\F = - \oD_{1122} + 3u \oD_{2222} + (1+v-u) \oD_{2233} \, .
}
Now using \relD\ to eliminate $u,v$ from \FDDD\ and also for
$\oD_{2233} - \oD_{1324}$ we get
\eqn\FDD{
\F = - 4\oD_{1122} + 5 \oD_{1133} - \oD_{2123} - \oD_{1223} 
+ \oD_{1324} + \oD_{3124} - \oD_{1144} \, .
}
{}From \relDa\ we may obtain $\oD_{1324} + \oD_{3124} - \oD_{1144}
= 3(\oD_{2123} + \oD_{1223} - \oD_{1133}) - 2 \oD_{2224}$ so that
\FDD\ becomes
\eqn\FD{
\F =  - 4\oD_{1122} + 2( \oD_{1133} + \oD_{2123} +  \oD_{1223} ) -
2 \oD_{2224} =  - 2 \oD_{2224} \, ,
}
using \relDa\ again. Thus we obtain \FDs\ where
$\F$ is reduced to a single $\oD$ function.
The result of \Arut\ for $c_1$ may similarly be expressed as
\eqn\cAr{\eqalign{
c_1 = {}& \hN \big ( uv (\oD_{2211} + 2 \oD_{2233} -3 \oD_{2222})
-v(1-v) \oD_{2222} \cr
&\qquad {} + \half( \oD_{1212} + \oD_{2121} - \oD_{2112}
- v \oD_{1221}) + u \leftrightarrow v \big ) \cr
={}&  \hN \big ( 2uv(\oD_{2233}+\oD_{3223})+ (v^2+u^2-v-u-6uv) \oD_{2222} \cr
&\qquad {} + (u+v) \oD_{1212} + (u-v)(\oD_{2112}- \oD_{1122} ) \big ) \, , \cr}
}
using \Dsym{d,e,f}. To simplify \cAr\ we first consider the terms with
$\Sigma=5$,
\eqn\sfive{\eqalign{
2uv(\oD_{2233}+\oD_{3223}) = {}& - u \oD_{3214} - v \oD_{1234} -(u+v)
\oD_{2123}+ u \oD_{3113} + v \oD_{1133} \, , \cr
- u \oD_{3214} - v \oD_{1234}  = {}& (u+v) \oD_{2224} +
u \oD_{2314} + v \oD_{1324} - 3u  \oD_{2213} - 3v \oD_{1223} \cr
= {}& 1 - (1-u-v) \oD_{2224} - 3u  \oD_{2213} - 3v \oD_{1223} \, , \cr}
}
where in the last line we have used \DDz\ for $\Delta_1=\Delta_3=1, \,
\Delta_2=2$. The $\Sigma=4$ terms may then be written using \relD\ as
\eqnn\sfour
$$\eqalignno{
-(u+v)& \oD_{2123}+ u\oD_{3113} + v\oD_{1133} - 3u \oD_{2213} - 3v \oD_{1223}
+ (v^2+u^2-v-u-6uv) \oD_{2222}\cr
=&{} -(u+v)(\oD_{2213}+\oD_{1223}+\oD_{1313}) +
(u-2v)(\oD_{1133}+\oD_{1223}+\oD_{2123}) \cr
&{}+ (v-2u)(\oD_{3113}+\oD_{2213}+\oD_{2123}) \cr
&{}- (u-3v)\oD_{1122} -(v-3u)\oD_{2112} + (u+v) \oD_{1212} \, . &\sfour  \cr}
$$
Using \relDa\ the $\Sigma=3$ terms cancel leaving just the result in
\Solac\ with \FDs\ again.

It is perhaps worth noting that if $\F$ is expressible as a single
$\oD$ function then the symmetry properties \Dsym{c,e} determine
\eqn\FD{
\F(u,v) = K \, \oD_{1{+s}\,1{+s}\,1{+s}\,1{+3s}}(u,v) \, ,
}
depending on a single parameter $s$,
which includes both the weak and strong coupling results for $s=0,1$.
\listrefs
\bye